\documentclass[11 pt,oneside]{article}
\usepackage{amsmath,amssymb,amsfonts,amsopn}
\usepackage{color,graphicx}
\usepackage[numbers]{natbib}

\usepackage{amsthm}
\usepackage{subfig}
\usepackage{float}
\usepackage{listings}
\usepackage{multirow}
\usepackage{paralist}
\usepackage{enumitem}
\usepackage{algorithm}
\usepackage{hyperref}
\usepackage[noend]{algpseudocode}
\usepackage{tikz}

\newcommand{\bone}{\boldsymbol{1}}
\newcommand{\br}{\boldsymbol{r}}

\newcommand{\bx}{\boldsymbol{x}}
\newcommand{\by}{\boldsymbol{y}}
\newcommand{\bz}{\boldsymbol{z}}

\newcommand{\bT}{\boldsymbol{T}}
\newcommand{\bV}{\boldsymbol{V}}
\newcommand{\bI}{\boldsymbol{I}}
\newcommand{\bX}{\boldsymbol{X}}

\newcommand{\bZ}{\boldsymbol{Z}}

\newcommand{\bB}{\boldsymbol{B}}

\newcommand{\bP}{\boldsymbol{P}}

\newcommand{\bal}{\boldsymbol{\alpha}}
\newcommand{\bbe}{\boldsymbol{\beta}}
\newcommand{\bmu}{\boldsymbol{\mu}}
\newcommand{\bves}{\boldsymbol{\varepsilon}}

\newcommand{\bth}{\boldsymbol{\theta}}
\newcommand{\bga}{\boldsymbol{\gamma}}

\newcommand{\hx}{\hat{x}}
\newcommand{\hy}{\hat{y}}

\newcommand{\hU}{\hat{U}}

\newcommand{\hmu}{\hat{\mu}}
\newcommand{\hsi}{\hat{\sigma}}

\newcommand{\hth}{\hat{\theta}}

\newcommand{\tsi}{\tilde{\sigma}}

\newcommand{\tbX}{\widetilde{\boldsymbol{X}}}
\newcommand{\tby}{\widetilde{\boldsymbol{y}}}

\newcommand{\tbves}{\tilde{\boldsymbol{\varepsilon}}}

\newcommand{\hbx}{\widehat{\boldsymbol{x}}}
\newcommand{\hby}{\widehat{\boldsymbol{y}}}
\newcommand{\hbz}{\widehat{\boldsymbol{z}}}
\newcommand{\hbB}{\widehat{\boldsymbol{B}}}
\newcommand{\hbX}{\widehat{\boldsymbol{X}}}

\newcommand{\hbZ}{\widehat{\boldsymbol{Z}}}

\newcommand{\hbmu}{\hat{\boldsymbol{\mu}}}

\newcommand{\hbga}{\hat{\boldsymbol{\gamma}}}
\newcommand{\hbth}{\hat{\boldsymbol{\theta}}}
\newcommand{\hbSi}{\hat{\boldsymbol{\Sigma}}}

\newcommand{\htbX}{\widehat{\widetilde{\boldsymbol{X}}}}
\newcommand{\htby}{\widehat{\widetilde{\boldsymbol{y}}}}

\newcommand{\sqdiamond}[1][]{\tikz [x=1ex,y=1 ex,line width=.1ex,line join=round, yshift=0.285ex] \draw  [#1]  (0,.71) -- (.71,1.42) -- (1.42,.71) -- (.71,0) -- (0,.71) -- cycle;}%
\newcommand{\MyDiamond}[1][]{\mathop{\raisebox{-0.275ex}{$\sqdiamond[#1]$}}}

\newenvironment{compactlist}{
	\begin{list}{}{
			\setlength{\partopsep}{6pt}
			\setlength{\parskip}{0pt}
			\setlength{\parsep}{4pt}
			\setlength{\topsep}{6pt}
			\setlength{\itemsep}{0pt}
			\setlength{\itemindent}{0pt}
			\setlength{\leftmargin}{0pt}
		}
	}{
	\end{list}
}

\title{{Robust variable screening for regression using factor profiling}}

\author{Yixin Wang\footnote{Email: Yixin.Wang@kuleuven.be}\ \ and Stefan Van Aelst\footnote{Corresponding author. Email: Stefan.VanAelst@kuleuven.be, Phone: +3216372383, Fax: +3216327998} \\
Department of Mathematics, KU Leuven,\\ Celestijnenlaan 200 B, 3001 Leuven, Belgium}

\usepackage{fancyhdr}
\begin{document}

	\maketitle
	{\bf Abstract:} Sure Independence Screening is a fast procedure for variable selection in ultra-high dimensional regression analysis. Unfortunately, its performance greatly deteriorates with increasing dependence among the predictors. To solve this issue, Factor Profiled Sure Independence Screening (FPSIS) models the correlation structure of the predictor variables, assuming that it can be represented by a few latent factors. The correlations can then be profiled out by projecting the data onto the orthogonal complement of the subspace spanned by these factors. 
However, neither of these methods can handle the presence of outliers in the data.  Therefore, we propose a robust screening method which uses a least trimmed squares method to estimate the latent factors and the factor profiled variables. Variable screening is then performed on factor profiled variables by using regression MM-estimators. Different types of outliers in this model and their roles in variable screening are studied. Both simulation studies and a real data analysis show that the proposed robust procedure has good performance on clean data and outperforms the two nonrobust methods on contaminated data.


	\section{Introduction}
	Advances in many areas, such as genomics, signal processing, image analysis and finance, call for new approaches to handle high dimensional data problems. 
	Consider the multiple linear regression model:
	\begin{equation}
	\by = \bX\bth+\bves,
	\label{RegY}
	\end{equation}
	where $\bX=(\bX_1,\ldots,\bX_p)\in\mathbb{R}^{n\times p}$ is the design matrix that collects $n$ independently and identically distributed (IID) observations $\bx_i\in\mathbb{R}^p$ ($i=1,\ldots,n$) as its rows, $\by\in\mathbb{R}^{n}$ collects the $n$ responses and  $\bves\in\mathbb{R}^n$ is the noise term. 
	The model is called ultra-high dimensional if the number of variables $p$ grows exponentially with the number of observations $n$ ($p\gg n$). In (ultra-)high dimensional settings it is common to assume that only very few predictors contribute to the response. In other words, the coefficient vector $\bth$ is assumed to be sparse, meaning that most of its elements are equal to zero. A major goal is then to identify all the important variables that actually contribute to the response. 
	
	Variable selection plays an essential role in modern statistics. Widely used classical variable selection techniques are based on the Akaike~\citep{Akaike1973,Akaike1974} and Bayesian information criteria~\citep{Schwarz1978}. However, they are unsuitable for high dimensional data due to their high computational cost. Penalized least squares (PLS) methods have gained a lot of popularity in the past decades, such as nonnegative garrote~\citep{Garrote1995,Garrote2007}, the least absolute shrinkage and selection operator (Lasso)~\citep{Lasso1996,Lasso2006}, adaptive Lasso~\citep{AdaptLasso2006}, bridge regression~\citep{Bridge1998,Bridge2007}, elastic net~\citep{Elastic2005} and smoothly clipped absolute deviation (SCAD)~\citep{SCAD2001,SCAD2007} among others. Many of these methods are variable selection consistent under the condition that the sample size $n$ is larger than the dimension $p$. Although it has been proven that lasso-type estimators can also select variables consistently for ultra-high dimensional data, this was studied under the irrepresentable condition on the design matrix~\citep{Lasso2006,Garrote2007}. As pointed out in~\citep{Lasso2006}, correct model selection for Lasso cannot be reached in ultra-high dimensions for all error distributions, e.g. when higher moments of the error distribution do not exist. Moreover, all these techniques have super-linear (in $p$) computational complexity which makes them computationally prohibitive in ultra-high dimensional settings~\citep{ExSIS}.
	
	Sure Independence Screening (SIS) is a very fast variable selection technique for ultra-high dimensional data~\citep{SIS}. 
	SIS has the sure screening property which means that under certain assumptions all the important variables can be selected with probability tending to 1. The basic idea is to apply univariate least squares regression for each predictor variable separately, to measure its marginal contribution to the response variable. Define $\mathcal{M}_F=\{1,\ldots,p\}$ as the full model, $\mathcal{M}_T=\{j:\theta_{j}\neq 0\}$ as the true model, and $\mathcal{M}_{q^*}=\{j_1,\ldots,j_{q^*}\} \subset \mathcal{M}_F$ as a candidate model of size $|\mathcal{M}|=q^*$. Denote by $\hth_j$ the $j$th simple regression coefficient estimate, i.e.
	\begin{equation}
	\hth_j = (\bX_j^\text{T}\bX_j)^{-1}\bX_j^\text{T}\by. \nonumber
	\end{equation}
	SIS then selects a model of size $q$ as
	\begin{equation}
	\mathcal{M}_{q}=\{1\leqslant j\leqslant p:|\hth_j|\text{ is among the first $q$ largest of all}\}. \nonumber
	\end{equation}
	The model size $q$ usually is of order $\mathcal{O}(n)$.  
	When the variables are  standardized componentwise, the regression coefficient estimate $\hth_j$ equals the marginal correlation between $\bX_j$ and $\by$. Hence, SIS is also called correlation screening. SIS can reduce the dimensionality from a large scale (e.g. $\mathcal{O}(\exp{(n^{\xi})})$ with $0<\xi<1$) to a moderate scale (e.g. $\mathcal{O}(n)$) while retaining all the important variables with high probability, which is called the sure screening property. Applying variable selection or penalized regression on this reduced set of variables rather than the original set then largely improves the variable selection results.
	
To guarantee the sure screening property for a reduced model of moderate size, SIS assumes that the predictors are independent, which is a strong assumption in high dimensional settings. In case of correlation among the predictors the number of variables that is falsely selected by SIS can increase dramatically. As shown in~\citet{Tilt}, in this case the estimate $\hth_j$ can be written as $\theta_j$ plus a bias term
	\begin{equation} 
	\hth_j = \bX_j^\text{T}\by = \bX_j^\text{T}\left( \sum_{k=1}^{p}\theta_k\bX_k + \bves \right) = \theta_j + \underset{\text{bias}}{\underbrace{\underset{k\in\mathcal{M}_T\backslash \{ j\}}{\sum} \theta_k \bX_j^\text{T}\bX_k +\bX_j^\text{T}\bves}}.
	\label{Bias}
	\end{equation}
	Hence, the higher the correlation of $\bX_j$ with other important predictors, the larger the bias of $\hth_j$. Moreover, correlation between $\bX_j$ and the error $\bves$ introduces bias on $\hth_j$ as well. Even when the predictors are IID Gaussian variables, so-called spurious correlations can be non-ignorable in high dimensional settings~\citep{SIS}. To handle correlated predictors, several methods have been developed, such as Iterative SIS~\citep{SIS}, Tilted Correlation Screening (TCS)~\citep{Tilt}, Factor Profiled Sure Screening (FPSIS)~\citep{FPSIS}, Conditional SIS~\citep{CSIS}, and High Dimensional Ordinary Least Squares Projection (HOLP)~\citep{HOLP}. A common feature shared by these methods is that they try to remove the correlation among the predictors before estimating their marginal contribution to the response.
	
	Although the aforementioned methods work well on clean data, none of these methods can resist the adverse influence of potential outliers. On the other hand, robust regression estimators, such as M-estimators~\citep{Huber1981}, S-estimators~\citep{RousseeuwYohai1984}, MM-estimators~\citep{Yohai1987} and the LTS-estimator~\citep{Rousseeuw1984} cannot be applied when $p>n$. To handle contamination in high dimensional regression problems, penalized robust estimators such as penalized M-estimators~\citep{Geer2008,Li2011}, penalized S-estimators~\citep{Maronna2011}, penalized MM-estimators~\citep{Maronna2011,Smucler2015}, LAD-Lasso~\citep{LAD-Lasso}, LTS-lasso~\citep{Alfons2013}, the enet-LTS estimator~\citep{Filzmoser2018}, and the Penalized Elastic Net S-Estimator (PENSE)~\citep{Smucler2018} have been proposed, as well as a robustified LARS algorithm~\citep{Khan2007}. Similarly as their classical counterparts, these methods cannot handle ultra-high dimensional problems. 
	
	
	To deal with ultra-high dimensional regression problem with outliers, more robust variable screening methods have been developed. Robust rank correlation screening (RRCS)~\citep{RRCS2012} replaces the classical correlation measure with Kendall's $\tau$ estimator in SIS. In~\citep{Filzmoser2014} A trimmed SIS-SCAD, called TSIS-SCAD, has been proposed which replaces the maximum likelihood and the penalized maximum likelihood estimator in SIS-SCAD with their trimmed versions. An iterative algorithm which combines SIS and the C-step for LTS regression estimator~\citep{Rousseeuw2006} has been developed in~\citep{TWang2018}. Although iterative versions of RRCS and TSIS-SCAD have been introduced for the case of correlated predictors, similarly to iterated SIS they may fail when a considerable proportion of the predictors are correlated. 
	
	In this paper, we propose a fast robust procedure for ultra-high dimensional regression analysis based on FPSIS, called Robust Factor Profiled Sure Independence Screening (RFPSIS). FPSIS can be seen as a combination of factor profiling and SIS. It assumes that the predictors can be represented by a few latent factors. If these factors can be obtained accurately, then the correlations among the predictors can be profiled out by projecting all the variables onto the orthogonal complement of the subspace spanned by the latent factors. Performing SIS on the profiled variables rather than the original variables then improves the screening results. FPSIS possesses the sure screening property and even variable selection consistency~\citep{FPSIS}. However, the method can break down with even a small amount of contamination in the data. Different types of outliers can be defined based on the factor model and regression model. To avoid the impact of potential outliers on the factor model, RFPSIS estimates the latent factors using a Least Trimmed Squares method proposed in~\citep{Maronna2005}. Based on the robustly estimated low-dimensional factor space we identify vertical outliers and four types of potential leverage points in the multiple regression model, and examine their roles in the marginal factor profiled regressions. After removing bad leverage points, the marginal regression coefficients are estimated using a 95\% efficient MM-estimator. Finally, a modified BIC criterion is used to determine the final model.
	
	The rest of this paper is organized as follows. In Section 2, we first review the factor profiling procedure and the LTS method to estimate the factor space. We study the effect of different types of outliers on the models and introduce the Robust FPSIS method. We then compare SIS, FPSIS and RFPSIS by simulation. We consider several modified BIC criteria for final model selection in Section~3 and compare their performance. Section 4 provides a real ultra-high dimensional dataset analysis while Section 5 contains conclusions.
	
\section{Robust FPSIS}
\label{sec:RFPSIS}
	\subsection{Factor profiling}
	FPSIS aims to construct decorrelated predictors. It assumes that the correlation structure of the predictors can be represented by a few latent factors. We now summarize the model proposed in~\citep{FPSIS}. The factor model for the predictors is given by
	\begin{equation}
	\bX =  \bZ\bB^\text{T} + \tbX,
	\label{RegX}
	\end{equation}
	under the constraint $\bZ^\text{T}\bZ=\bI_d$, where $\bZ\in\mathbb{R}^{n\times d}$ collects the $d$-dimensional latent factor scores as its rows, $\bB\in\mathbb{R}^{p\times d}$ is the factor loading matrix which specifies the linear combinations of the factors involved in each of the predictors  $\bX_j$.
	Finally,
	$\tbX=(\tbX_{1},\ldots,\tbX_{p})\in\mathbb{R}^{n\times p}$ contains the information in $\bX$ which is missed by $\bZ$. It is assumed that $E(\by)=E(\bX_j)=E(\tbX_j)=0$ and $\text{var}(\by)=\text{var}(\bX_j)=1\geqslant\tsi_j^2=\text{var}(\tbX_j)$. Moreover, it is assumed that $\text{cov}(\tbX)$ is a diagonal matrix, so $\text{cov}(\tbX_{j_1},\tbX_{j_2})=0$ for $j_1\neq j_2\in\{1,\ldots,p\}$. The error term is allowed to be correlated with the predictors, but only through the latent factors. It is modeled by
	\begin{equation}
	\bves = \bZ\bal+\tbves,
	\label{RegE}
	\end{equation}
	where $\boldsymbol{\alpha}\in\mathbb{R}^d$ is a $d$-dimensional vector and $\tbves$ is independent of both $\bZ$ and $\tbX$. The two factor models \eqref{RegX} and \eqref{RegE} allow us to profile out the correlations introduced by the latent factors, both among the predictors and with the error term. The resulting $\tbX_j$'s and $\tbves$ are called profiled predictors and error variable, respectively.   
	
	By writing $\bga=\bB^\text{T}\bth+\bal\in\mathbb{R}^d$, one can define the profiled response variable as $\tby=\by-\bZ^\text{T}\bga$. Using equations \eqref{RegX}-\eqref{RegE}, the regression model~\eqref{RegY} can then be modified to
\begin{equation}
	\tby=\by-\bZ\bga=\tbX\boldsymbol{\theta}+\tbves,
	\label{RegFP}
	\end{equation}
	which has uncorrelated predictors and error term. 

	\subsection{Robustly fitting the factor model}
	\label{sec:FS}
To estimate the latent factors $\bZ$, in~\citep{FPSIS} the least squares type objective function
	\begin{equation}
	\mathcal{O}(\bZ,\bB) = \|\bX-\bZ\bB^\text{T}\|_E^2\,,
	\label{LSFP}
	\end{equation}
is minimized under the constraint $\bZ^\text{T}\bZ=\bI_d$, where $\|\cdot\|_E$ denotes the Euclidean norm. Let $\hbZ$ and $\hbB$ denote minimizers of \eqref{LSFP}. Then $\bX$ can be approximated by $\hbX = \hbZ\hbB^\text{T}$ which is a low-dimensional approximation of $\bX$ in a $d$-dimensional  subspace. The optimal solution to (6) is not unique, but one solution is given by $\hbZ =(\hU_1,\ldots,\hU_d )^\text{T}\in\mathbb{R}^{n\times d}$, where $\hU_j$ is the $j$th leading eigenvector of the matrix $\bX\bX^\text{T}$~\citep[see][]{FPSIS}.		

Note that minimization of objective function~\eqref{LSFP} is closely related to dimension reduction by principal component analysis.
Indeed, the $d$ first principal components of the centered matrix $\bX$ are obtained by minimizing 
\begin{equation}
\mathcal{O}(\bT,\bV) = \|\bX-\bT\bV^\text{T}\|_E^2\,,
\label{LSPCA}
\end{equation}
under the constraint $\bV^\text{T}\bV = \bI_{d}$, where $\bV\in \mathbb{R}^{p\times d}$ contains the PC loadings as its columns and $\bT\in\mathbb{R}^{n\times d}$ is the corresponding PC score matrix. Clearly, the objective functions in~\eqref{LSFP} and~\eqref{LSPCA} are the same, but this objective function is optimized under different constraints in both cases. 
The constraint $\bZ^\text{T}\bZ=\bI_d$ for~\eqref{LSFP} yields uncorrelated latent factors while for~\eqref{LSPCA} the constraint $\bV^\text{T}\bV = \bI_{d}$ yields uncorrelated principal components. Both solutions can immediately be derived from a singular value decomposition of the matrix $\bX$ and yield the same approximation $\hbX$ of $\bX$.


It is well-known that LS-estimation is very sensitive to outliers. Observations that lie far away from the true subspace may pull the estimated subspace toward them if least squares is applied. Using the notation $r_{ij}=x_{ij}-\hx_{ij}$, the objective function \eqref{LSFP} can be written as
\begin{equation}
\mathcal{O}(\bZ,\bB)= \sum_{i=1}^{n}(\|\br_i\|_E^2),
\label{ResCl}
\end{equation}
with $\br_i = (r_{i1},\ldots,r_{ip})^\text{T}$.
To downweight the influence of potential outliers, the LS objective function in \eqref{ResCl} can be replaced by a Least Trimmed Squares (LTS) objective function~\citep{Maronna2005}. The LTS objective function is the sum of squared residuals over the observations with the $h$ smallest residuals $\|\br_i\|_E$. That is, 
\begin{equation} 
\mathcal{O}(\bZ,\bB,\bmu)=\sum_{i=1}^{h}(\|\br_i\|_E^2)_{i:n}=\sum_{i=1}^{h}(\|\bx_i-\bB\bz_i-\bmu\|_E^2)_{i:n},
\label{ResMLTS1}
\end{equation}
with $[(n-d+2)/2]\leqslant h < n$, where $\bz_i$ is the $i$th row of $\bZ$, $\bmu$ is a robust location estimator, and $(\cdot)_{i:n}$ means the $i$th smallest value of an ordered sequence. To obtain the latent factors, we first minimize \eqref{ResMLTS1} without constraint and then orthogonalize $\hbZ$ afterwards. 

To solve~\eqref{ResMLTS1}, we use a computationally efficient algorithm that has been developed recently, see~\citep{Cevallos2016,MLTS-PCA}. A brief summary of this LTS algorithm can be found in the Supplemental Material. Similarly as in~\citep{robpca}, to further speed up the procedure we use singular value decomposition to represent the data matrix $\bX$ in the subspace spanned by the $n$ observations before estimating the factor subspace using the LTS algorithm. We thus first reduce the data space $\bX$ to the affine subspace of dimension $r=\text{rank}(\bX-\bone_n\bar{\bx}^\text{T})$ where $\bar{\bx}$ is the columnwise mean of $\bX$. Denote the new matrix as $\bX^* \in\mathbb{R}^{n\times r}$. By applying the LTS algorithm on $\bX^*$, we obtain estimates  $(\hbZ^*,\hbB^*, \hbmu^*)$, with $\hbZ^*\in\mathbb{R}^{n\times d}$, $\hbB^*\in\mathbb{R}^{r\times d}$ and $\hbmu^*\in\mathbb{R}^r$. The final solution is given by $(\hbZ^*, \bP\hbB^*, \bP\hbmu^*+\bar{\bx})$, where $\bP\in \mathbb{R}^{p\times r}$ is the projection matrix from the initial singular value decomposition. To simplify the notation, we write the final output of the LTS algorithm as $(\hbZ, \hbB, \hbmu)$.

To refine the estimation of the factor model, we apply two reweighting steps to the initial solution obtained by the LTS algorithm. The first step improves the estimation of the low dimensional subspace spanned by the latent factors and the second step increases the accuracy of the robust location estimate $\hbmu$. For these reweighting steps, we need to identify outliers in the data with respect to the assumed factor model~\eqref{RegX}. Therefore, following~\citet{robpca} we first introduce two distances of an observation with respect to a given subspace. The orthogonal distance (OD) of an observation $\bx_i$ measures the distance of that observation to the subspace. It is thus given by 
$\text{OD}_i = \|\br_i\|_E$.
On the other hand, the score distance (SD) of an observation $\bx_i$ measures the distance between its approximation $\hbx_i$ in the subspace to the center of the subspace and is given by
$\text{\small SD}_i =\|\bz_i\|_E$.

\begin{figure}[ht!]
	\centering
	\begin{minipage}{0.45\textwidth}
		\centering
		\footnotesize
		(a) PC outlier
	\end{minipage}
	\begin{minipage}{0.45\textwidth}
		\centering
		\footnotesize
		(b) OC outlier
	\end{minipage}\\
	\begin{minipage}{0.45\textwidth}
		\centering
		\includegraphics[width=5.5 cm]{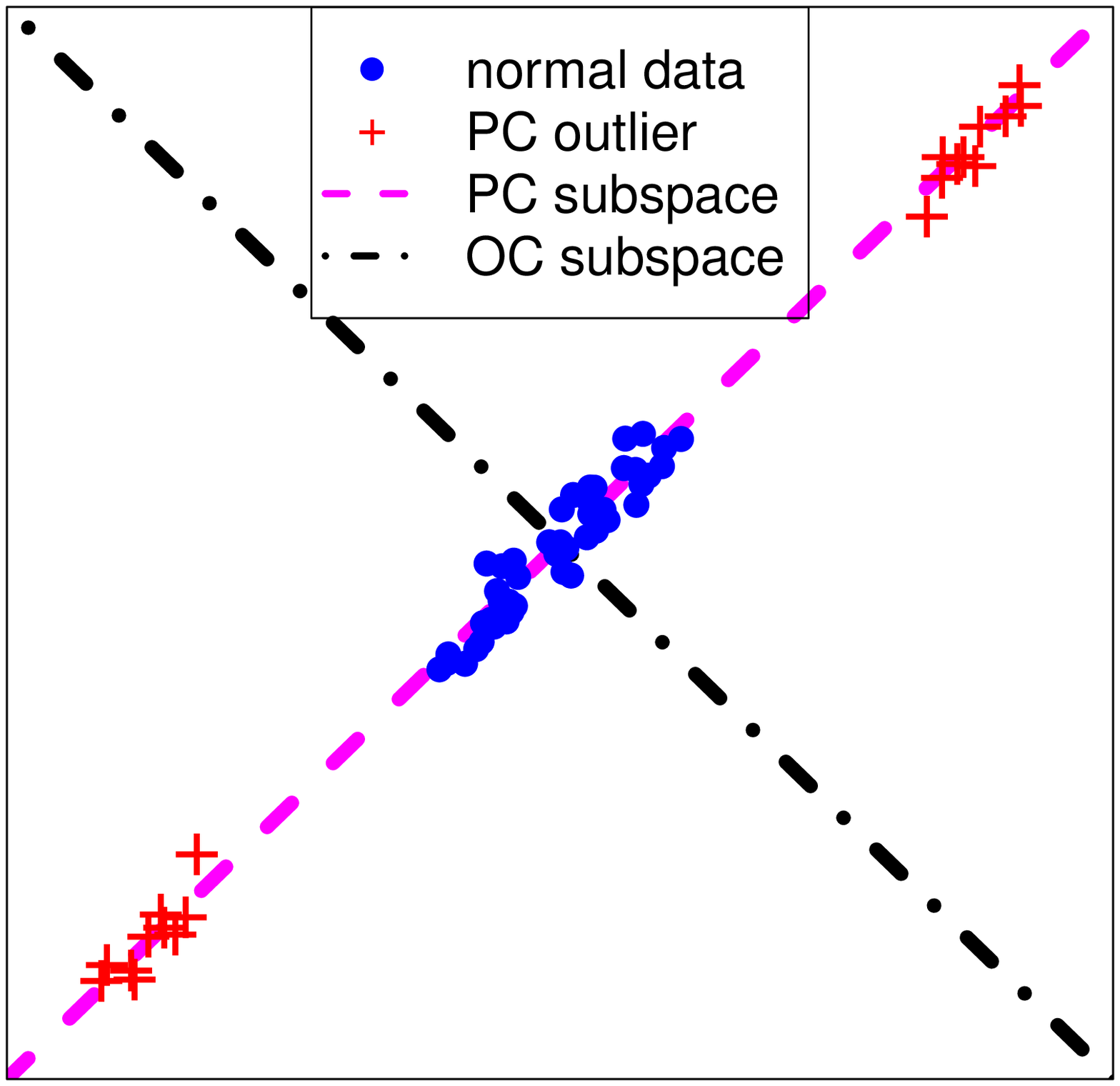}
	\end{minipage}
	\begin{minipage}{0.45\textwidth}
		\centering
		\includegraphics[width=5.5 cm]{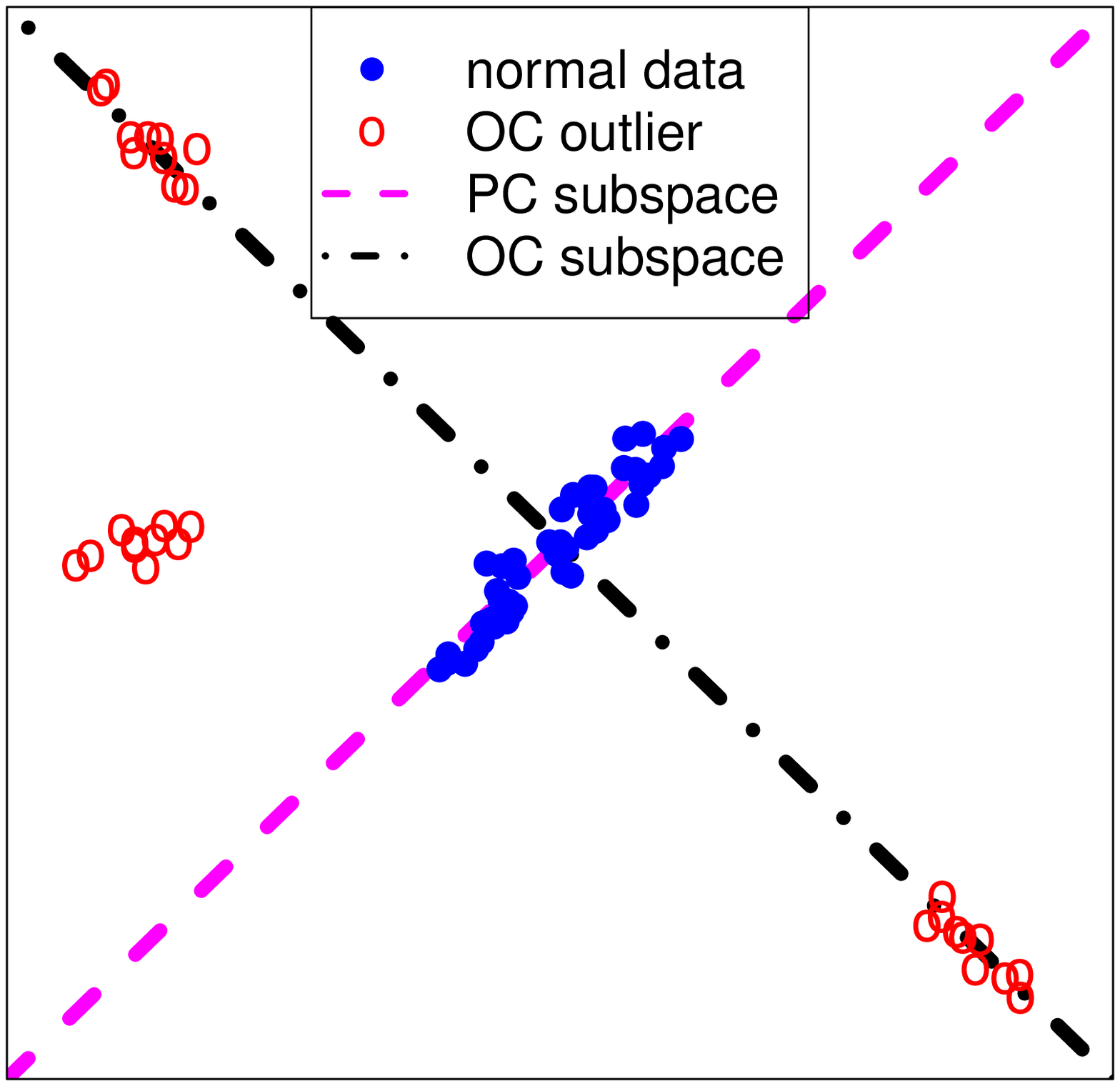}
	\end{minipage}
	\caption{\small PC outliers and OC outliers: (a) Normal data ($\protect \bullet$) and PC outliers (“+”); (b) Normal data ($\protect \bullet$) and OC outliers (“o”).}
	\label{PC_OC_outlier}
\end{figure}

Based on the orthogonal distance we can identify {\it OC outliers} which are observations that lie far from the subspace and thus are outlying in the orthogonal complement (OC) of the subspace, i.e. the OC subspace~\citep{She2016}. Based on the score distance within the subspace we can identify {\it score outliers}, also called  {\it PC outliers} in~\citep{She2016}, which are observations that lie far from the center within the subspace. Following~\citep{robpca} we call a score outlier a {\it good leverage point} if it is outlying within the subspace, but does not lie far from the subspace. A score outlier is called a {\it bad leverage point} if it is not only outlying within the subspace, but at the same time is an OC outlier. The plots in Figure~\ref{PC_OC_outlier} show examples of PC and OC outliers in case of bivariate data and a one-dimensional subspace.

{\bf Reweighted subspace estimation}. The $n-h$ observations with the largest squared residuals are excluded in the least trimmed squares objective function~\eqref{ResMLTS1}. Smaller values of $h$ yield more robustness, but also a lower efficiency because many observations are excluded. To increase the statistical efficiency, we identify the OC outliers, and re-estimate the factor subspace by applying least squares on the subset of observations that is obtained by removing the OC outliers. 
Unfortunately, the distribution of the orthogonal distances for the regular data is generally not known, so it is not straightforward to define a cutoff value to identify OC outliers. 
To overcome this issue we use a robust version of the Yeo-Johnson transformation~\citep{Yeo2000}, proposed in~\citep{RobT2010}. The orthogonal distances are first standardized robustly by using their median and $Q_n$ scale estimate, that is
\[ d_i=\frac{\|\br_i\|_E-\text{med}(\|\br_i\|_E)}{Q_n(\|\br_i\|_E)}. \]
 Then, we apply the Yeo-Johnson transformation
		 \begin{equation}
		 \psi(\lambda,d) = \scalebox{2}{\Bigg\{ }
		 \begin{tabular}{cc}
		 $((d+1)^\lambda - 1)/\lambda$  \quad & if $\lambda \neq 0$ and $d\geqslant 0 $ \\
		 $\log(d+1)$ \quad & if $\lambda = 0$ and $d\geqslant 0$ \\
		 $-((-d+1)^{2-\lambda}-1)/(2-\lambda)$ & if $\lambda \neq 2$ and  $d <0$ \\
		 $-\log(-d+1)$  & if $\lambda =2$ and $d<0$
		 \end{tabular}
		 \end{equation}
	to the standardized orthogonal distances $d_i$ for a grid of $\lambda$ values. The optimal value of $\lambda$ is selected by maximizing the trimmed likelihood
		 \begin{equation}
		 L_\text{Trim}(\lambda) = \sum_{i=1}^{h}l(\lambda;d_i)_{i:n},
		 \end{equation}
		 where $l(\lambda;d_i)$ measures the contribution of the $i$th observation to the likelihood, given by
		 \begin{equation}
		 l(\lambda;d_i) = -\frac{1}{2}\log(2\pi)-\log(\hat{\sigma}_{\lambda}) - \frac{1}{2\hat{\sigma}_{\lambda}^2}(\psi(\lambda,d_i)-\hat{\mu}_{\lambda})^2 + (\lambda - 1)\text{sign}(d_i)\log(|d_i|+1),
		 \end{equation}
	where	$\hat{\mu}_{\lambda}$ and $\hat{\sigma}_{\lambda}$ are the median and $Q_n$ estimates of the transformed observations $\psi(\lambda,d_i)$ ($i=1,\ldots,n$), respectively. Here, we use the same value of $h$ as in the LTS estimation of the factor space. The optimal value of $\lambda$ is searched over the grid $[0,1]$ with step size 0.02. $\lambda$ values exceeding 1 are not considered to avoid a swamping effect when the chosen contamination level (through $h$) in the LTS algorithm is much larger than the actual level in the data. Finally, observations whose transformed orthogonal distance $\psi(\lambda_{\text{opt}},d_i)$ exceeds the cutoff $\Phi^{-1}(0.975)$ are flagged as OC outliers. After re-estimating the factor subspace, we update the orthogonal distance of each observation and flag the OC outliers.

	{\bf Reweighting within the subspace}. The LTS method is designed to downweight the adverse influence of OC outliers when estimating the low-dimensional subspace. However, there may be score outliers as well. These outliers do not influence the estimation of the subspace, but they affect the factor scores and the estimate of the subspace center. Therefore, we re-estimate the location and scatter of the scores and update the estimates of $\bmu$ and $\bZ$ accordingly. 
Similarly as in~\citep{robpca}, we first estimate the location and scatter of the scores $\hat{\bz}_i$ using the reweighted MCD estimator~\citep{Rousseeuw1984} and calculate the corresponding robust distances $\text{\footnotesize RD}_i$ of the observations $\hat{\bz}_i$, that is the Mahalanobis distances of the scores $\hat{\bz}_i$  with respect to these reweighted MCD estimates. The reweighted estimate of the center of the scores then becomes $\hbmu_{\bz}=\sum_{i=1}^{n}\tilde{w}_i\hat{\bz}_i/\sum_{i=1}^{n}\tilde{w}_i$, where $\tilde{w}_i = I(\text{\footnotesize RD}_i\leqslant  c_\text{\tiny SD} \text{ and } \text{\footnotesize OD}_i \leqslant c_\text{\tiny OD})$ with \scalebox{0.9}{$c_\text{\tiny RD}=\sqrt{\chi_{d;0.975}^2}$}, and $I(\cdot)$ denotes the indicator function.  Similarly, the scatter estimate $\hbSi_{\bz}$ of the scores is given by the covariance matrix of the scores with weight $\tilde{w}_i =1$. Note that to minimize the bias due to outlying observations, both the PC and OC outliers are downweighted when re-estimating the location and scatter of the scores. Finally, we update the location estimate in the original space and the score matrix, i.e. $\hbmu \gets \hbmu+\hat{\bB}\hbmu_{\bz}$, $\hat{\bZ} \gets (\hat{\bZ} - \bone_d\hbmu_{\bz}^\text{T}) \hbSi_{\bz}^{-1/2}$ and  $\hat{\bB}=\hat{\bB} \hbSi_{\bz}^{1/2}$. 
Then, we recompute the score distance for each observation $i$ by  $\text{\footnotesize SD}_i = \|\hbz_i\|_E$, and flag it as a score outlier if $\text{\footnotesize SD}_i > c_\text{\tiny SD}$.

	{\bf Estimating $d$.}  In practice, the dimension $d$ of the factor subspace is unknown. To estimate the dimension $d$, we use the criterion in~\citep{Bai2002} which determines the number of factors by minimizing
	\begin{eqnarray}
	\label{PC_Select}
	\text{\footnotesize PC}(d) & = & \tilde{n}_d^{-1}p^{-1}tr\{(\bX-\bone_{\tilde{n}_d}\hbmu_d^\text{T}-\hbZ_d\hbB_d^\text{T})^\text{T}W^d(\bX-\bone_{\tilde{n}_d}\hbmu^\text{T}_d-\hbZ_d\hbB_d^\text{T})\}  \\
	&+& {\tilde{n}_d}^{-1} p ^{-1}tr((\bX-\bone_{\tilde{n}_d}\hbmu_d^\text{T})^\text{T}W^d(\bX-\bone_{\tilde{n}_d}\hbmu_d^\text{T}))\{d(\frac{\tilde{n}_d+p}{\tilde{n}_d p})\log(\frac{\tilde{n}_d p}{\tilde{n}_d+p})\}, \nonumber
	\end{eqnarray}
	with respect to $d$. Here, $\hbmu_d$, $\hbZ_d$ and $\hbB_d$ are the estimates obtained by the procedure outlined above when the number of factors is fixed at $d$. $W^d$ is a diagonal matrix with on the diagonal the weights $w_i^d = I(\text{\footnotesize OD}^d_i \leqslant c_\text{\tiny OD} \text{ and } \text{\footnotesize SD}^d_i \leqslant c_\text{\tiny SD})$ where $\text{\footnotesize OD}_i^d$ and $\text{\footnotesize SD}_i^d$ are computed with $\hbmu_d$, $\hbZ_d$ and $\hbB_d$. Finally, $\tilde{n}_d = \sum_{i=1}^{n}w_i^d$. To control the computation time we fix $d_{\max}$, the maximal number of factors. The estimated dimension of the subspace is then given by $\hat{d} = \arg\min_{1\leqslant d \leqslant d_{\max}}\text{\footnotesize PC}(d)$ which yields the final estimates
	  $\hbmu_{\hat{d}}$, $\hbZ_{\hat{d}}$ and $\hbB_{\hat{d}}$. To simplify notation we will drop the subscript $\hat{d}$ in the remainder of the paper.

	\subsection{Robust Variable Screening}
	\label{sec:robust screening}
	In FPSIS, the profiled variables are obtained by projecting the original variables onto the orthogonal complement of the subspace spanned by the latent factors. However, each profiled observation is then a linear combination of all the original observations. If there are outliers in the data, this implies that all the profiled observations would become contaminated which would make them useless. To avoid this, we  instead calculate the profiled variables directly by using \eqref{RegX}-\eqref{RegE}. The profiled predictors are obtained as
	\begin{equation}
	\htbX = \bX - \bone_n\hbmu^\text{T} - \hbZ\,\hbB^\text{T}.
	\end{equation}
	To obtain the profiled response variable, we robustly regress $\by$ on $\hbZ$. 
	We use the 95\% efficient MM-estimator~\citep{Yohai1987} with bisquare loss function for this purpose.
	The resulting slope estimates are denoted by $\hbga$ while the estimated intercept is denoted by $\hmu_y$ since it provides a robust estimate of the center of $\by$.  The corresponding profiled response is given by
	\begin{equation}
	\htby = \by - \hmu_y\bone_n-\hbZ^\text{T}\hat{\gamma}.
	\end{equation}

Variable screening is conducted on the profiled variables by using marginal regression models. 
Before applying variable screening, we first investigate which types of outliers can occur in the data
with respect to the different regression and factor models. As discussed in Section \ref{sec:FS}, in the factor model for the predictors, we may have two types of outliers: OC outliers and PC outliers. Since PC outliers are only outlying in $\hbZ$ rather than $\hbX$, by profiling out the effect of $\hbZ$ they become non-outlying observations in $\hbX$. However, OC outliers are outlying with respect to the low-dimensional subspace, it is unable to remove their outlyingness by factor profiling. Therefore, these observations remain outliers in the profiled predictor matrix $\hbX$. 

For the multiple regression model~\eqref{RegY} based on the original variables, there can be vertical outliers, good leverage points, and bad leverage points. Vertical outliers are only outlying in the response variable $\by$. Good leverage points are outlying in the predictor space $\bX$, but do follow the regression model, while bad leverage points are not only outlying in $\bX$ but also have responses that deviate from the regression model of the majority.  
	
By combining the types of outliers that can occur in the multiple linear regression model (LM) and the factor model for the predictors, we can have the following 5 types of outliers:
	\begin{itemize}[topsep= -6 pt,itemsep=-0.5 ex,partopsep=-0.5 ex, parsep= 1ex]
		\item[1.] LMV: vertical outlier in the multiple regression;		
		\item[2.] PC+LMG: good leverage point due to PC outlier in the predictors;
		\item[3.] PC+LMB: bad leverage point due to PC outlier in the predictors;
		\item[4.] OC+LMG: good leverage point due to OC outlier in the predictors;
		\item[5.] OC+LMB: bad leverage point due to OC outlier in the predictors.
	\end{itemize}

Each outlier type may affect the multiple regression model for the profiled variables as well as the corresponding marginal regression models. 	
To illustrate the effect of the different types of leverage points on these regression models, we consider a regression example with only 2 predictors and 1 factor. 
A set of clean observations ($\bX_\text{clean}$, $\by_\text{clean}$) is generated according to $\bX_\text{clean} = \bz\bB^\text{T} + \tilde{\bX}$ and $\by_\text{clean} = \bX_\text{clean}\bbe + \bves$, where $\bbe = (2,1)^\text{T}$, $\bz \sim N(0,1)$, $\tilde{\bX}\sim N_2(0,\bI_2)$ $\bB = (1/\sqrt{2},1/\sqrt{2})^\text{T}$, and $\bves\sim N(0,1)$. 
For the factor model we generate PC outliers by  $\bX_\text{\tiny PC} = \bz_\text{\tiny PC}\bB^\text{T} + \tilde{\bX}$ with $\bz_\text{\tiny PC} \sim N(10,1)$ and OC outliers by $\bX_\text{\tiny OC}= \bz_\text{\tiny OC}\bB_\text{\tiny OC}^\text{T} + \tilde{\bX}$ with  $\bz_\text{\tiny OC} \sim N(10,1)$ and $\bB_\text{\tiny OC} = (-1/\sqrt{2},1/\sqrt{2})^\text{T}$.

Observations according to the 4 types of leverage points are then obtained as follows.
	\begin{itemize}[topsep= -6 pt,itemsep=-0.5 ex,partopsep = -0.5 ex,parsep= 1 ex]		
		\item[1.] PC+LMG: ($\bX_\text{\tiny PC}$,$\by_\text{\tiny PC+LMG}$) where $\by_\text{\tiny PC+LMG}$ = $\bX_\text{\tiny PC}\bbe + \bves$ ;
		\item[2.] PC+LMB: ($\bX_\text{\tiny PC}$,$\by_\text{\tiny PC+LMB}$), where $\by_\text{\tiny PC+LMB}\sim N(50,1)$;
		\item[3.] OC+LMG: ($\bX_\text{\tiny OC}$,$\by_\text{\tiny OC+LMG}$), where $\by_\text{\tiny OC+LMG}$ = $\bX_\text{\tiny OC}\bbe + \bves$;
		\item[4.] OC+LMB: ($\bX_\text{\tiny OC}$,$\by_\text{\tiny OC+LMB}$), where $\by_\text{\tiny OC+LMB}\sim N(50,1)$;	
	\end{itemize}
	
\begin{figure}[H]
	\centering
	\begin{minipage}{0.24\textwidth}
		\centering
		\footnotesize
		(a) PC + LMG
	\end{minipage}
	\begin{minipage}{0.24\textwidth}
		\centering
		\footnotesize
		(b) PC + LMB
	\end{minipage}
	\begin{minipage}{0.24\textwidth}
		\centering
		\footnotesize
		(c) OC + LMG
	\end{minipage}
	\begin{minipage}{0.24\textwidth}
		\centering
		\footnotesize
		(d) OC + LMB
	\end{minipage}\\
	\begin{minipage}{0.24\textwidth}
		\centering
		\includegraphics[width=3.5 cm]{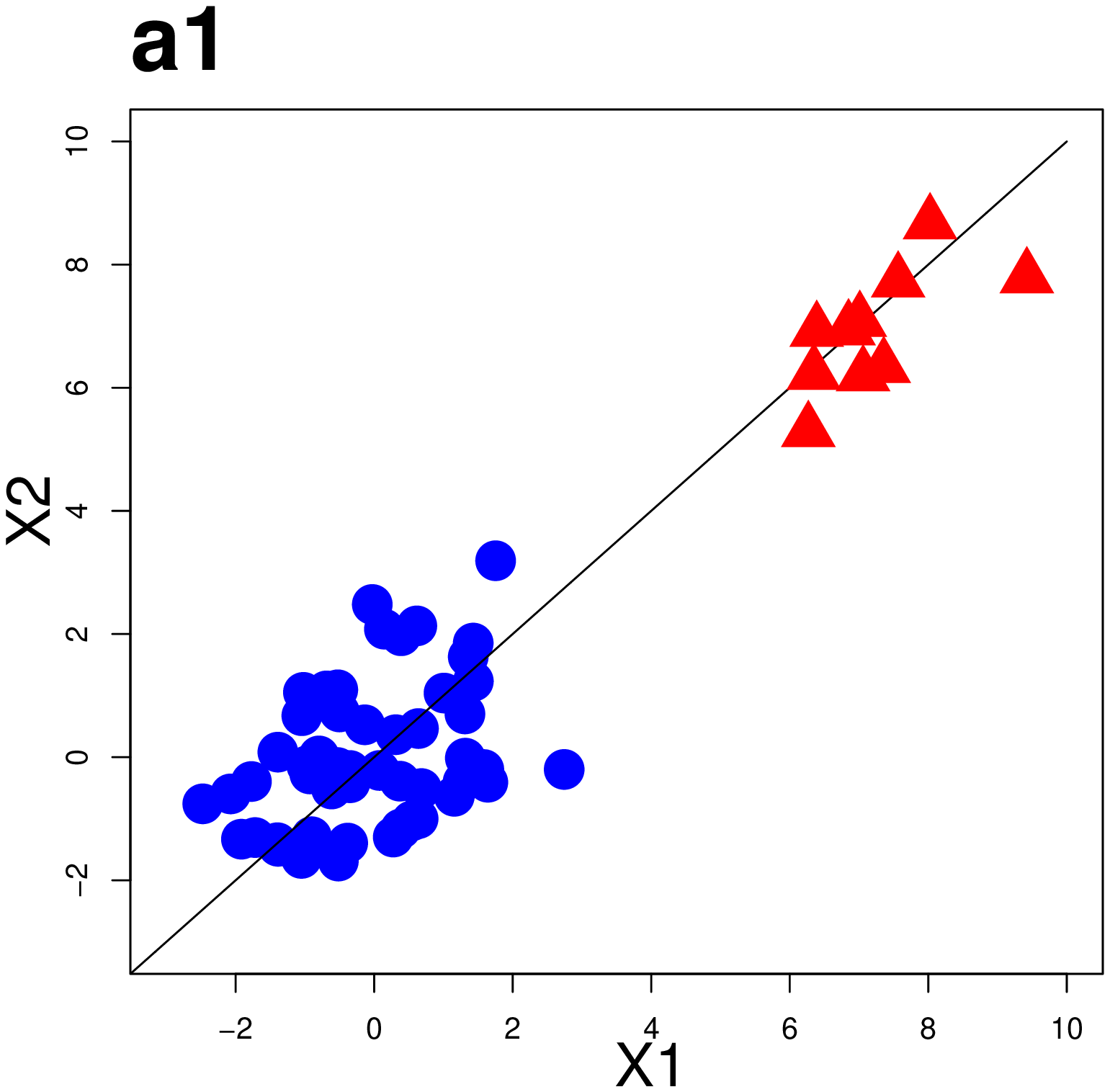}
	\end{minipage}
	\begin{minipage}{0.24\textwidth}
		\centering
		\includegraphics[width=3.5 cm]{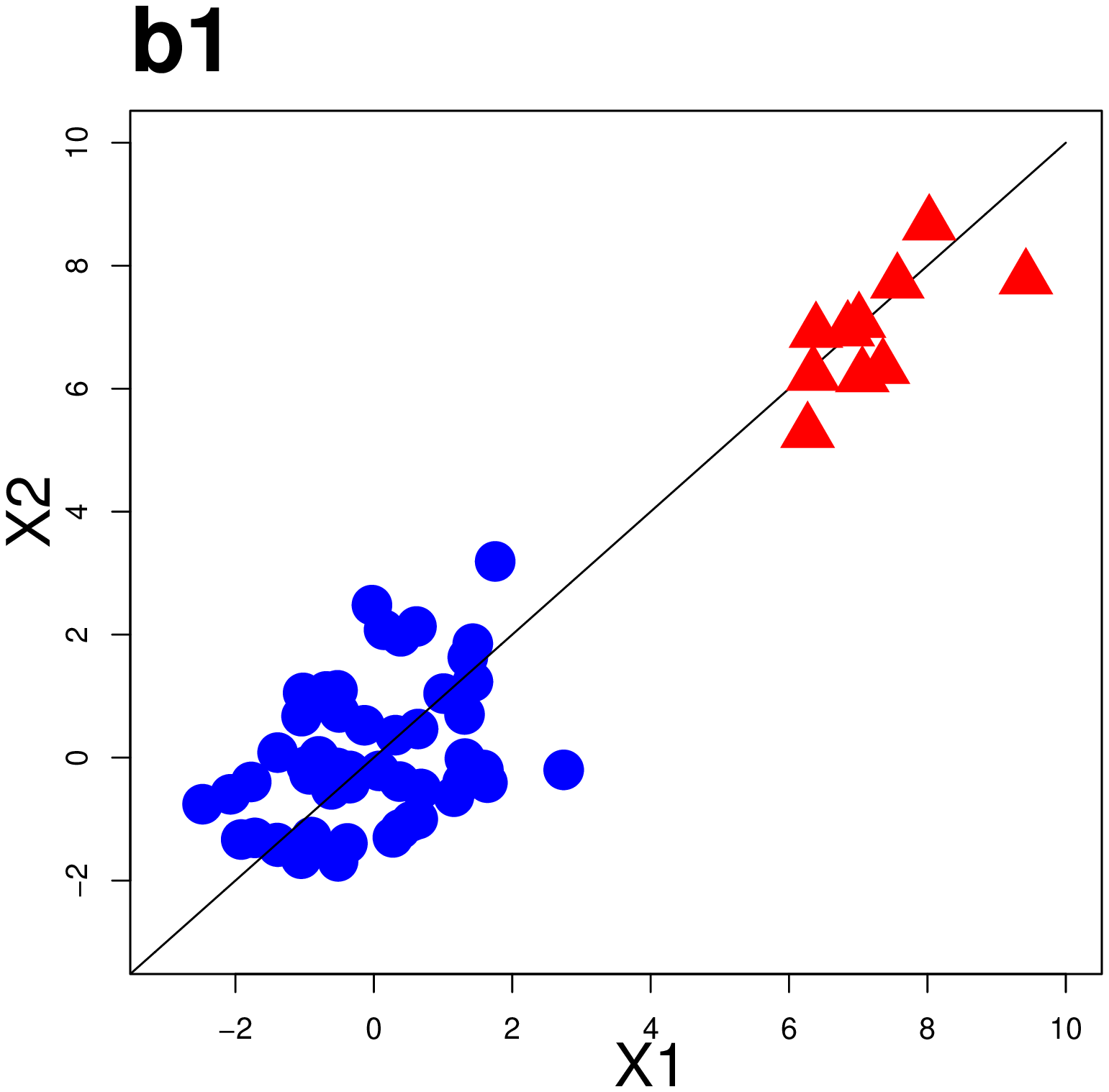}
	\end{minipage}
	\begin{minipage}{0.24\textwidth}
		\centering
		\includegraphics[width=3.5 cm]{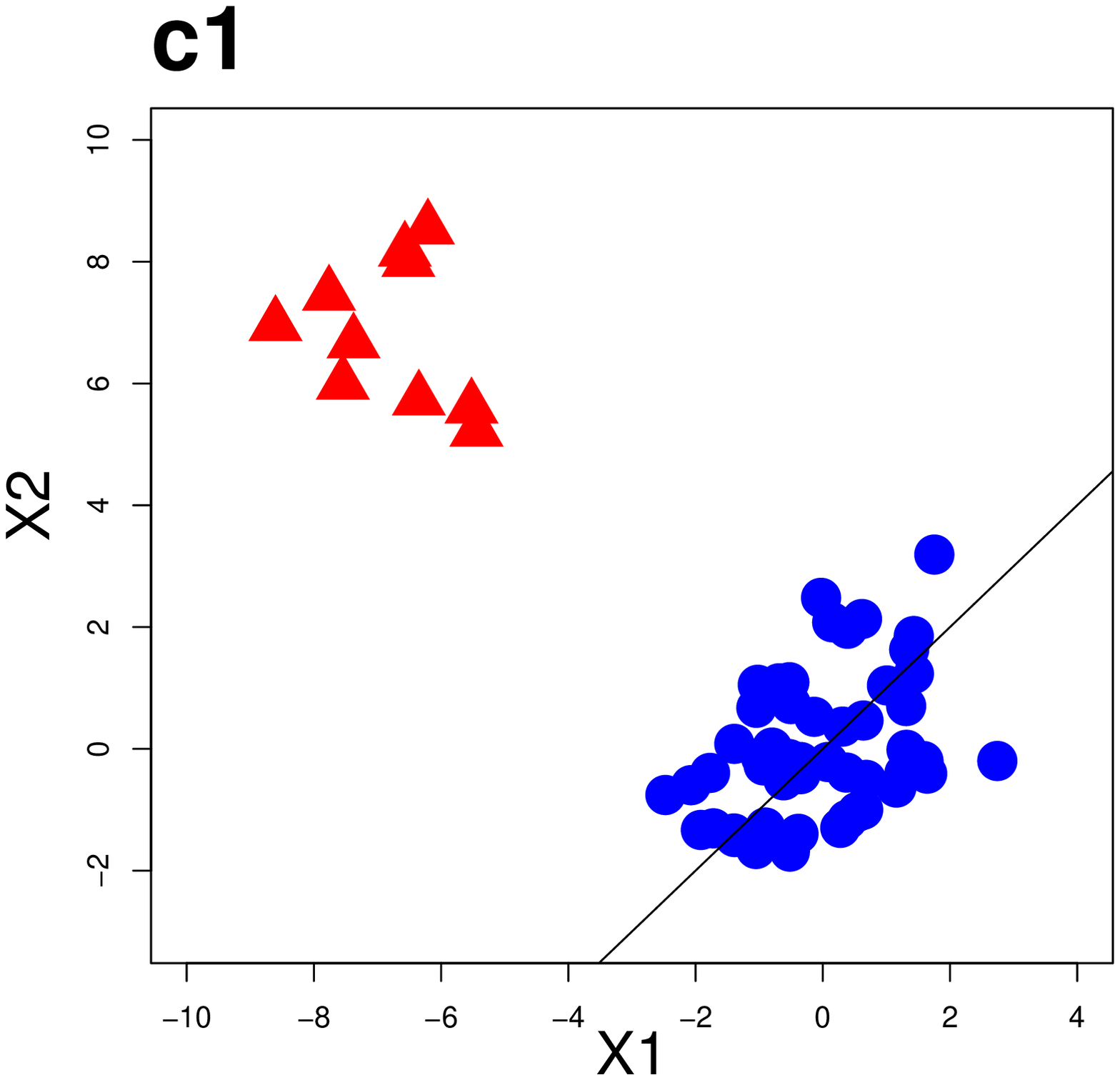}
	\end{minipage}
	\begin{minipage}{0.24\textwidth}
		\centering
		\includegraphics[width=3.5 cm]{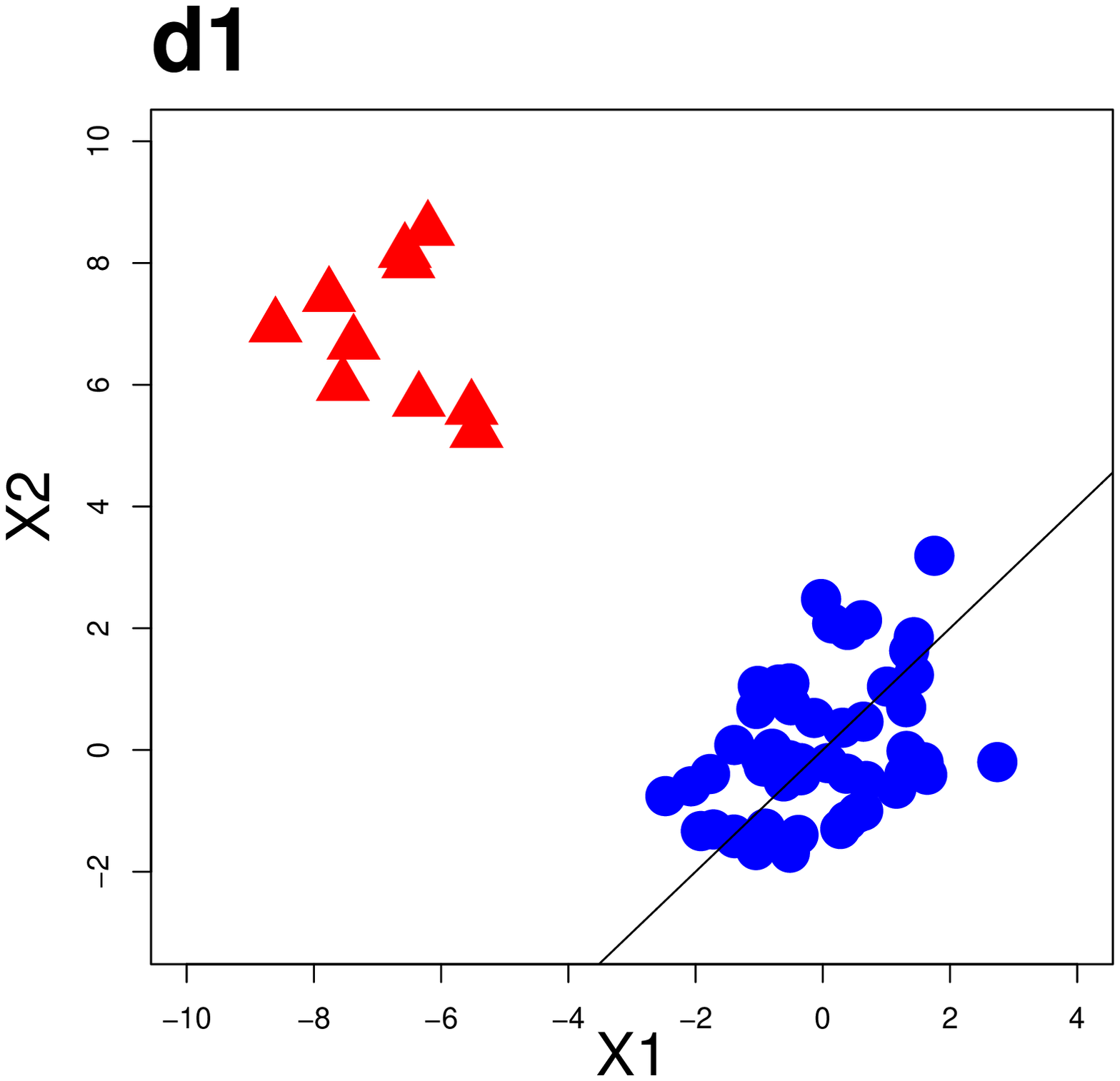}
	\end{minipage}\\
	\begin{minipage}{0.24\textwidth}
		\centering
		\includegraphics[width=3.5 cm]{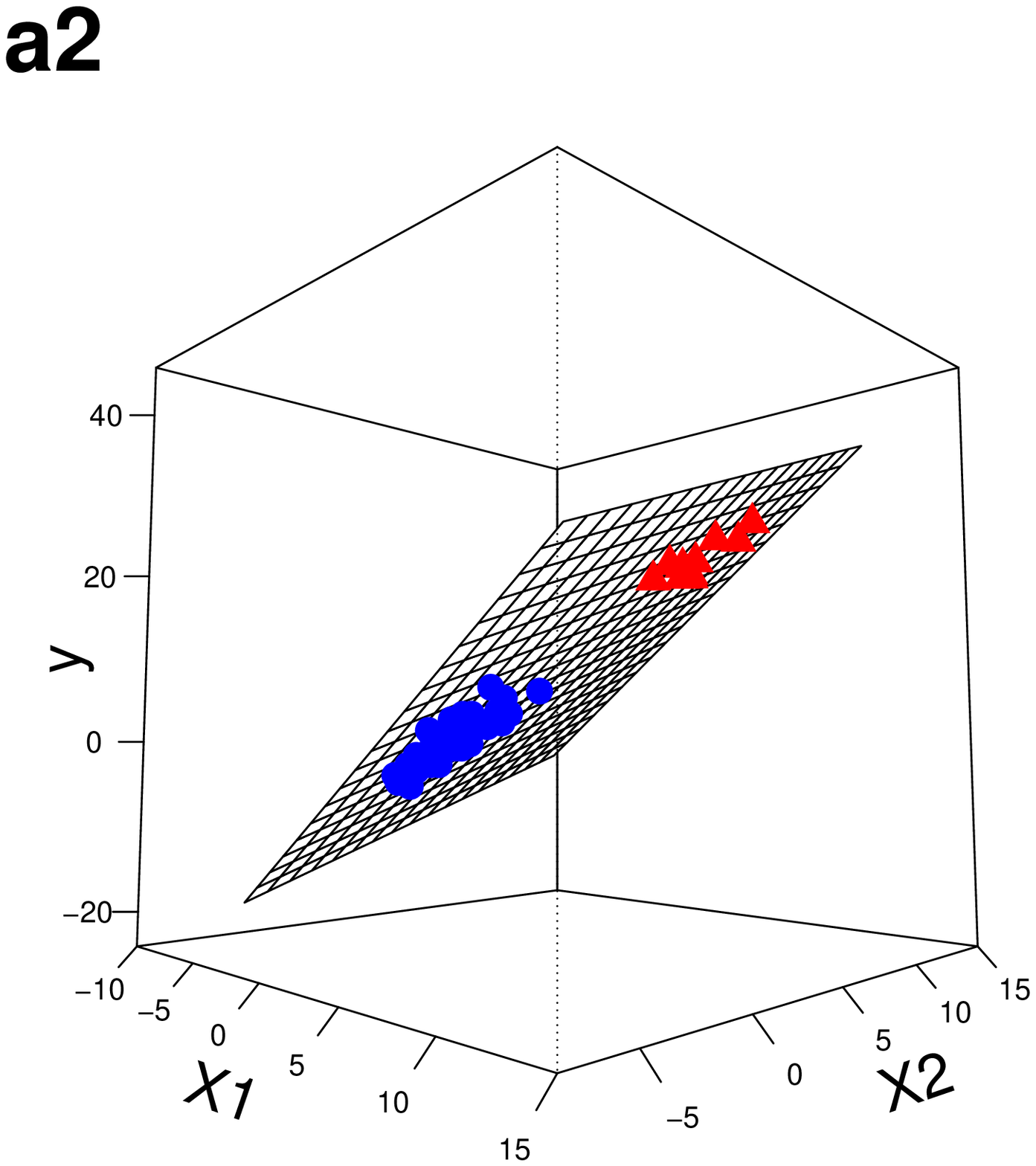}
	\end{minipage}
	\begin{minipage}{0.24\textwidth}
		\centering
		\includegraphics[width=3.5 cm]{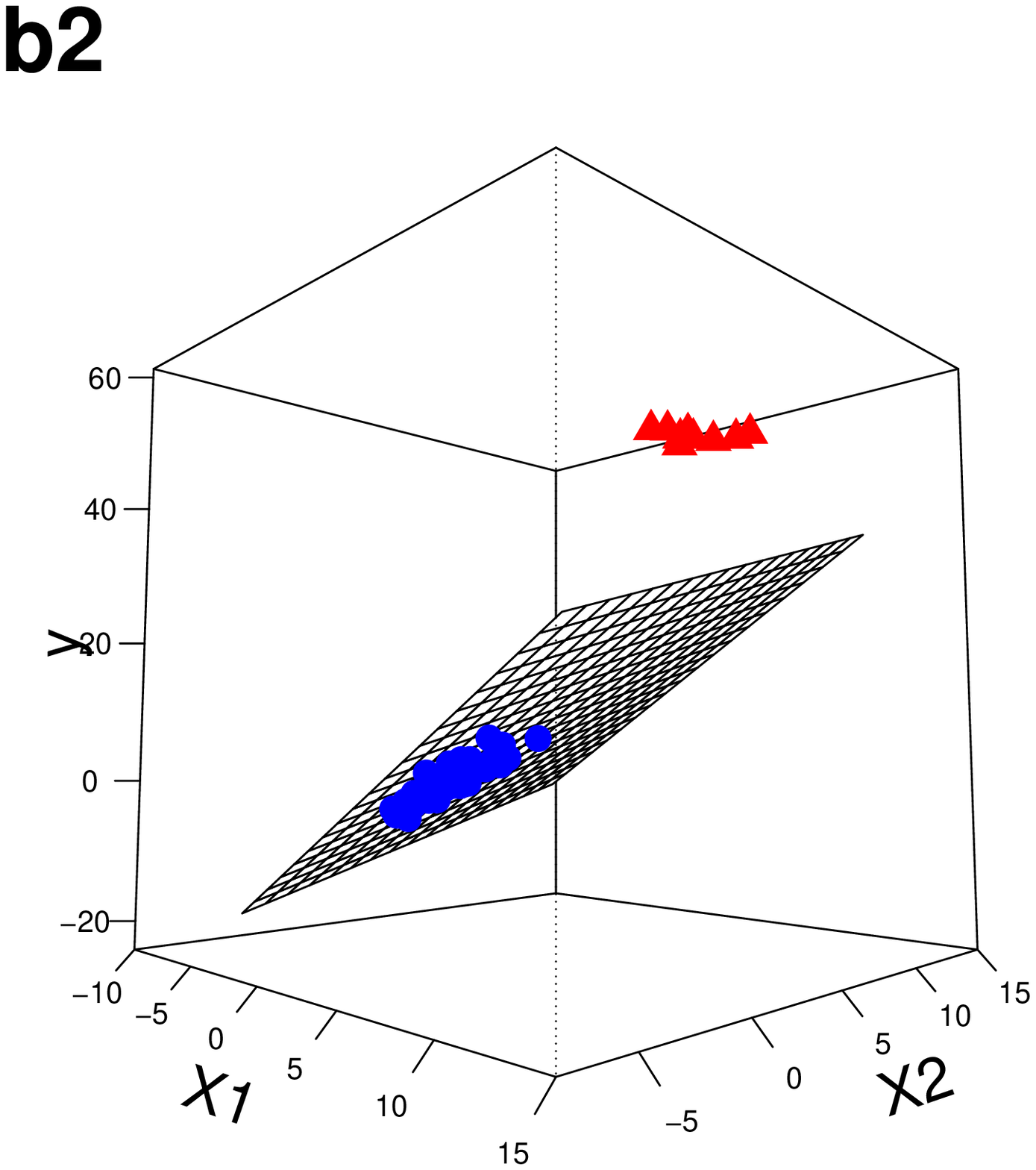}
	\end{minipage}
	\begin{minipage}{0.24\textwidth}
		\centering
		\includegraphics[width=3.5 cm]{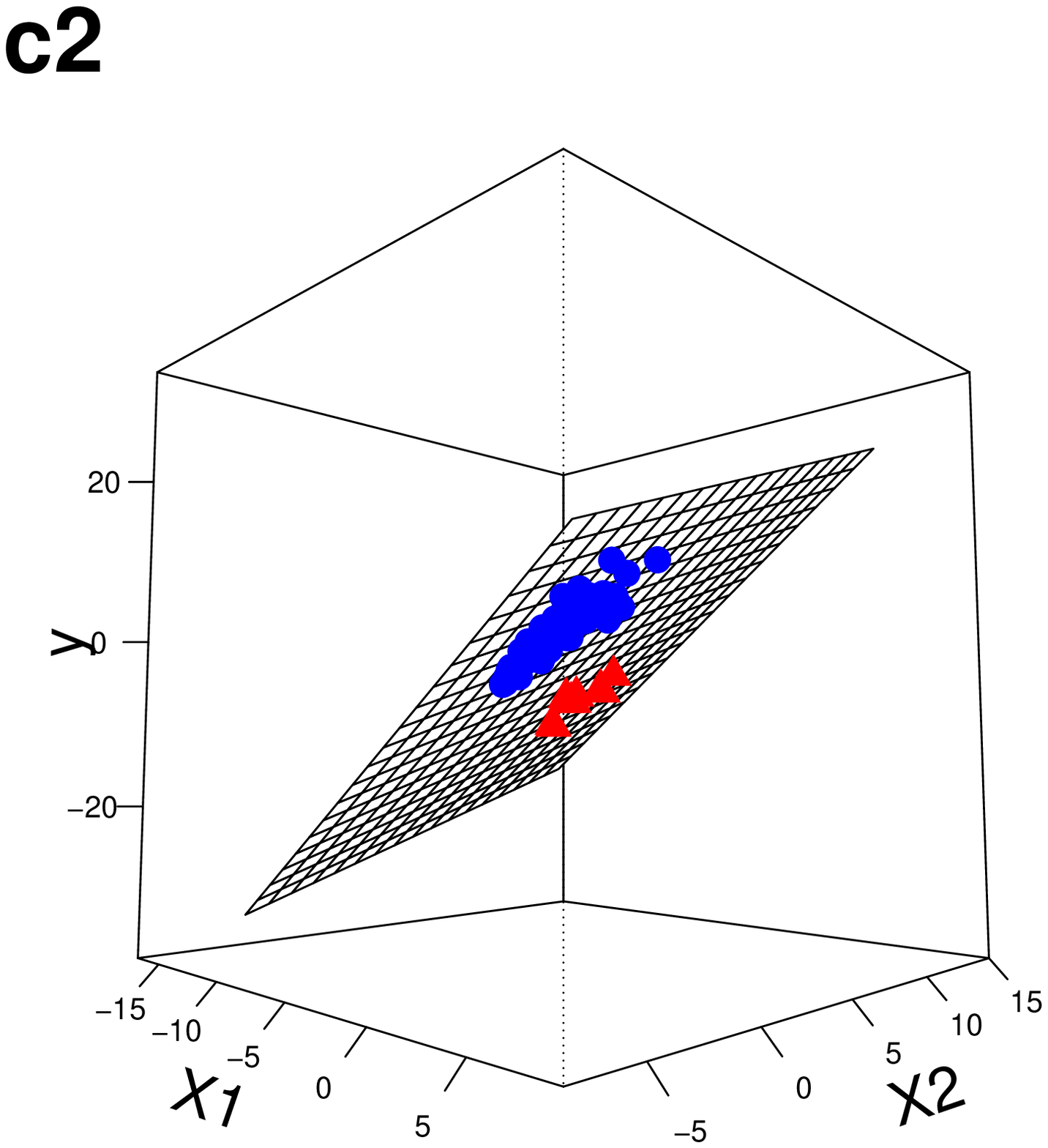}
	\end{minipage}
	\begin{minipage}{0.24\textwidth}
		\centering
		\includegraphics[width=3.5 cm]{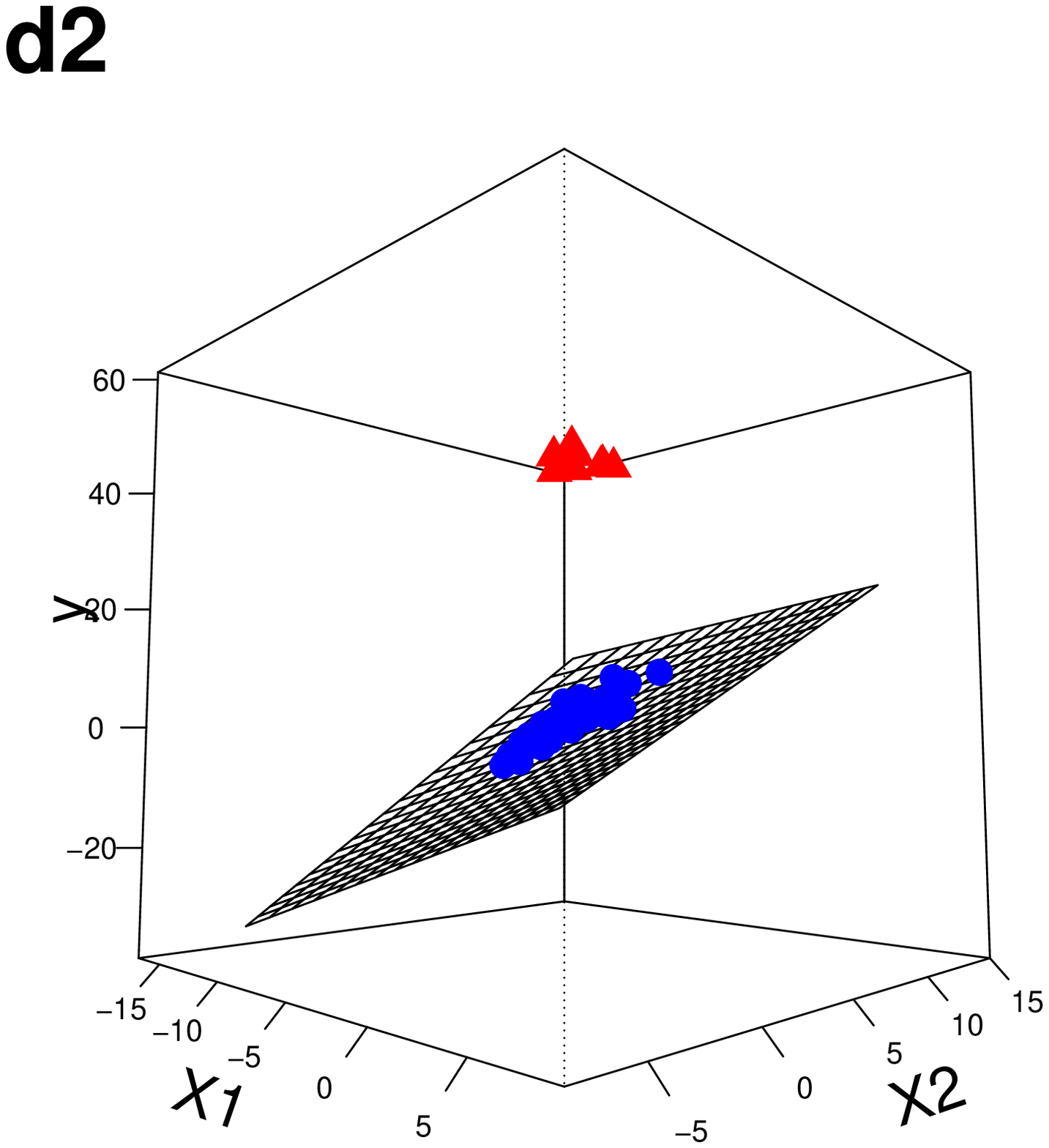}
	\end{minipage}\\
	
	\begin{minipage}{0.24\textwidth}
		\centering
		\includegraphics[width=3.5 cm]{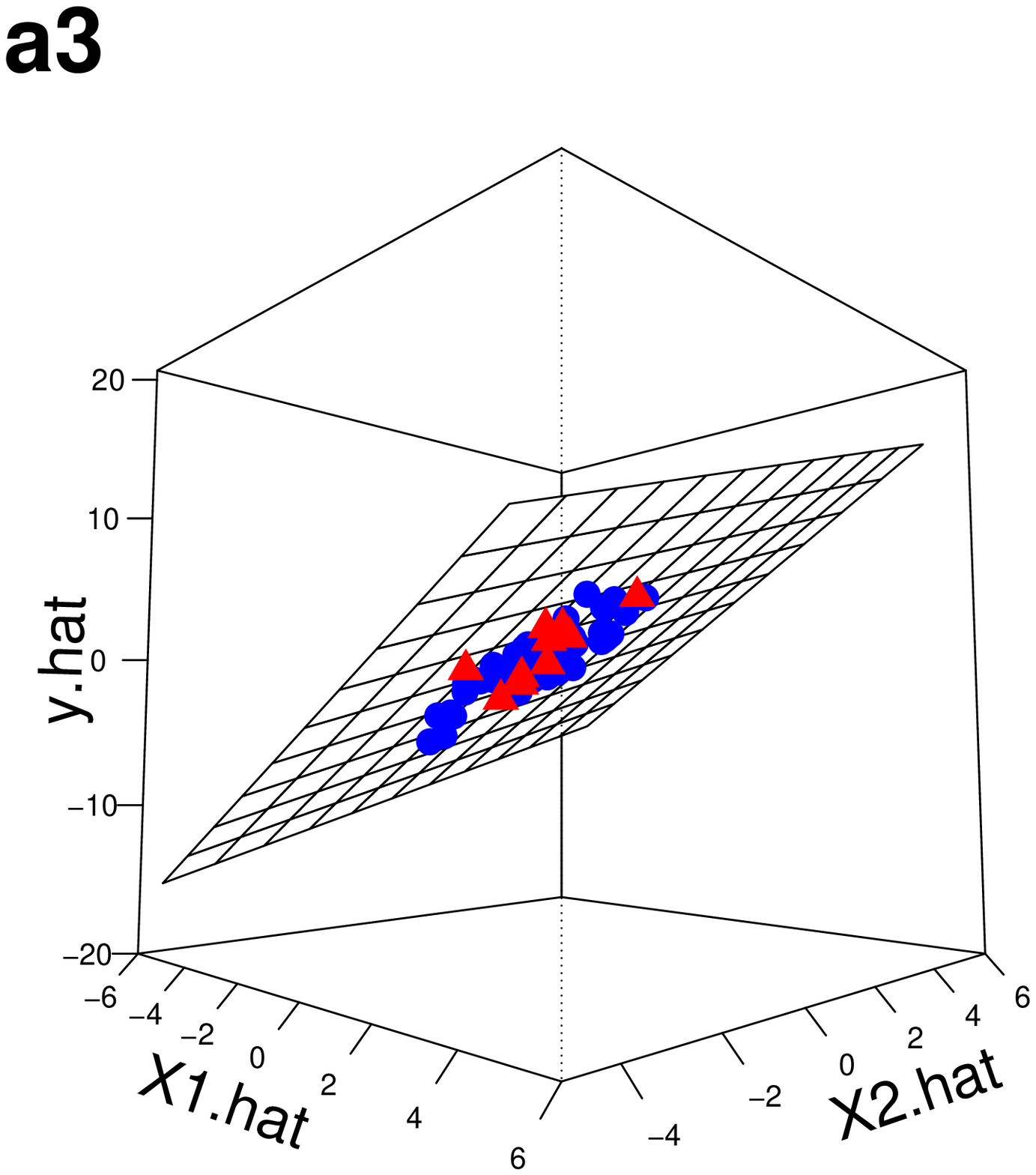}
	\end{minipage}
	\begin{minipage}{0.24\textwidth}
		\centering
		\includegraphics[width=3.5 cm]{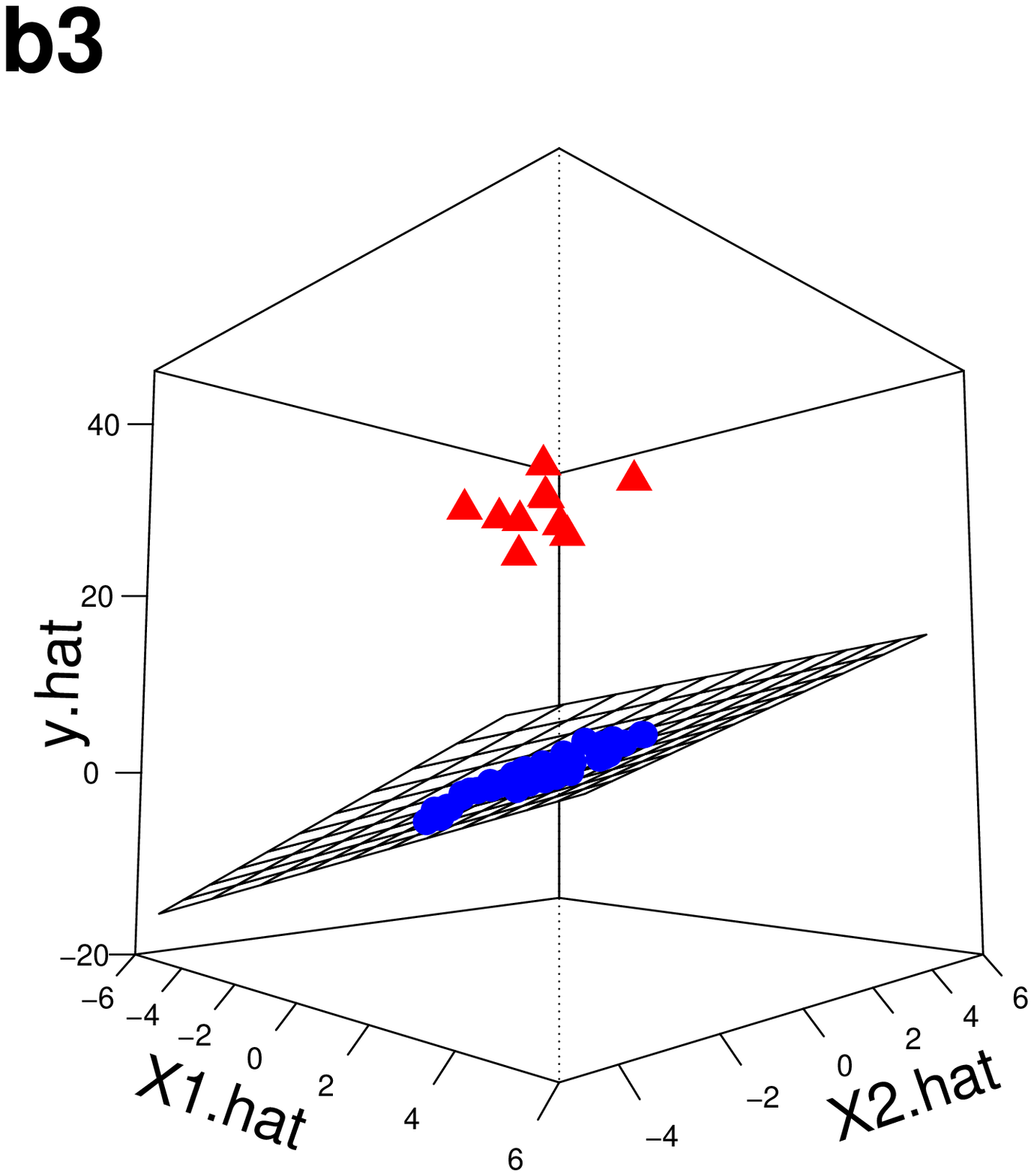}
	\end{minipage}
	\begin{minipage}{0.24\textwidth}
		\centering
		\includegraphics[width=3.5 cm]{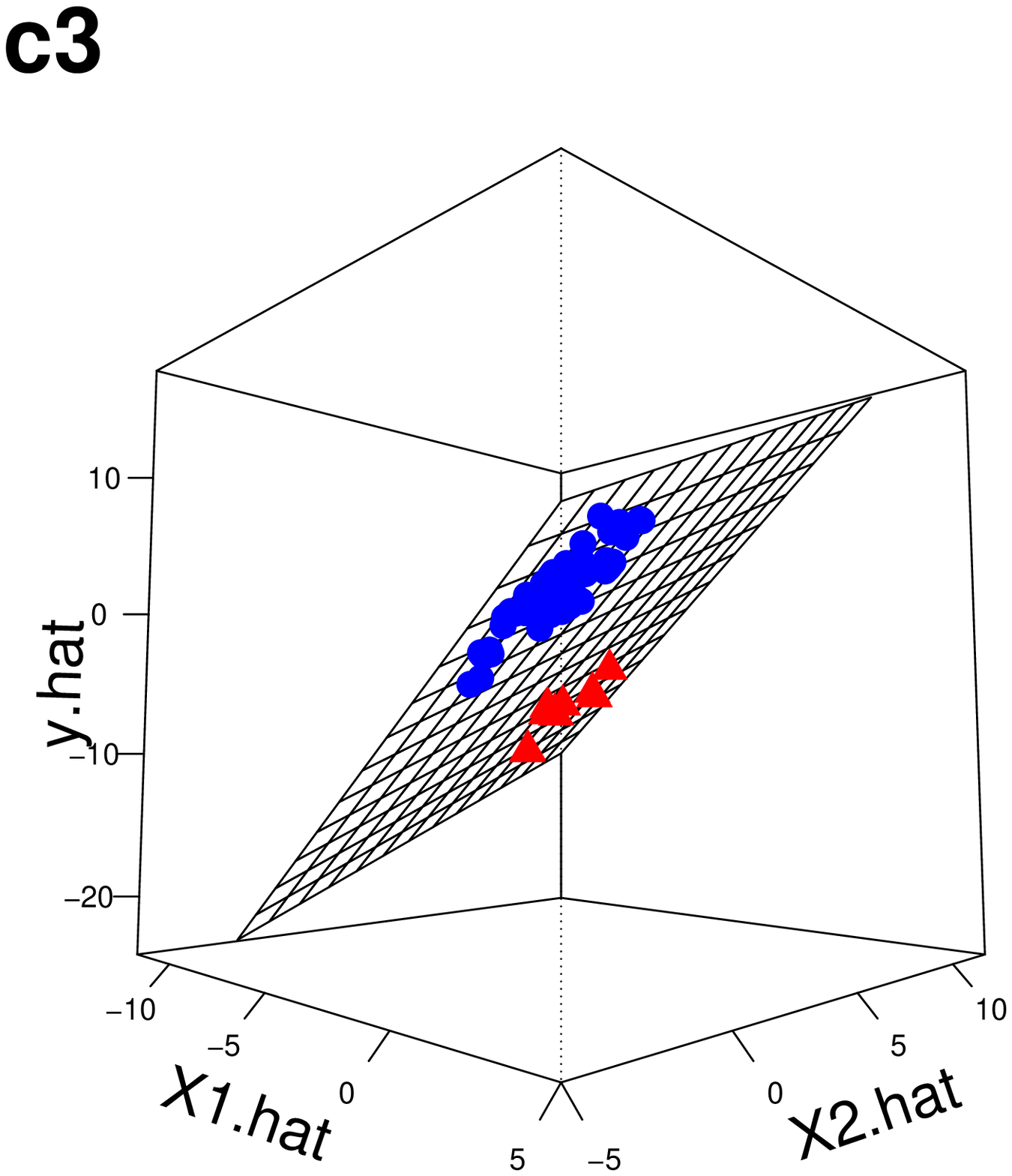}
	\end{minipage}
	\begin{minipage}{0.24\textwidth}
		\centering
		\includegraphics[width=3.5 cm]{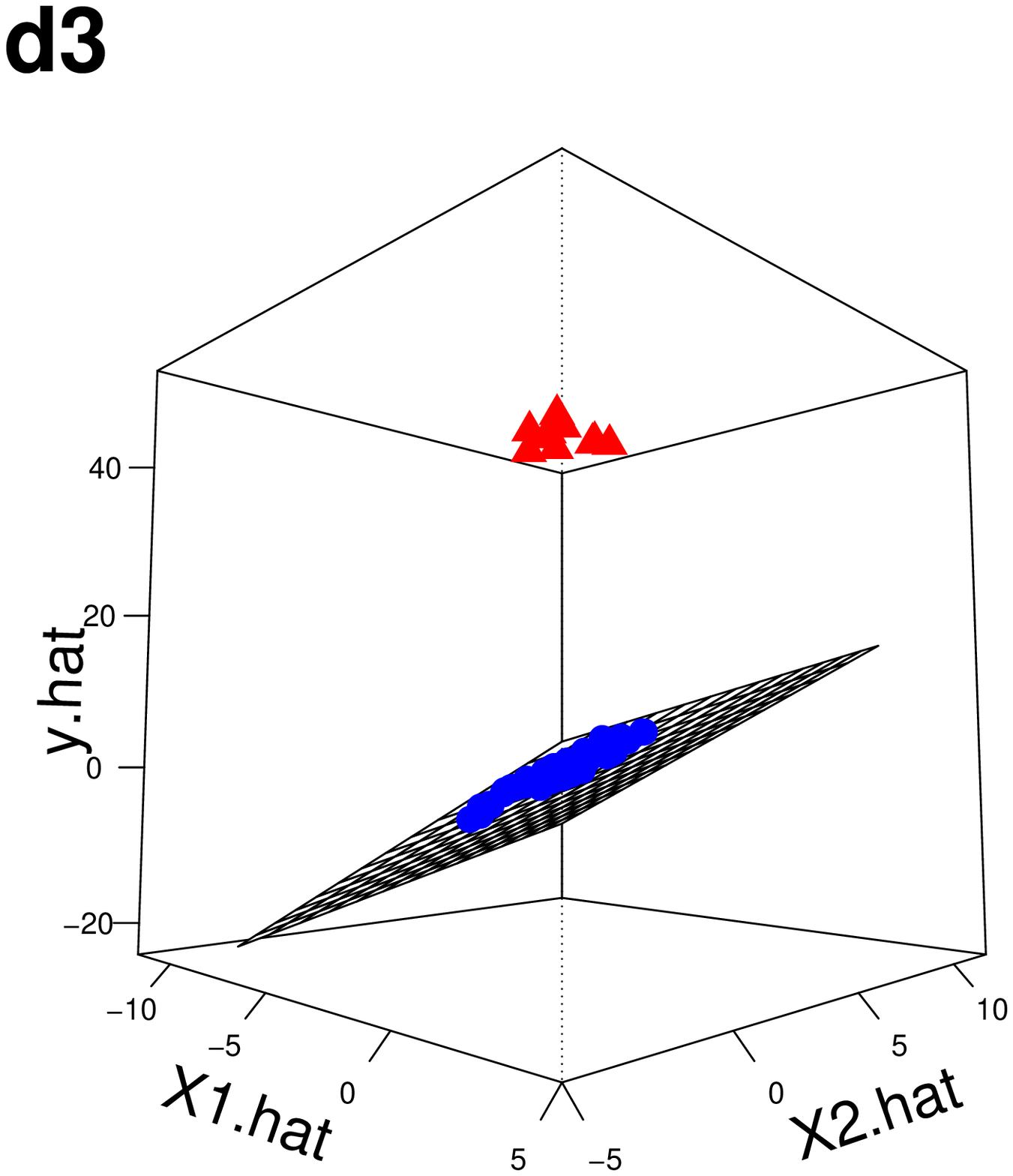}
	\end{minipage}\
	
	\begin{minipage}{0.24\textwidth}
		\centering
		\includegraphics[width=3.5 cm]{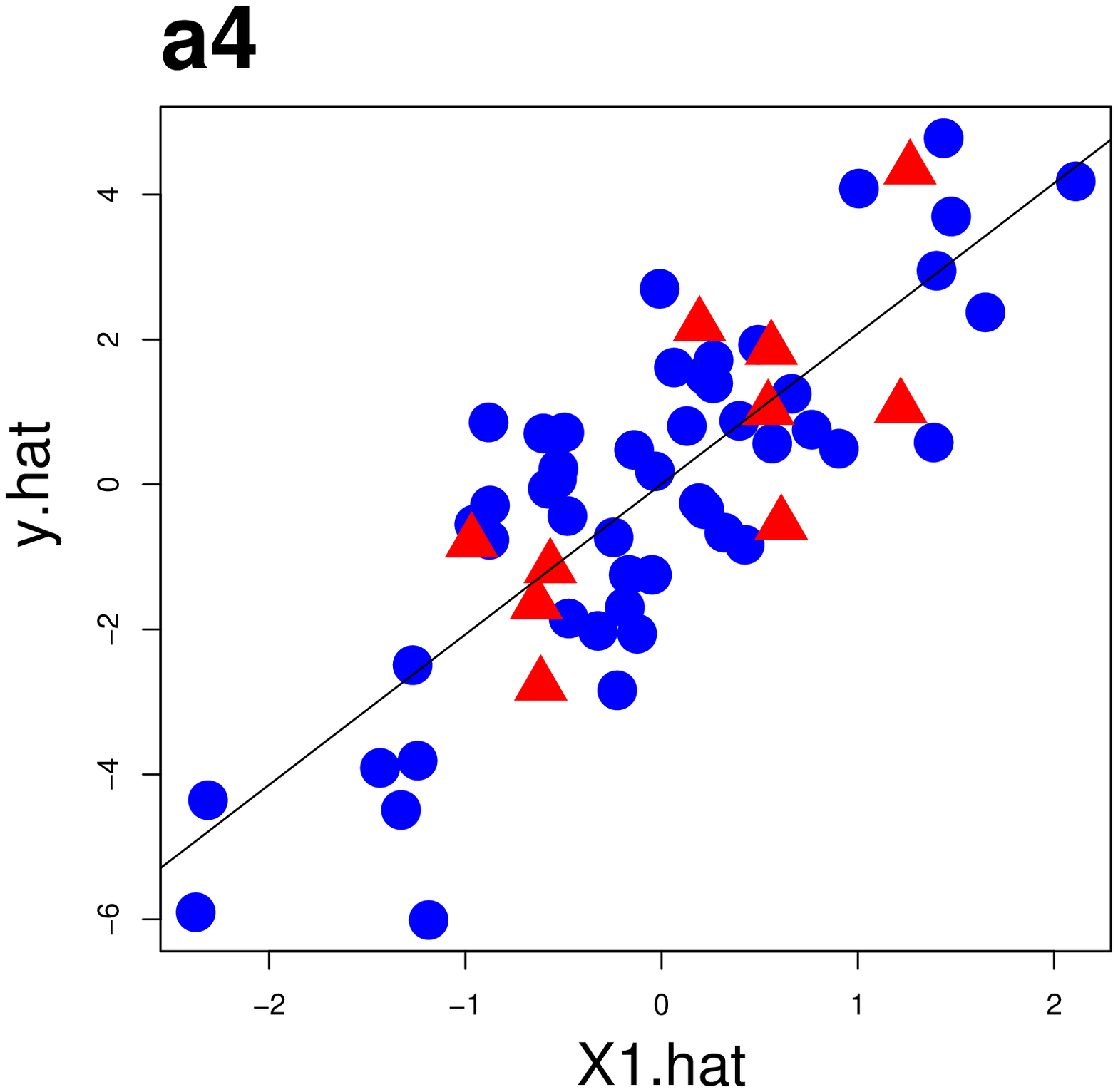}
	\end{minipage}
	\begin{minipage}{0.24\textwidth}
		\centering
		\includegraphics[width=3.5 cm]{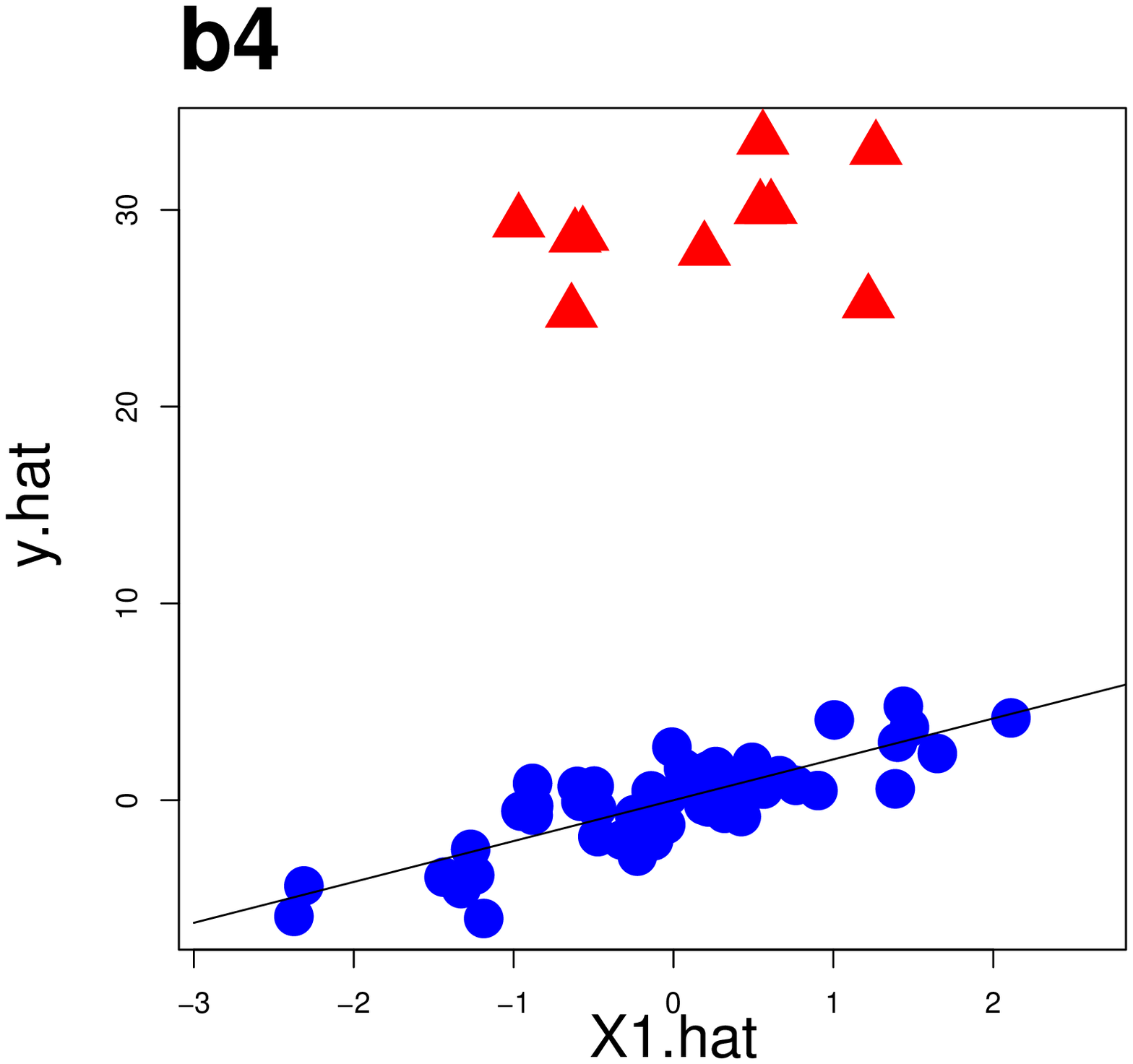}
	\end{minipage}
	\begin{minipage}{0.24\textwidth}
		\centering
		\includegraphics[width=3.5 cm]{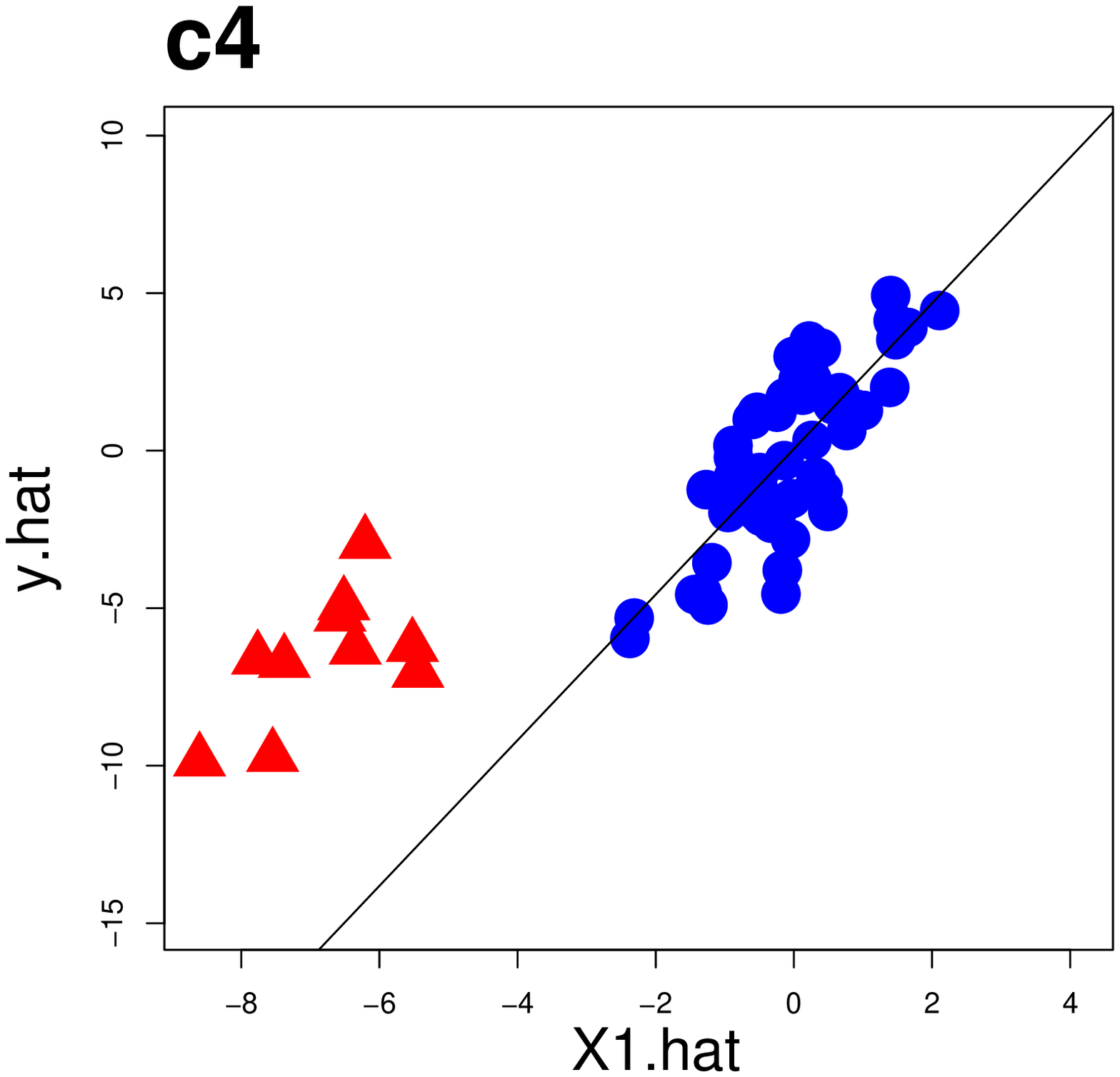}
	\end{minipage}
	\begin{minipage}{0.24\textwidth}
		\centering
		\includegraphics[width=3.5 cm]{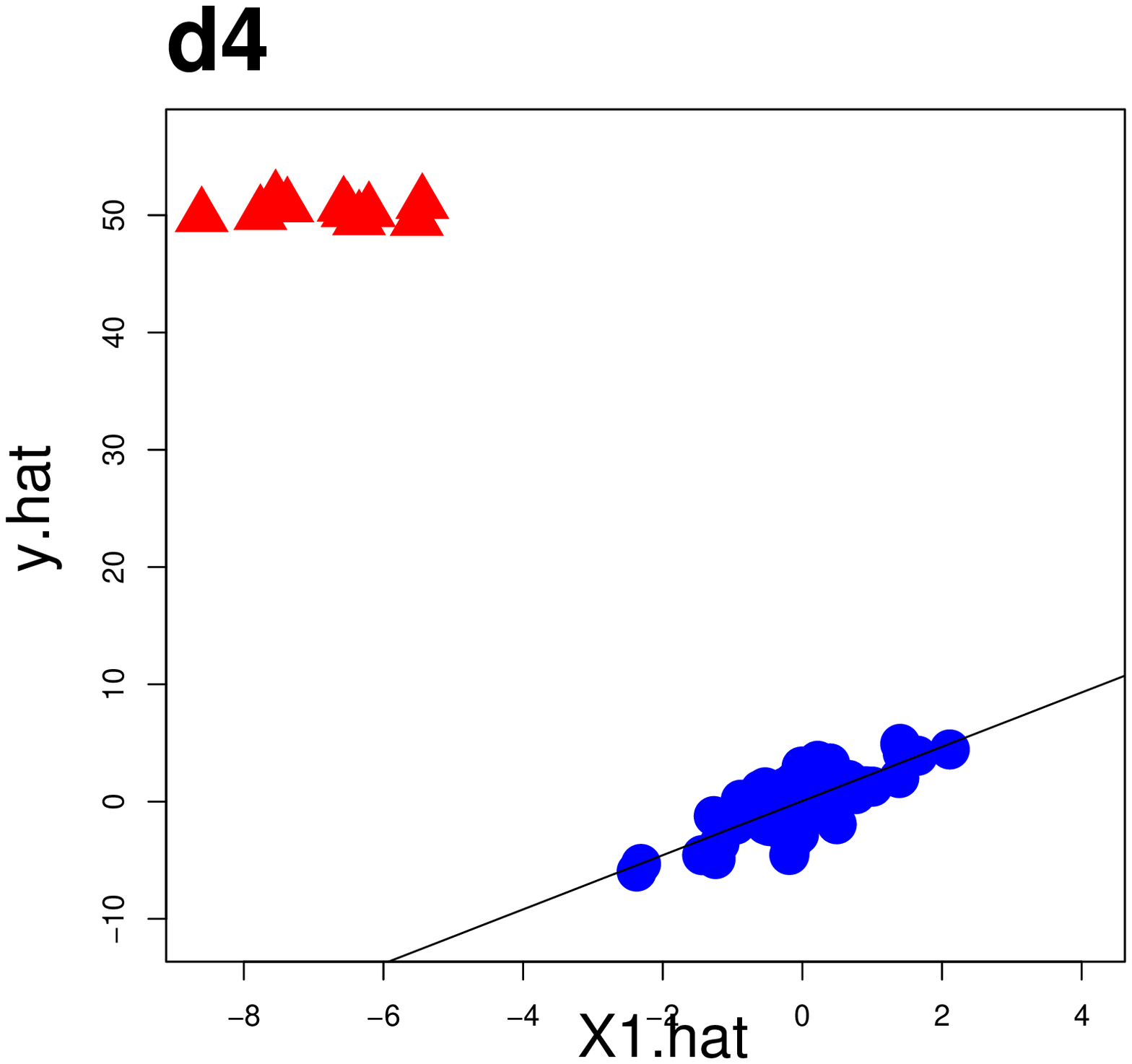}
	\end{minipage}\\
	
	\begin{minipage}{0.24\textwidth}
		\centering
		\includegraphics[width=3.5 cm]{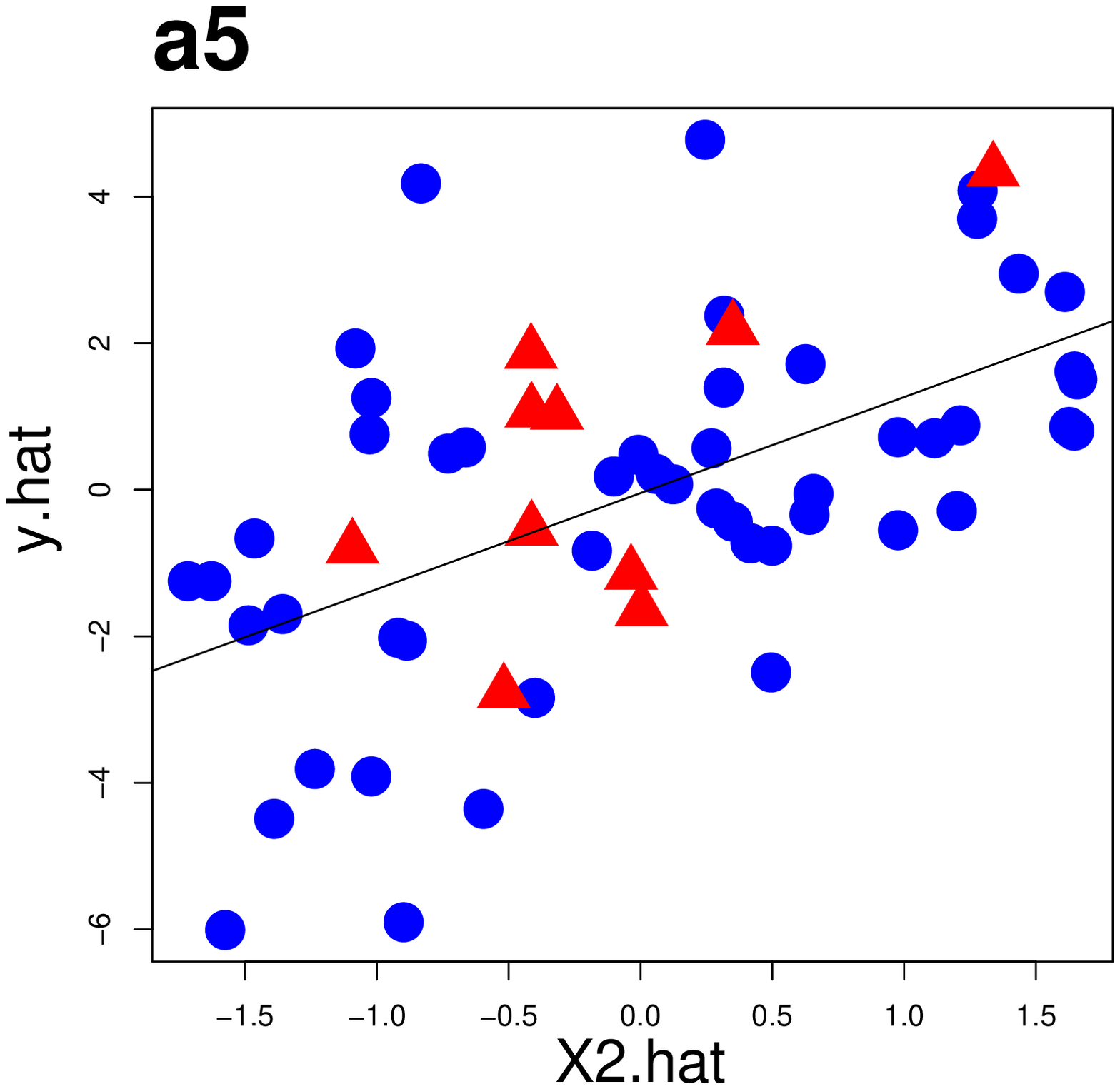}
	\end{minipage}
	\begin{minipage}{0.24\textwidth}
		\centering
		\includegraphics[width=3.5 cm]{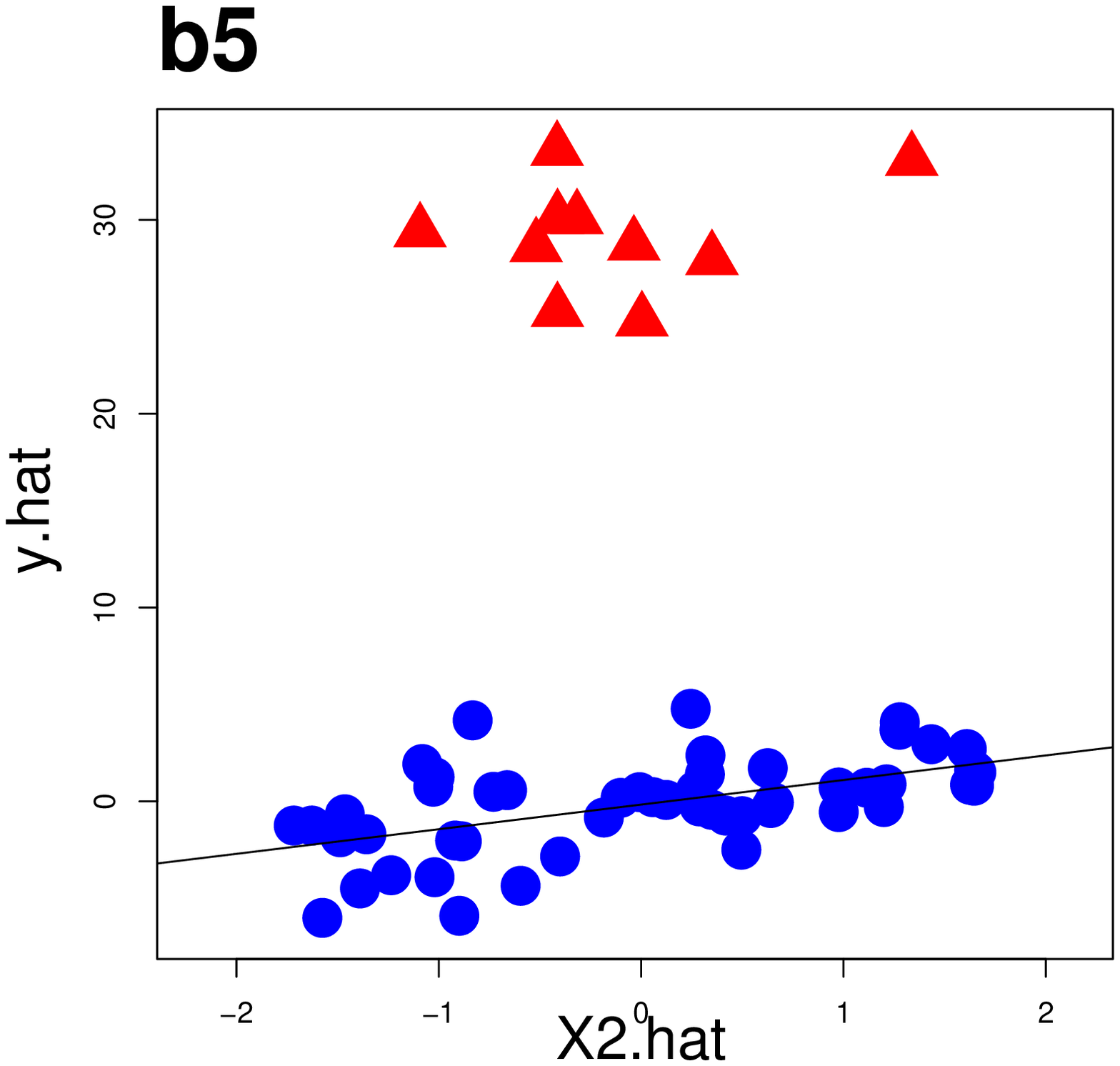}
	\end{minipage}
	\begin{minipage}{0.24\textwidth}
		\centering
		\includegraphics[width=3.5 cm]{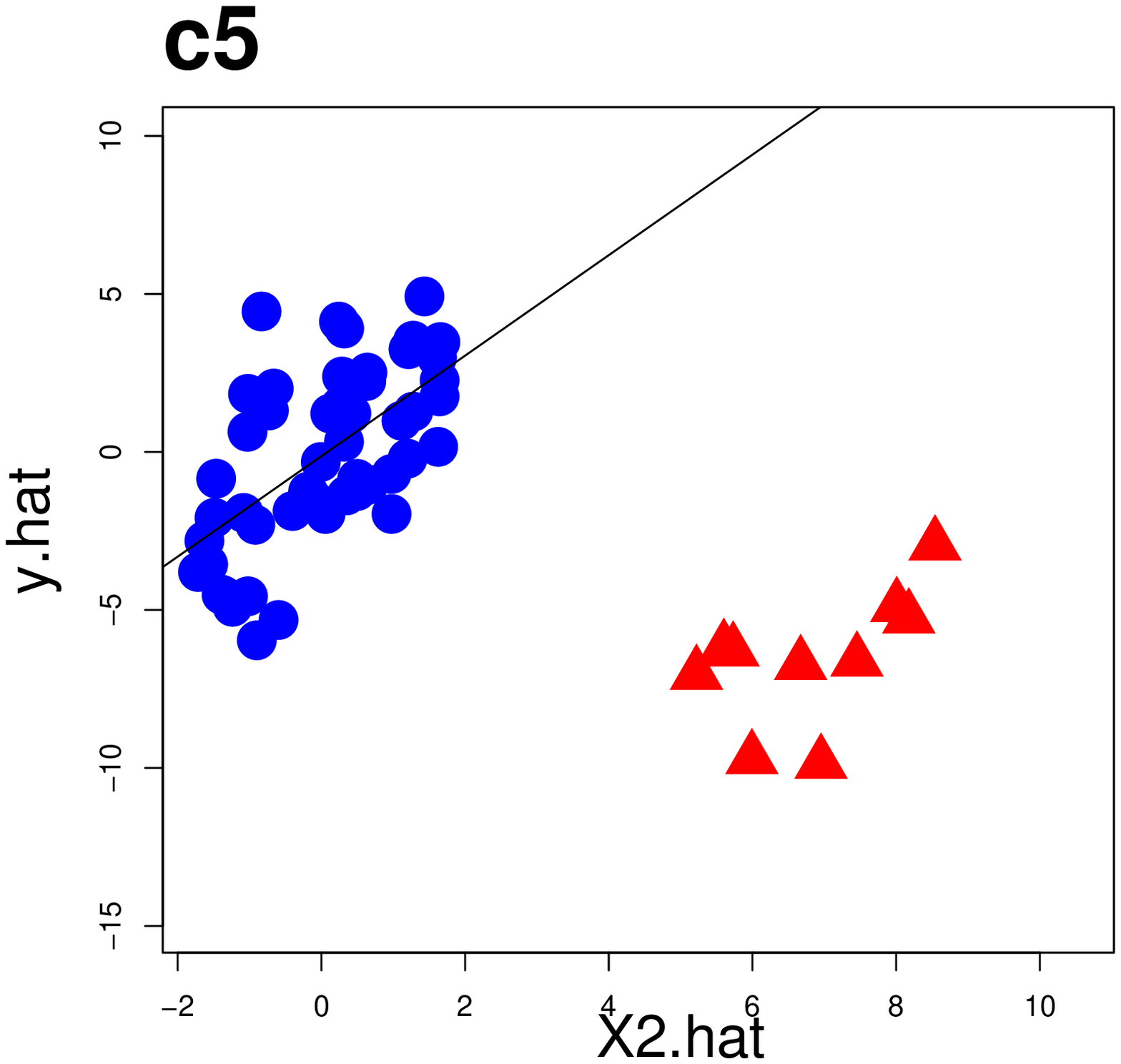}
	\end{minipage}
	\begin{minipage}{0.24\textwidth}
		\centering
		\includegraphics[width=3.5 cm]{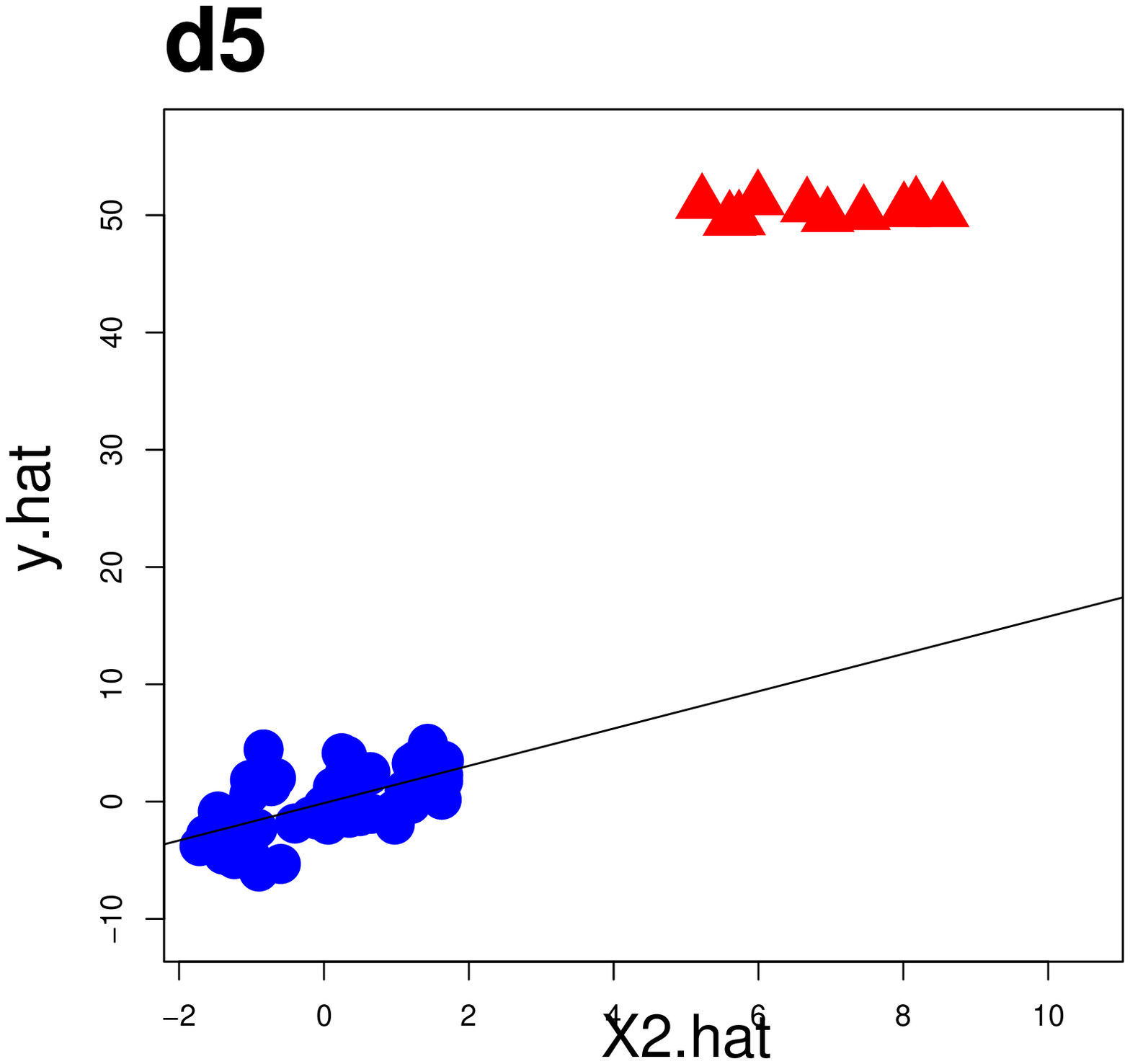}
	\end{minipage}
	\caption{\small The scatter plot of the original variables  ($\protect \bX_1,\protect \bX_2$) and ($\protect \bX,\protect \by$), the profiled variables ($\protect \hbX,\protect \hat{\protect\by}$), ($\protect \hbX_1,\protect\hat{\protect \by}$) and ($\protect \hbX_2,\protect\hat{\protect\by}$) for the datasets with the PC+LMG, PC+LMB, OC+LMG and OC+LMG leverage points. The regular observations and the leverage points are plotted by {\color{blue} $\protect\bullet$} and {\color{red} $\protect\blacktriangle$}, respectively.}
	\label{leverage2}
\end{figure}

For a generated dataset ($\bX$,$\by$) with $\bX = (\bX_1,\bX_2)$ we can obtain $\bz$ by $(\bX-\tbX)\bB$ in this case because $\bB$ is known and there are only two predictors. It follows that the profiled predictors and response are given by: ${\hbX} = (\hbX_1,\hbX_2) = \bX - \hat{\bz}\bB^\text{T}$ and ${\hat{\by}} = \by-\hat{\bz}\bB^\text{T}\bbe$. The scatter plots of the original variables ($\bX_1,\bX_2)$ and ($\bX,\by$), the profiled variables ($\hbX,\hat{\by}$) as well as ($\hbX_1,\hat{\by}$) and ($\hbX_2,\hat{\by}$) are shown in the five rows of Figure \ref{leverage2}. The four columns in Figure \ref{leverage2} correspond to the cases PC+LMG, PC+LMB, OC+LMG and OC+LMB leverage points, respectively.

Since PC outliers become regular observations after factor profiling, i.e. they are non-outlying in the factor profiled predictors, PC+LMG leverage points become regular observations in the multiple regression model~\eqref{RegFP} based on the factor profiled variables, as can be seen in panel a3 of Figure \ref{leverage2}; while PC+LMB leverage points become vertical outliers in this model (see Figure \ref{leverage2}, b3). On the other hand, OC outliers remain outlying in the factor profiled predictors. Hence, OC+LMG leverage points remain good leverage points (see Figure \ref{leverage2}, c3); while OC+LMB leverage points remain bad leverage points (see Figure \ref{leverage2}, d3) in the multiple regression model with factor profiled variables. Let us now look at the marginal regression models based on the profiled variables. The PC+LMG leverage points became regular observations in the multiple model and thus remain regular observations for the marginal models (see Figure \ref{leverage2}, a4 and a5). Similarly, the PC+LMB leverage points remain vertical outliers in the marginal models (see Figure \ref{leverage2}, b4 and a5). On the other hand, while the OC+LMG leverage points remain good leverage points for the multiple model~\eqref{RegFP}, they in general become bad leverage points in the marginal models (see Figure \ref{leverage2}, c4 and c5). Finally, the OC+LMB leverage points remain bad leverage points in the marginal models as well  (see Figure \ref{leverage2}, d4 and d5).

To avoid the adverse effect of outliers, our procedure downweights all types of leverage points in an initial variable screening step. Since outlying scores will affect the estimates of the profiled response variable, we first estimate the profiled response variables based on the observations with non-outlying predictors. Then, we check whether a PC outlier is outlying in the profiled response as well or not, i.e. whether it is a good or a bad leverage point in the regression models. The PC+LMG leverage points will not be downweighted anymore, and both the profiled response and the marginal coefficients will be re-estimated by including these good leverage points to increase efficiency.

Finally, we give an overview of the proposed robust factor profiled sure independence screening (RFPSIS) procedure. The RFPSIS procedure consists of the following steps:

\textbf{Step 1. Profiled predictors.}\\
Standardize each of the original variables using its median and $Q_n$ estimates. 
Fit the factor model to the scaled data robustly by using the least trimmed squares method discussed in Subsection~\ref{sec:FS} to obtain the factor profiled predictors $\htbX$. Then, identify the PC and OC outliers. Denote by $\mathcal{I}_1$ the index set of the regular observations, i.e. the observations with non-outlying predictors according to the factor model. Let $\hbZ_{\mathcal{I}_1}$ denote the sub-matrix of $\hbZ$ which collects the observations corresponding to $\mathcal{I}_1$.

\textbf{Step 2. Profiled response.}
\begin{itemize}
\item[2a.] \textit{Initial profiling.}\\
Regress $\by_{{ \mathcal{I}_1}}$ on $\hbZ_{\mathcal{I}_1}$ robustly to obtain the estimated slope $\hat{\bga}^o$  and intercept $\hmu_y^o$. By default we use a $95\%$ efficient  regression MM-estimator. Let $\hsi_y^o$ denote the estimated error scale. An initial estimate of the profiled response is then obtained by $\hy_i^o = y_i - \hmu^o_y- {\hat{\bz}_i}^\text{T}\hat{\gamma}^o$, $i=1,\ldots,n$.

\item[2b.] \textit{Outlier identification.}\\
Denote $\mathcal{I}_\text{s}$ as the index set of the PC outliers identified in Step 1. 
For each of these PC outliers, check whether it is a vertical outlier or a regular observation based on its standardized residual $\hat{t}_i = {\hy_i^o}/{\hat{\sigma}^o_{\hy}}$ corresponding to the regression model in Step 2a. Define an enlarged index set $\mathcal{I}_2 =\mathcal{I}_1\cup \{ i \in \mathcal{I}_s: \hat{t}_i^2\leqslant \chi^2_{0.975,1}\}$.

\item[2c.] \textit{Updated profiling.}\\
Calculate updated estimates $\hmu_y$, $\hat{\bga}$ and $\hby_{\mathcal{I}_2}$ by 
regressing $\by_{{ \mathcal{I}_2}}$ on $\hbZ_{\mathcal{I}_2}$ using the MM-estimator (by default). The updated estimate of the profiled response is then given by $\hat{y}_i = y_i - \hmu_y-\hat{\bz}_i^\text{T}\hat{\gamma}$, $i=1,\ldots,n$. 
\end{itemize}

\textbf{Step 3. Variable screening.}\\
Regress $\hby_{\mathcal{I}_2}$ robustly on each of the corresponding profiled predictors $(\hbX_{\mathcal{I}_2})_j$ ($j=1,\ldots,p$), using a $95\%$ efficient simple regression MM-estimator by default. Let $\hbth=(\hth_1,\ldots,\hth_p)^\text{T}\in\mathbb{R}^p$ be the marginal slope estimates. Sort these estimates according to decreasing absolute value to obtain the solution path $\mathbb{M} = \{\mathcal{M}_{(k)}: k=0,\ldots,p\}$, with $\mathcal{M}_{(0)}=\emptyset$ and $\mathcal{M}_{(k)}=\{j: |\hth_{j}| \text{ belongs to the largest } k \text{ values }\}$, $k=1,\ldots,p$.

\subsection{Empirical performance study}
\label{sec:perf}

To investigate the performance of RFPSIS, we generate regular data as in~\citep{FPSIS}. The predictors are obtained by \scalebox{1}{$\bX = \bZ\bB^\text{T} + \tbX$}, where $\bZ\sim N_{d}(\boldsymbol{0},\bI_d)$, $\bB\sim N_{d}(\boldsymbol{0},\bI_d)$, $\tbX\sim N_p(\boldsymbol{0},\bI_p)$. The response is generated as \scalebox{1}{$\by = \bX\bth_0+\bves$} with coefficients 
	\begin{equation}
	\label{theta}
	\theta_{0j}=    \begin{cases}
	(-1)^{R_{aj}}(4n^{-1/2}\log n +|R_{bj}|) & \text{for}\ j=1,\ldots,|\mathcal{M}_\text{T}| \\
	0, & \text{otherwise}
	\end{cases}
	\end{equation}
	where $R_{aj}\sim B(1,0.4)$, $R_{bj}\sim N(0,1)$ and $|\mathcal{M}_\text{T}|=8$. 
	Hence, there are 8 important variables in the model. Moreover, the errors are generated according to $\bves = \bZ^\text{T}\bal_0+\tbves$, with
	  $\bal_0=0.8\sigma_\varepsilon(\sqrt{2},\sqrt{2})^\text{T}\in\mathbb{R}^2$ and $\tbves\sim N(0,\tilde{\sigma}_\varepsilon^2)$, where $\tilde{\sigma}_\varepsilon=0.6\sigma_\varepsilon$ and $\sigma_\varepsilon^2=\text{var}(\boldsymbol{X}^\text{T}\bth_0)/c$, with $c$ the signal-to-noise ratio.

To study the robustness of our method, we replace a fraction of the observations by outliers. 
Let  $y_{\min}$ and $y_{\max}$ be the minimal and maximal value of the regular responses, respectively. Then, we simulate outlying responses by replacing the original response $y_i$ of the observation by
 $y_{\text{\tiny LMV}}\sim N(\mu_c^y ,1)$, where $\mu_c^y = y_{\max}\cdot\mathcal{I}(y_i\leqslant \frac{y_{\min} + y_{\max}}{2}) + y_{\min}\cdot\mathcal{I}(y_i >\frac{y_{\min} + y_{\max}}{2})$ with $I(\cdot)$ the indicator function. In this way we generate a set of vertical outliers which lie at the tails of the response distribution. These outliers are extreme vertical outliers while they are hard to detect by inspecting the empirical distribution of $\by$. 

Next to vertical outliers we also consider leverage points. Leverage points are generated as either PC outliers or OC outliers.
PC outliers are generated as 
$\bX_\text{\tiny PC} = \bZ_\text{\tiny PC}\bB^\text{T} + \tbX$, where $\bZ_\text{\tiny PC} \sim N_{d}(\bmu_\text{\tiny PC},\bI_{d})$ and $\bmu_\text{\tiny PC}=5\cdot\bone_{d}^\text{T}$. 
OC outliers are generated as $\bX_\text{\tiny OC}\sim N_p(\bmu_\text{\tiny OC},\bI_p)$, where $\bmu_\text{\tiny OC}=10\cdot(\underbrace{1,\ldots,1}_{0.2p},0\ldots,0)_p^\text{T}$. Both good leverage points and bad leverage points for the linear model are considered: for good leverage points, the response is generated according to the true regression model; and for bad leverage points, the response is simulated in the same way as vertical outliers.

	The following five contamination levels are considered:
	\begin{compactlist}
		\item {\emph Case 1.} $\epsilon = 0\%$, no contamination;
		\item {\emph Case 2.} $\epsilon =5\%$ (good/bad) leverage points, no vertical outliers;
		\item {\emph Case 3.} $\epsilon =5\%$ (good/bad) leverage points + $5\%$ extra vertical outliers;
		\item {\emph Case 4.} $\epsilon = 20\%$ (good/bad) leverage points, no vertical outliers;
		\item {\emph Case 5.} $\epsilon = 20\%$ (good/bad) leverage points + $10\%$ extra vertical outliers.
	\end{compactlist}

The simulations are performed for different combinations of  $p$ ($1000$ or $10000$), $n$ ($200$ or $400$) and $d$ ($2$ or $5$). We also consider three levels for the signal-to-noise ratio, by setting $c=1$, $3$ or $5$. Screening performance is measured by the minimal model size that is required to cover $m$ ($m=1,\ldots,\mathcal{M}_\text{T}$) of the important variables. 
For each setting, we use 200 simulated datasets and report both the median and the $95\%$ quantile of the minimal model size. Here, we only present the simulation results for $d=2$,  $n=200$ or $400$, and $p=10000$. The results for the other settings lead to similar conclusions and can be found in the Supplemental Material.
 
\begin{figure}[ht!]
 	\centering
 	\begin{minipage}{0.01\textwidth}
 		\hfill
 	\end{minipage}
 	\begin{minipage}{0.32\textwidth}
 		\centering
 		(a) $n=200$, $c=1$ 
 	\end{minipage}
 	\begin{minipage}{0.32\textwidth}
 		\centering
 		(b) $n=200$, $c=3$
 	\end{minipage}
 	\begin{minipage}{0.32\textwidth}
 		\centering
 		(c) $n=200$, $c=5$
 	\end{minipage}\\
 	
 	\begin{minipage}{0.01\textwidth}
 		\rotatebox[]{90}{\footnotesize Median}
 	\end{minipage}
 	\begin{minipage}{0.32\textwidth}
 		\centering
 		\includegraphics[width=3.8 cm]{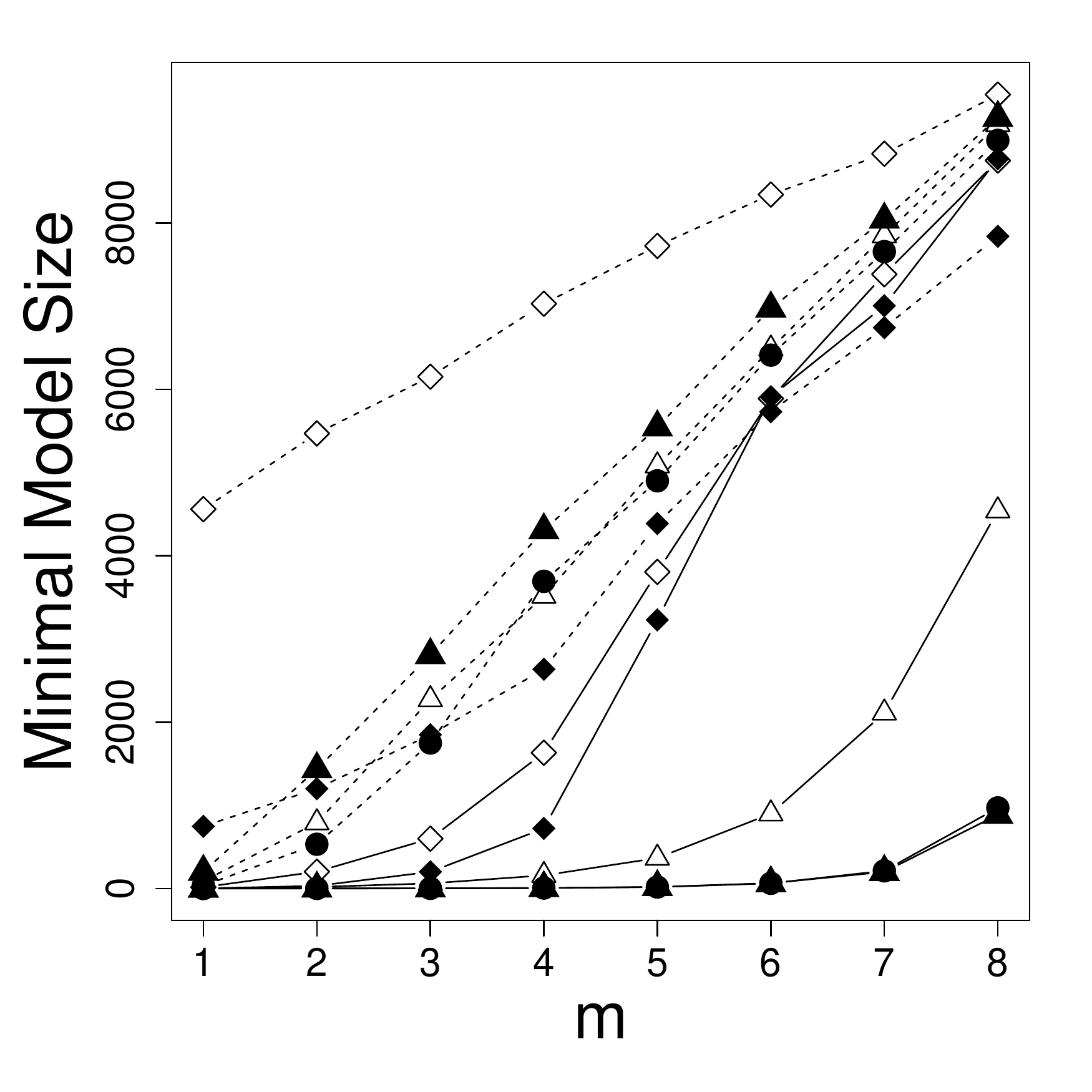}
 	\end{minipage}
 	\begin{minipage}{0.32\textwidth}
 		\centering
 		\includegraphics[width=3.8 cm]{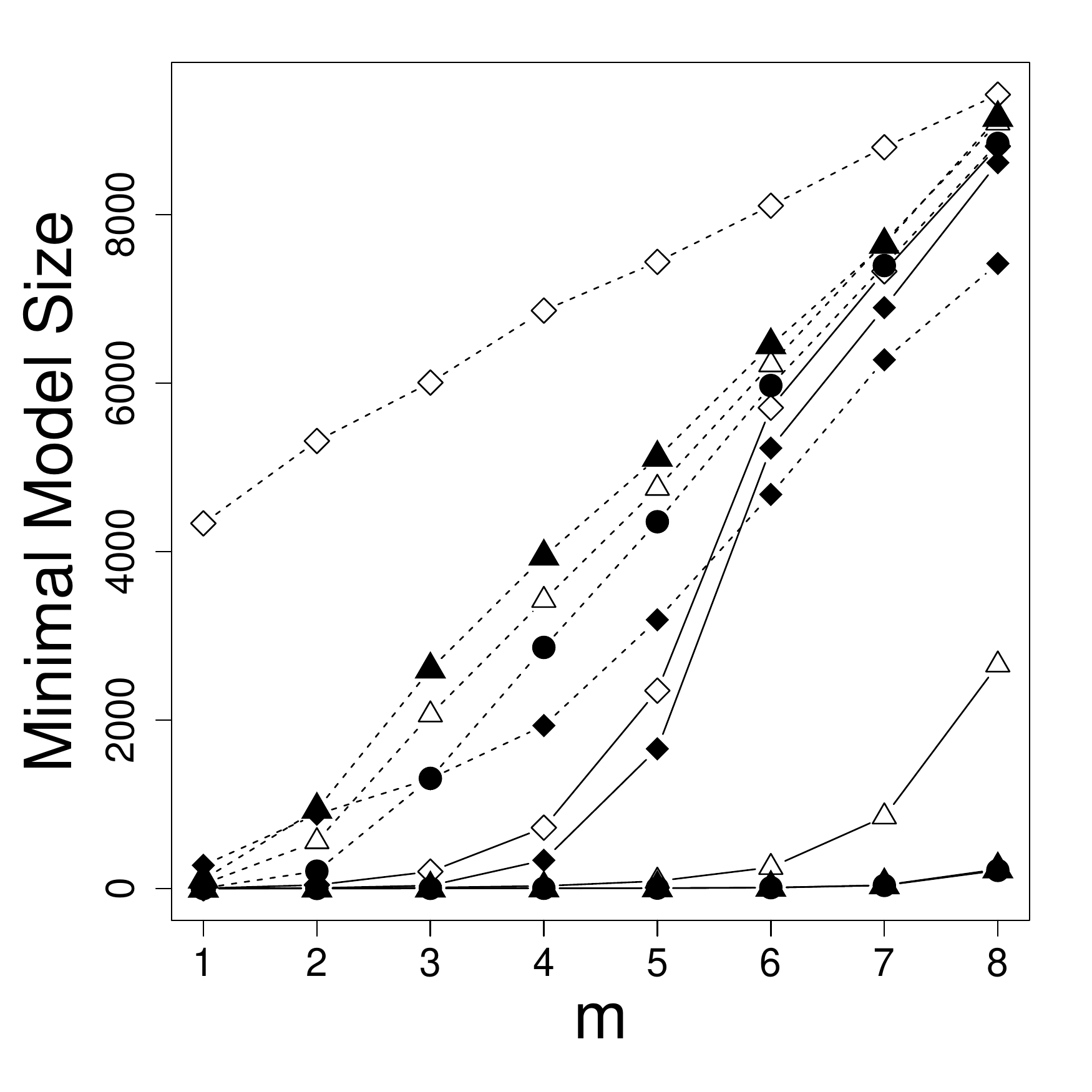}
 	\end{minipage}
 	\begin{minipage}{0.32\textwidth}
 		\centering
 		\includegraphics[width=3.8 cm]{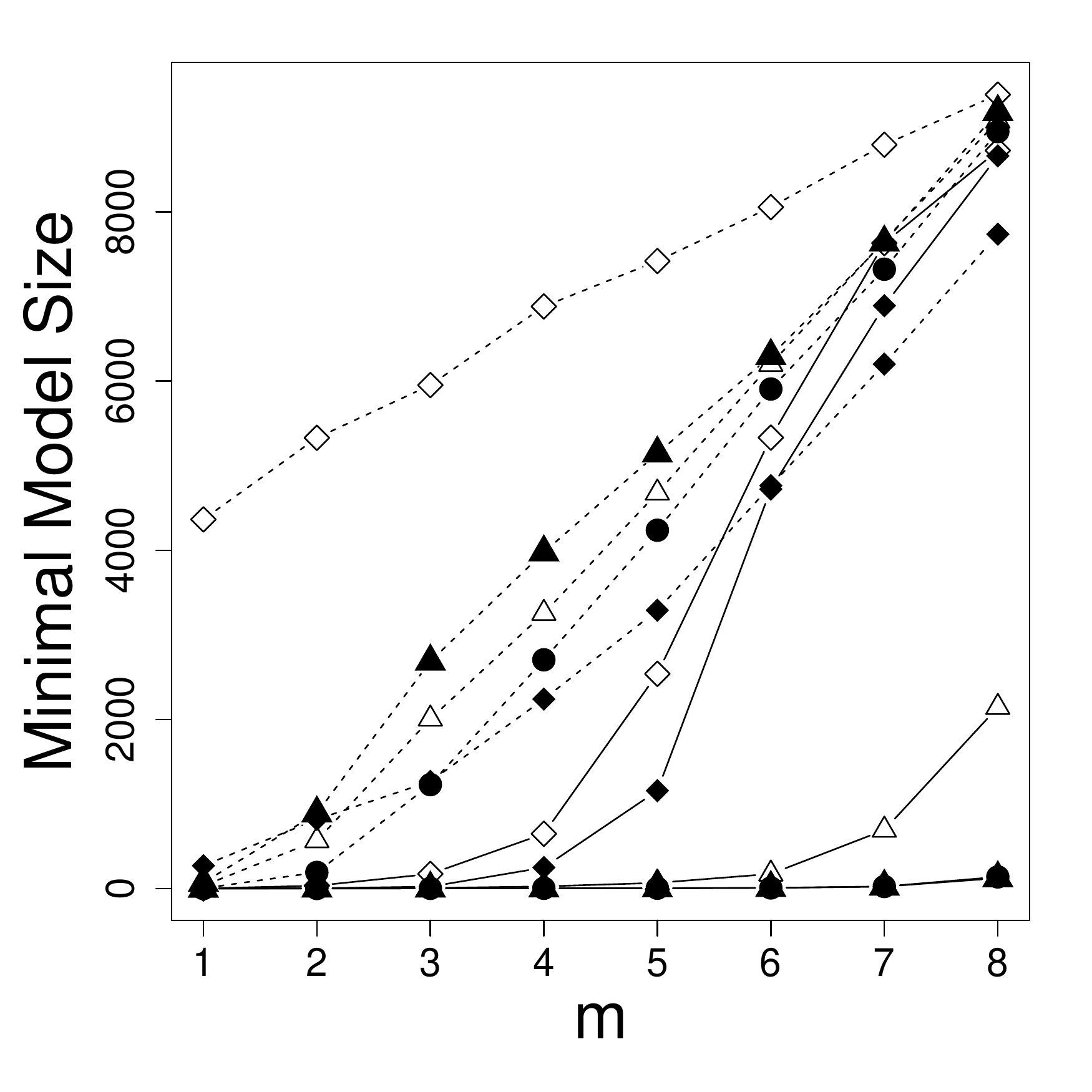}
 	\end{minipage}\\
 	
 	\begin{minipage}{0.01\textwidth}
 		\rotatebox[]{90}{\footnotesize $95\%$ Quantile}
 	\end{minipage}
 	\begin{minipage}{0.32\textwidth}
 		\centering
 		\includegraphics[width= 3.8 cm]{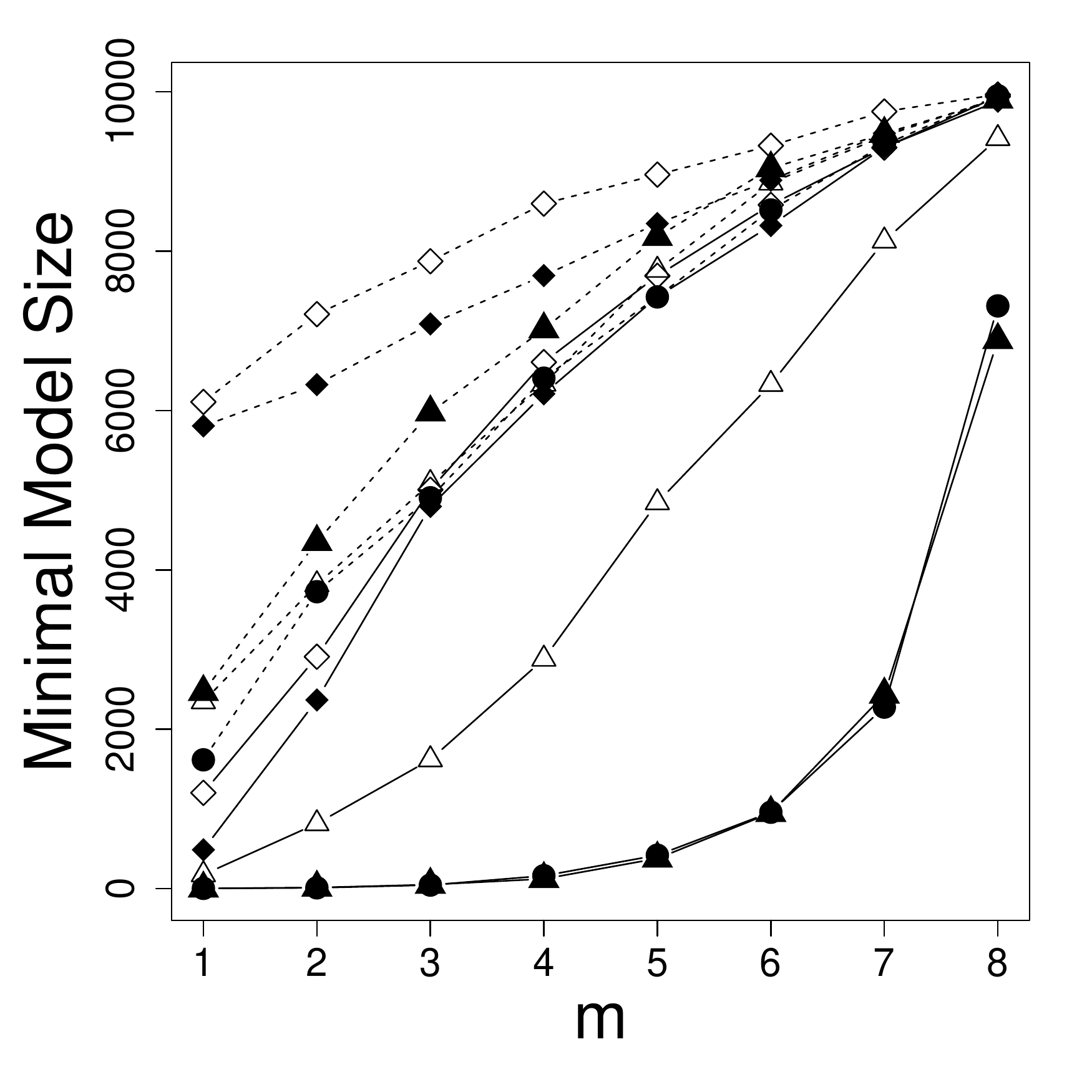}
 	\end{minipage}
 	\begin{minipage}{0.32\textwidth}
 		\centering
 		\includegraphics[width= 3.8 cm]{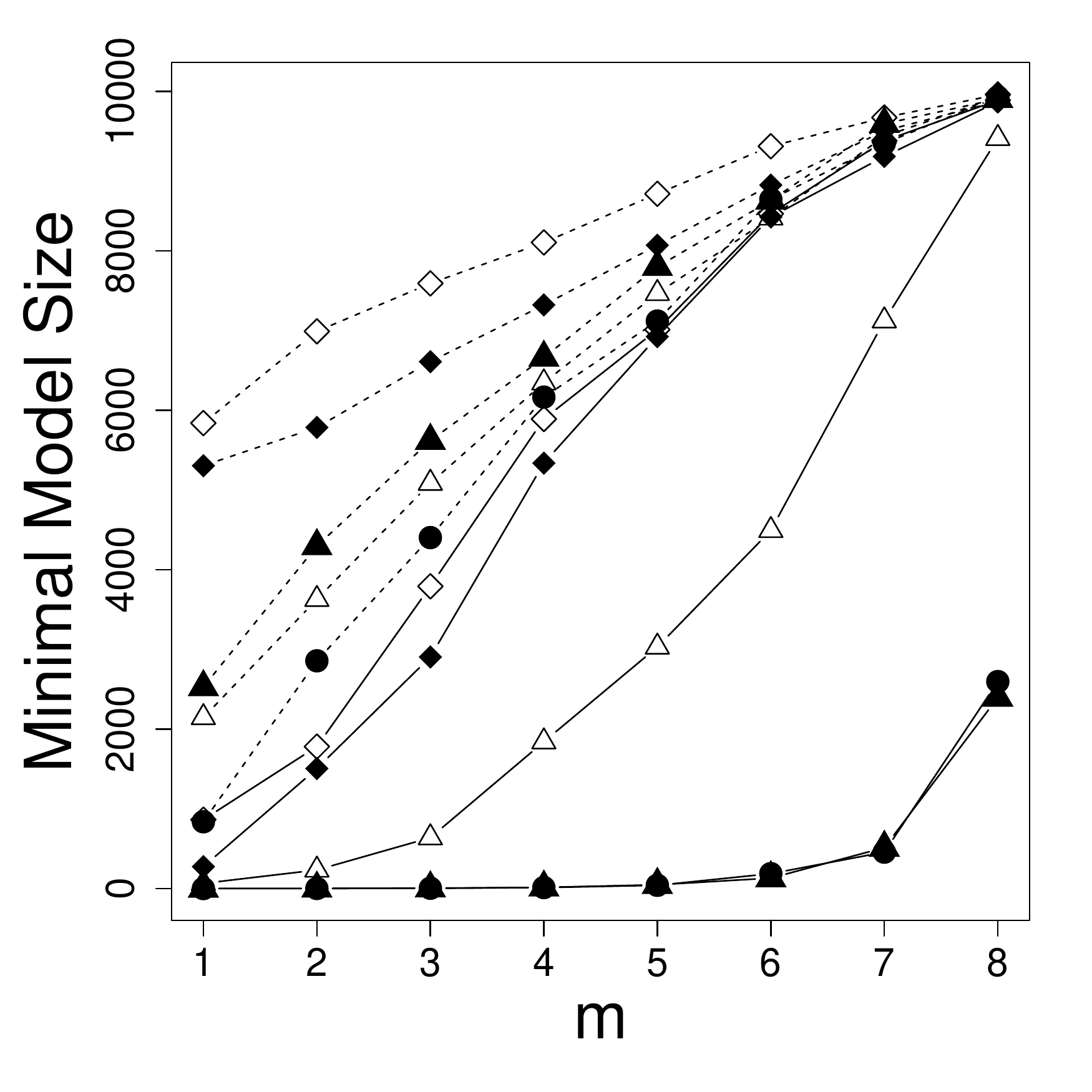}
 	\end{minipage}
 	\begin{minipage}{0.32\textwidth}
 		\centering
 		\includegraphics[width= 3.8 cm]{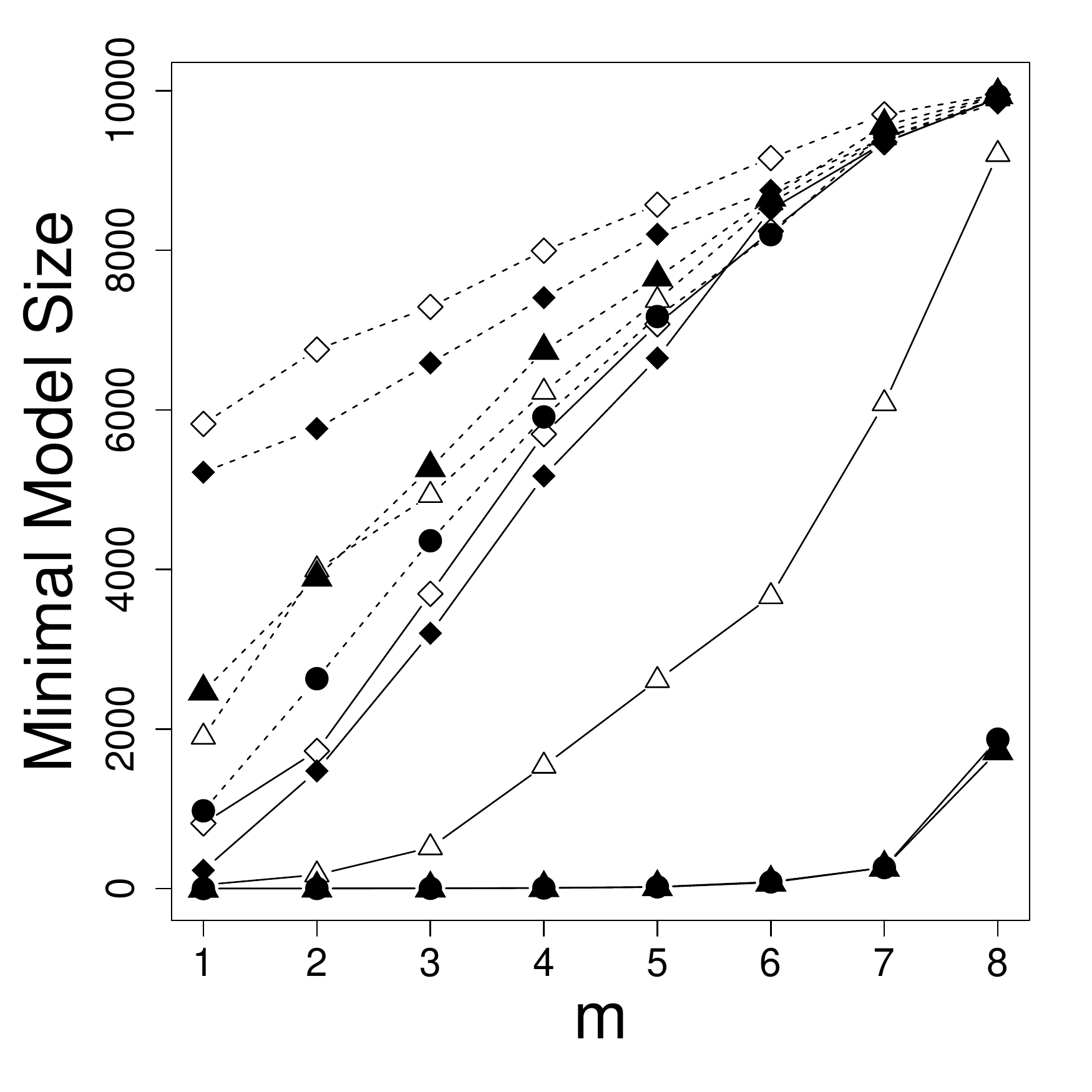}
 	\end{minipage}\\

 	\begin{minipage}{\textwidth}
 		\hfill
 	\end{minipage}\\
 	
 	\begin{minipage}{0.01\textwidth}
 		\hfill
 	\end{minipage}
 	\begin{minipage}{0.32\textwidth}
 		\centering
 		(e) $n=400$, $c=1$
 	\end{minipage}
 	\begin{minipage}{0.32\textwidth}
 		\centering
 		(f) $n=400$, $c=3$
 	\end{minipage}
 	\begin{minipage}{0.32\textwidth}
 		\centering
 		(g) $n=400$,  $c=5$
 	\end{minipage}\\
 	
 	\begin{minipage}{0.01\textwidth}
 		\rotatebox[]{90}{\footnotesize Median}
 	\end{minipage}
 	\begin{minipage}{0.32\textwidth}
 		\centering
 		\includegraphics[width=3.8 cm]{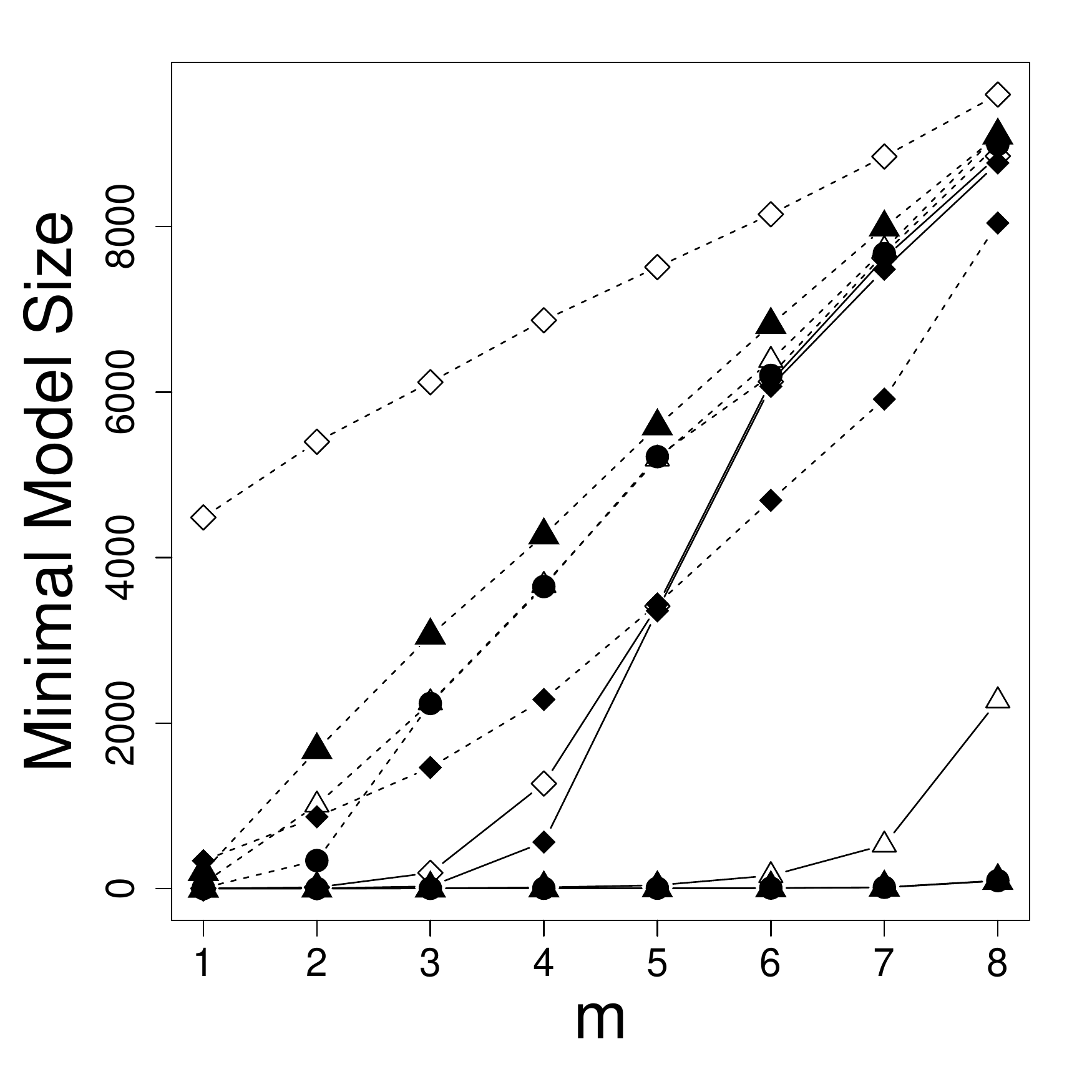}
 	\end{minipage}
 	\begin{minipage}{0.32\textwidth}
 		\centering
 		\includegraphics[width=3.8 cm]{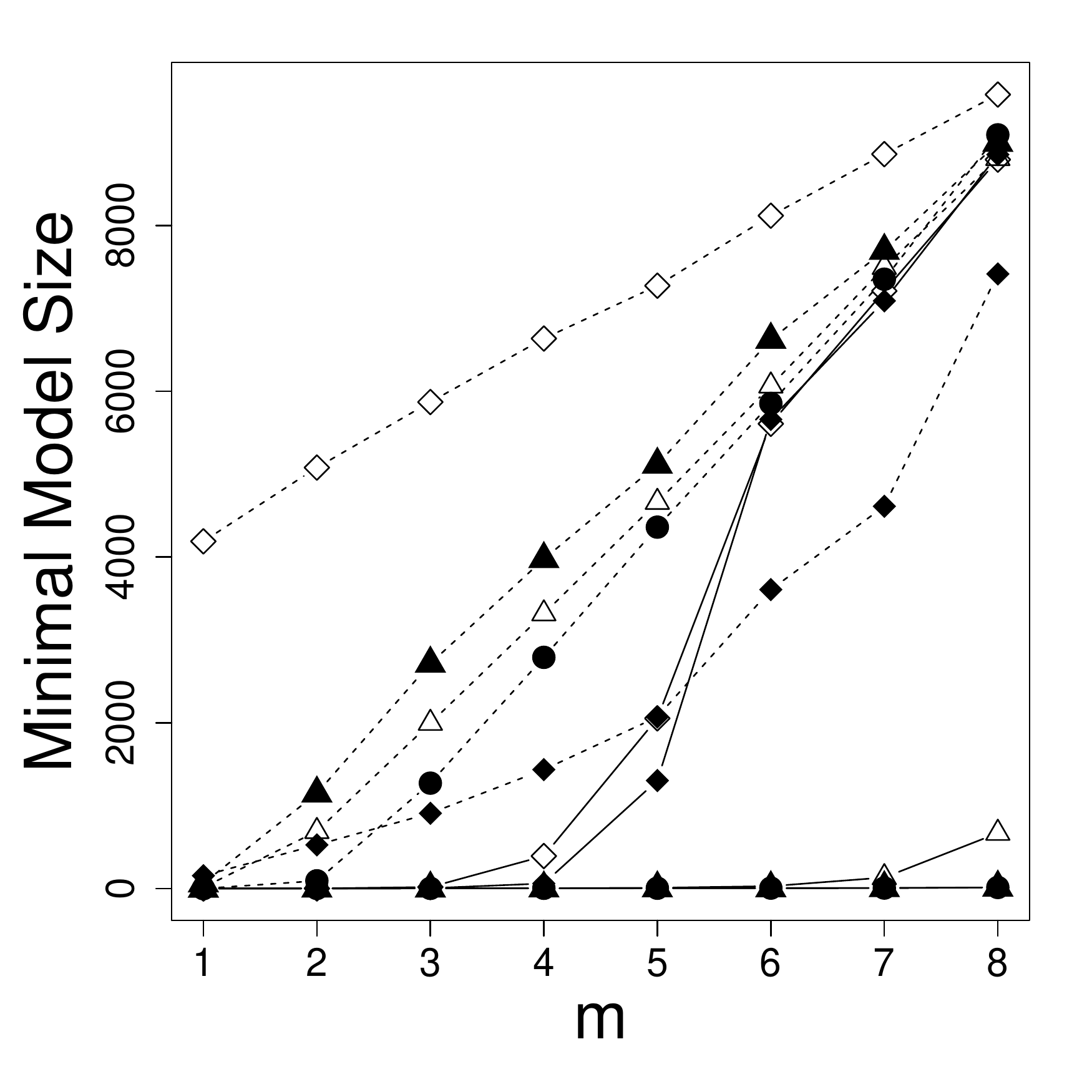}
 	\end{minipage}
 	\begin{minipage}{0.32\textwidth}
 		\centering
 		\includegraphics[width=3.8 cm]{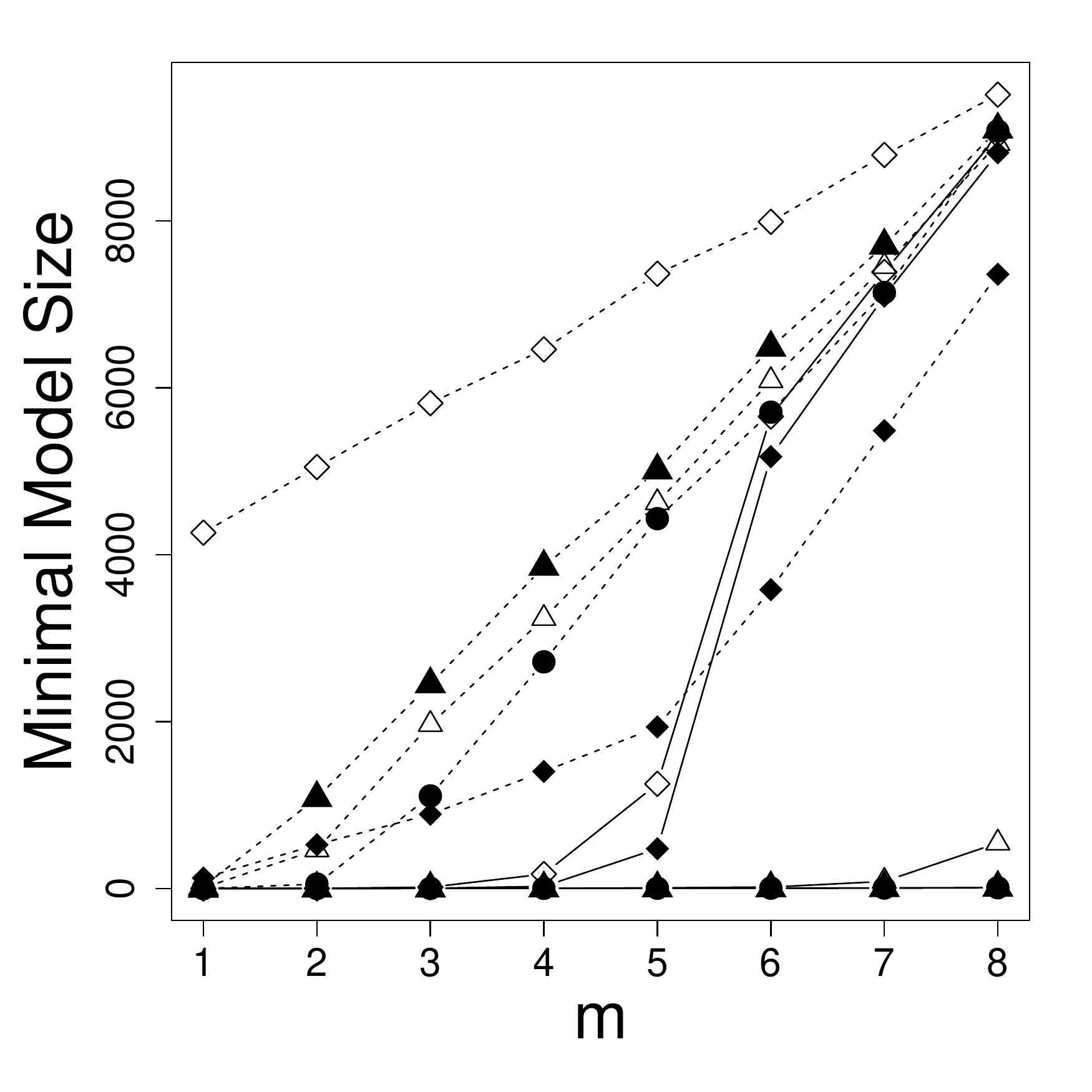}
 	\end{minipage}\\
 	
 	\begin{minipage}{0.01\textwidth}
 		\rotatebox[]{90}{\footnotesize $95\%$ Quantile}
 	\end{minipage}
 	\begin{minipage}{0.32\textwidth}
 		\centering
 		\includegraphics[width= 3.8 cm]{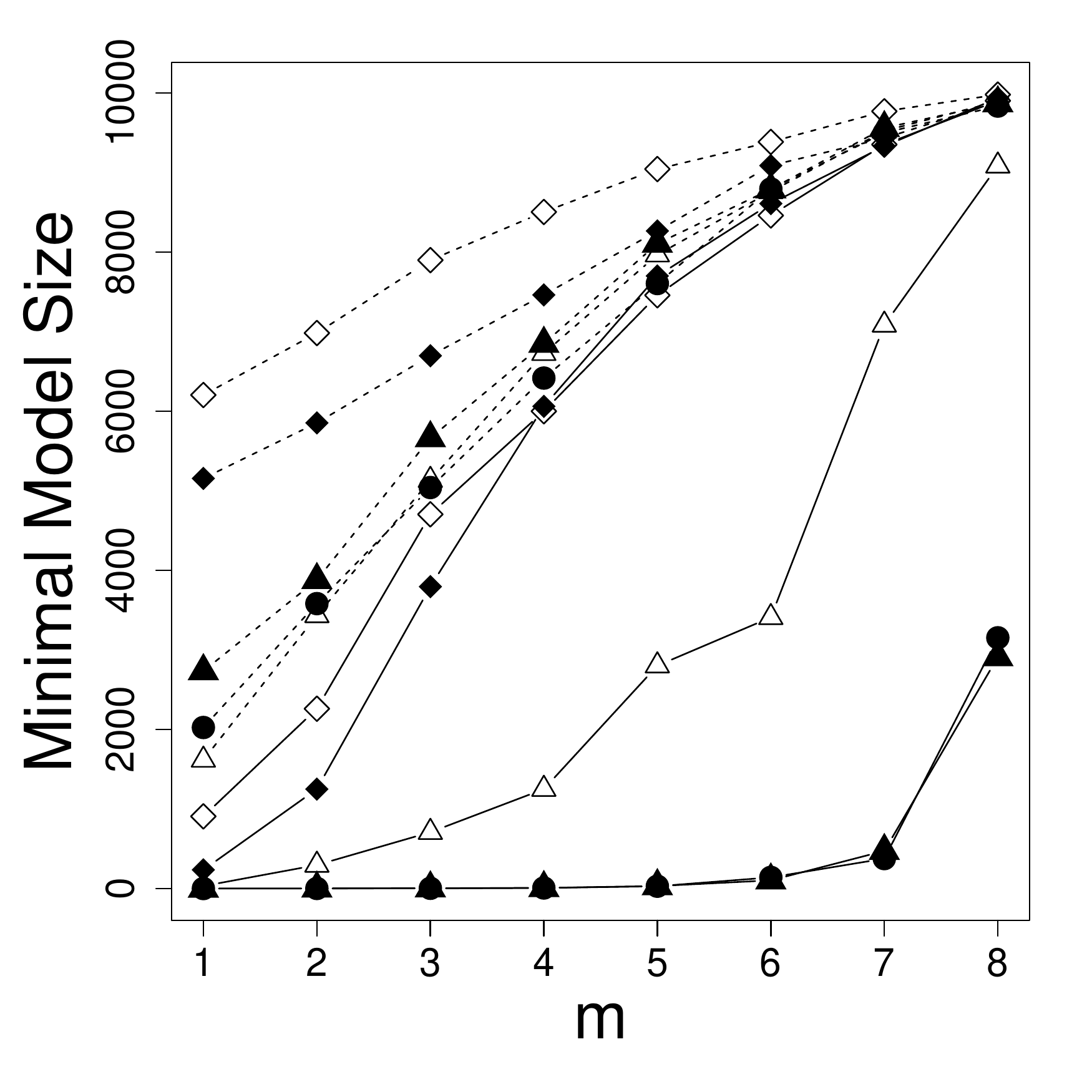}
 	\end{minipage}
 	\begin{minipage}{0.32\textwidth}
 		\centering
 		\includegraphics[width= 3.8 cm]{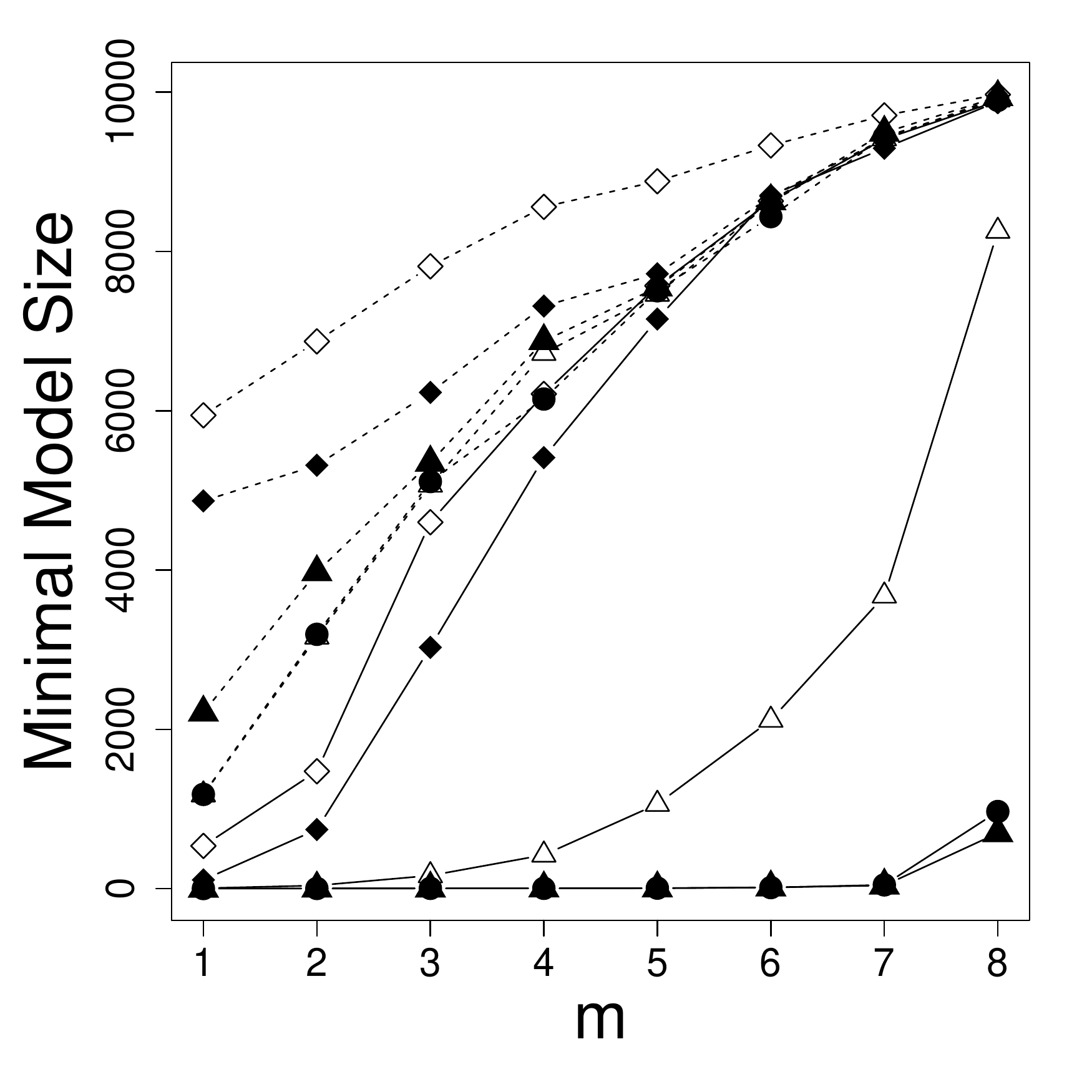}
 	\end{minipage}
 	\begin{minipage}{0.32\textwidth}
 		\centering
 		\includegraphics[width= 3.8 cm]{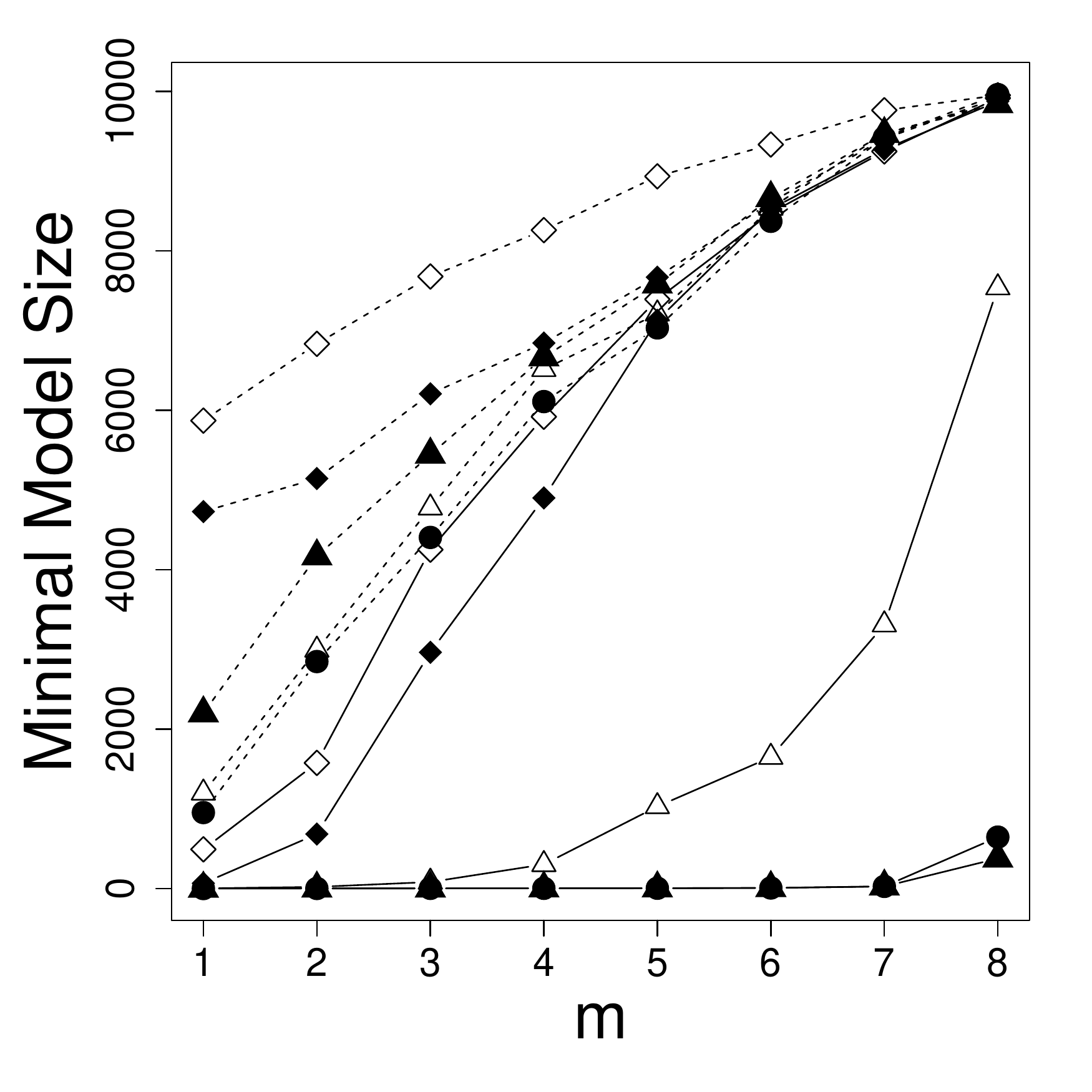}
 	\end{minipage}\\
 	\caption{
 		\small Median and $\protect 95\%$ of the minimal model size needed to capture $m$ important variables by SIS (dotted lines) and FPSIS (solide lines) 
 		for data containing regular observations (bullets), PC+LMG (black triangles),  PC+LMB (triangles),  OC+LMG (black diamonds) and  OC+LMB (diamonds) with $p=10000$ and $d=2$.}
 	\label{location_classical}
 \end{figure}

The results for SIS and FPSIS on regular data and data with $5\%$ leverage points are shown in Figure~\ref{location_classical}. We can see that the SIS curves increase quickly, even for regular data. SIS can only detect the first two important predictors with a reasonable model size. Clearly, SIS fails in all cases due to the correlation in the data. On the other hand, FPSIS which takes the correlation structure into account performs well on clean data. FPSIS shows nearly optimal performance on regular data with a moderate sample size and signal-to-noise ratio. The decreasing sample size or the signal-to-noise ratio only affect FPSIS in the model size required to screen out the last few important variables. Interestingly, FPSIS can obtain equally good results for data with good PC leverage points as for regular data. However, when the data contains bad PC leverage points or (good/bad) OC leverage outliers, FPSIS can at best pick up 3 to 4 important predictors in the beginning of its solution path in case of a large sample size and a high signal-to-noise ratio, but the model size required to include the remaining ones increases dramatically. 
 
\begin{figure}[ht!]
	\centering
	\begin{minipage}{0.01\textwidth}
		\hfill
	\end{minipage}
	\begin{minipage}{0.32\textwidth}
		\centering
		(a) $n=200$, $c=1$ 
	\end{minipage}
	\begin{minipage}{0.32\textwidth}
		\centering
		(b) $n=200$, $c=3$
	\end{minipage}
	\begin{minipage}{0.32\textwidth}
		\centering
		(c) $n=200$, $c=5$
	\end{minipage}\\
	
	\begin{minipage}{0.01\textwidth}
		\rotatebox[]{90}{\footnotesize Median}
	\end{minipage}
	\begin{minipage}{0.32\textwidth}
		\centering
		\includegraphics[width=3.8 cm]{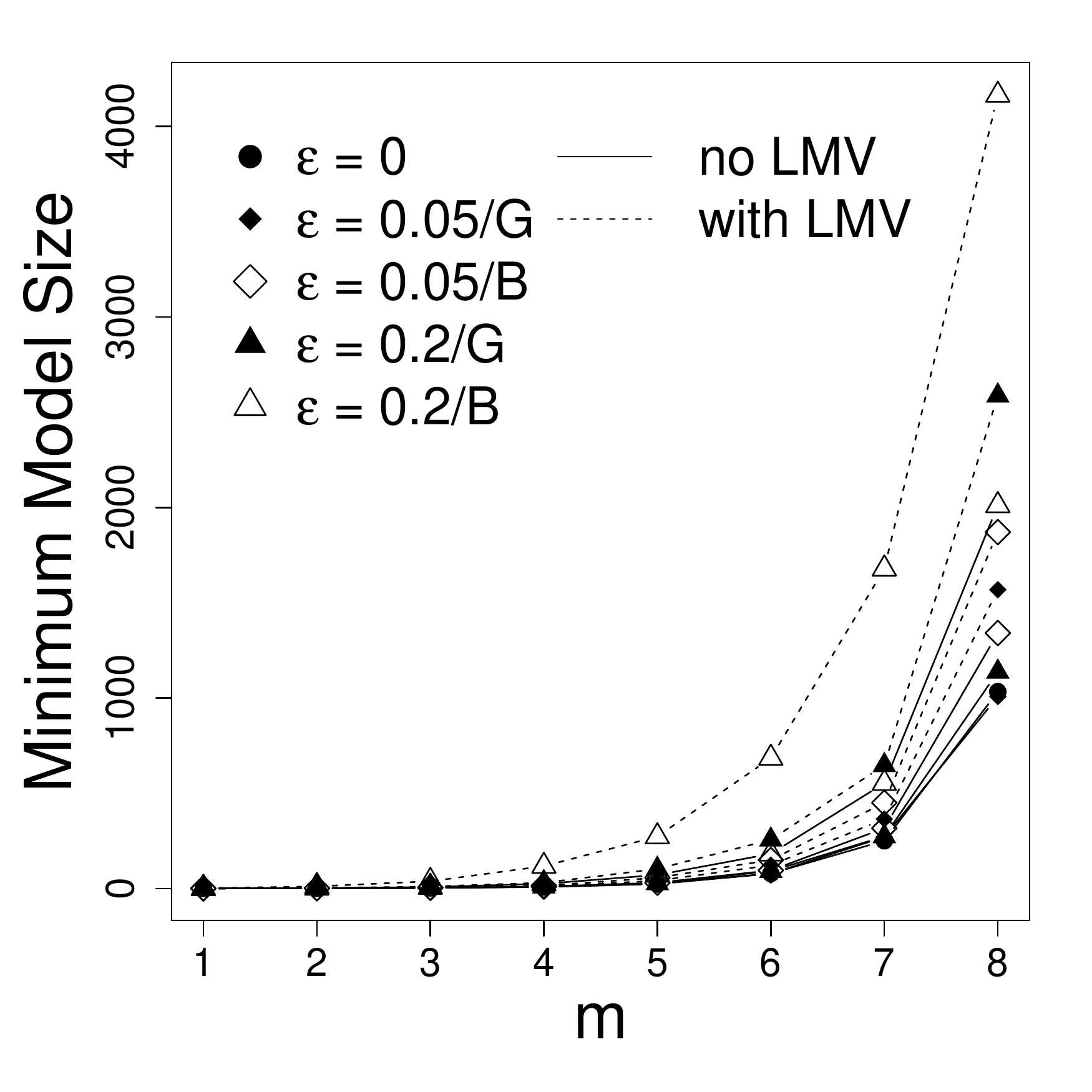}
	\end{minipage}
	\begin{minipage}{0.32\textwidth}
		\centering
		\includegraphics[width=3.8 cm]{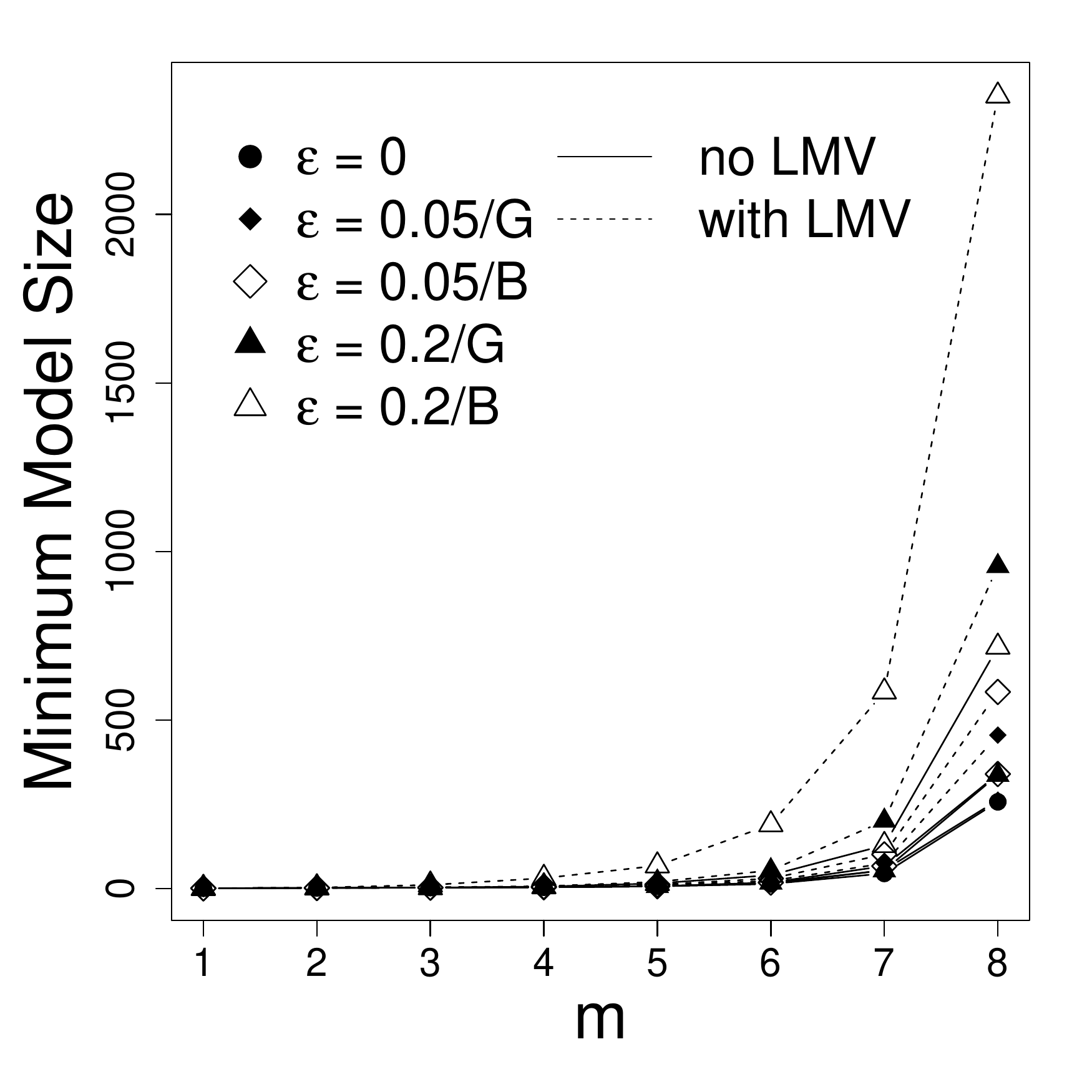}
	\end{minipage}
	\begin{minipage}{0.32\textwidth}
		\centering
		\includegraphics[width=3.8 cm]{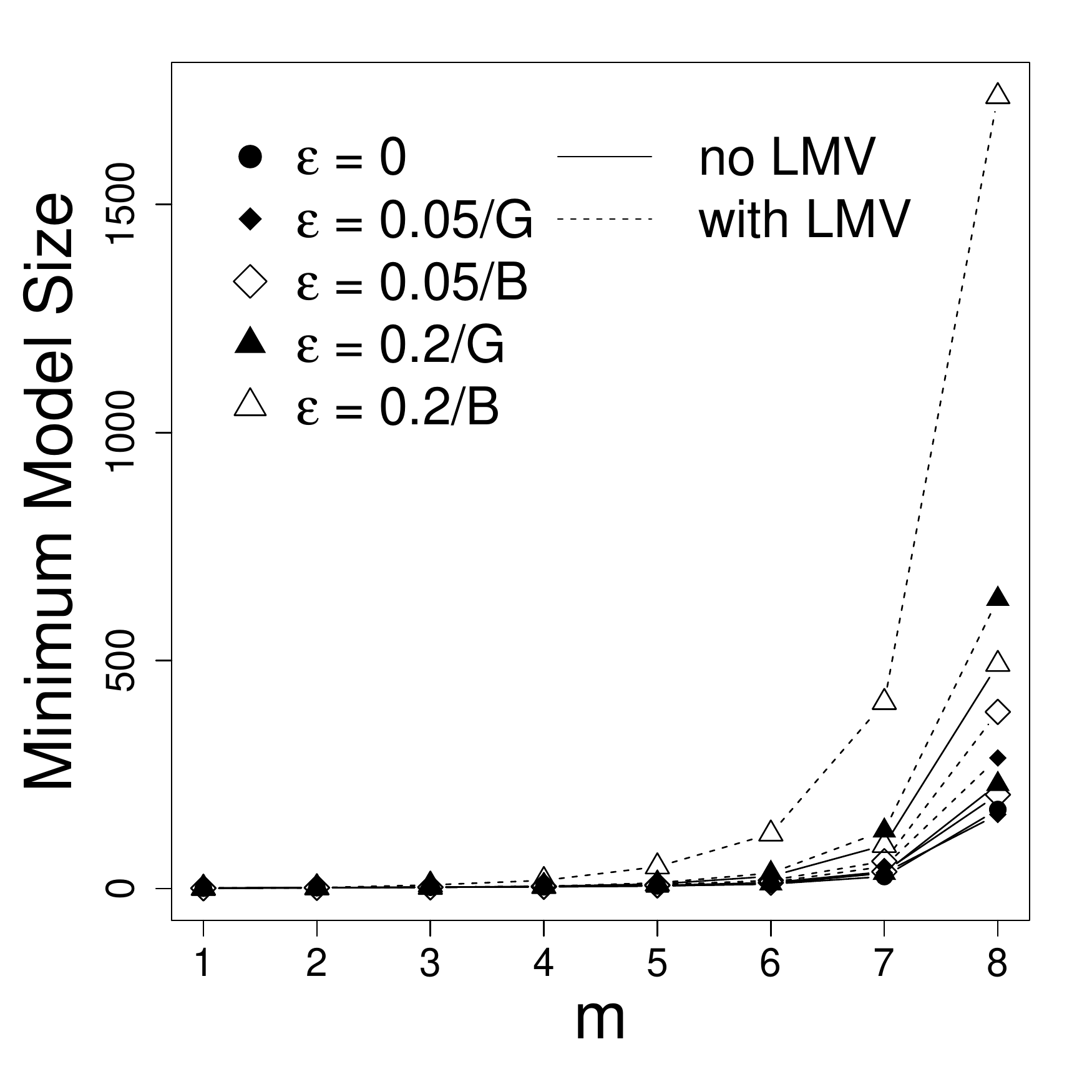}
	\end{minipage}\\
	
	\begin{minipage}{0.01\textwidth}
		\rotatebox[]{90}{\footnotesize $95\%$ Quantile}
	\end{minipage}
	\begin{minipage}{0.32\textwidth}
		\centering
		\includegraphics[width= 3.8 cm]{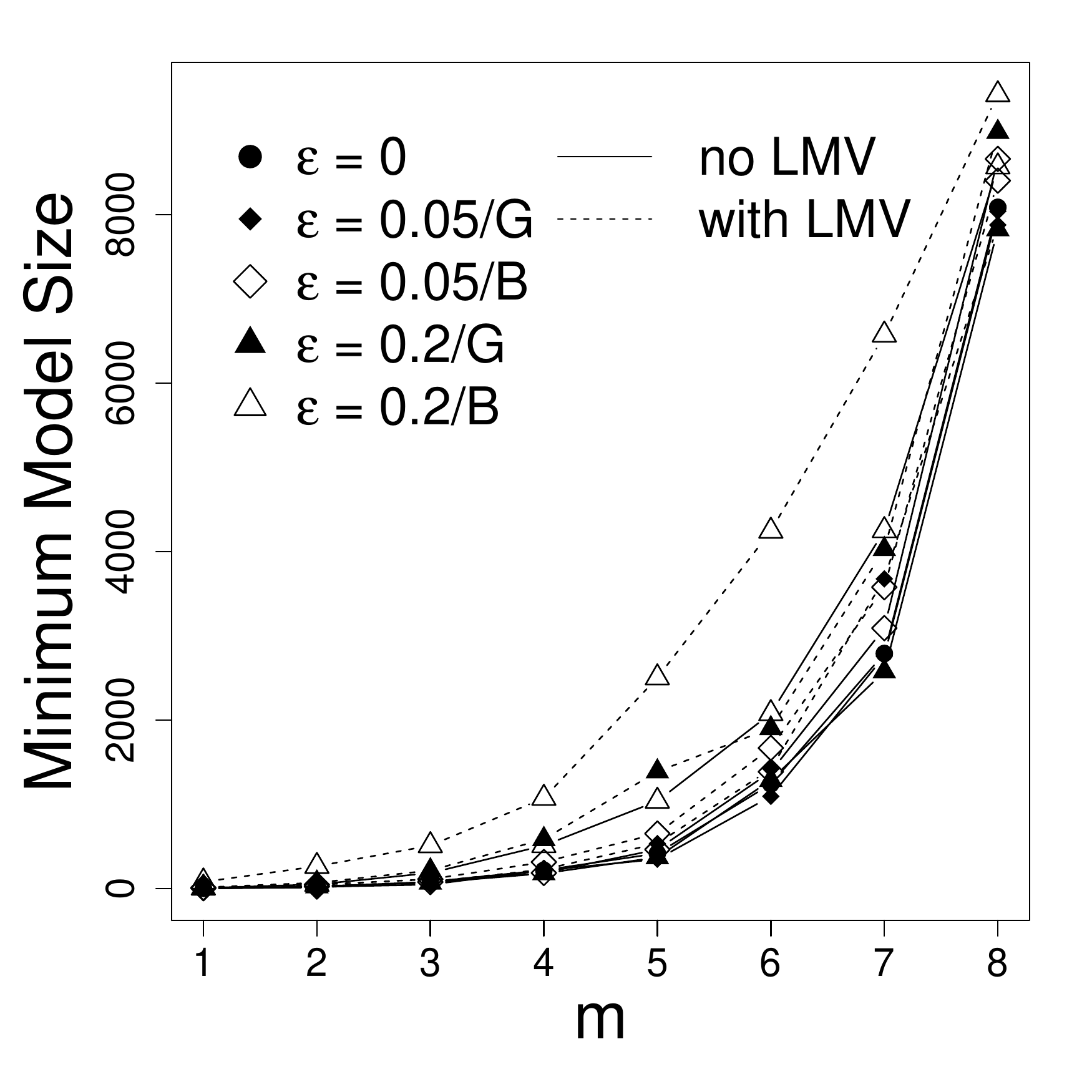}
	\end{minipage}
	\begin{minipage}{0.32\textwidth}
		\centering
		\includegraphics[width= 3.8 cm]{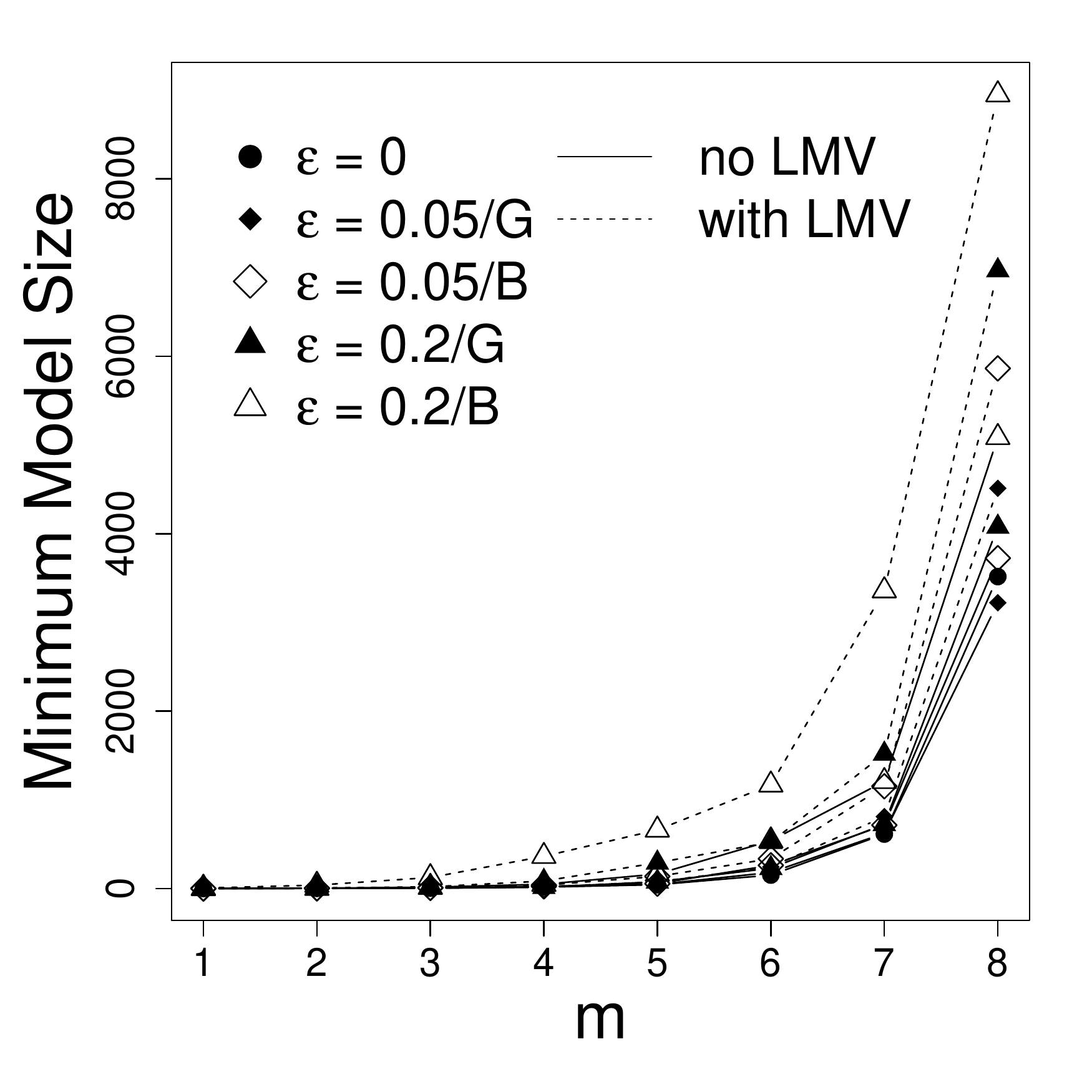}
	\end{minipage}
	\begin{minipage}{0.32\textwidth}
		\centering
		\includegraphics[width= 3.8 cm]{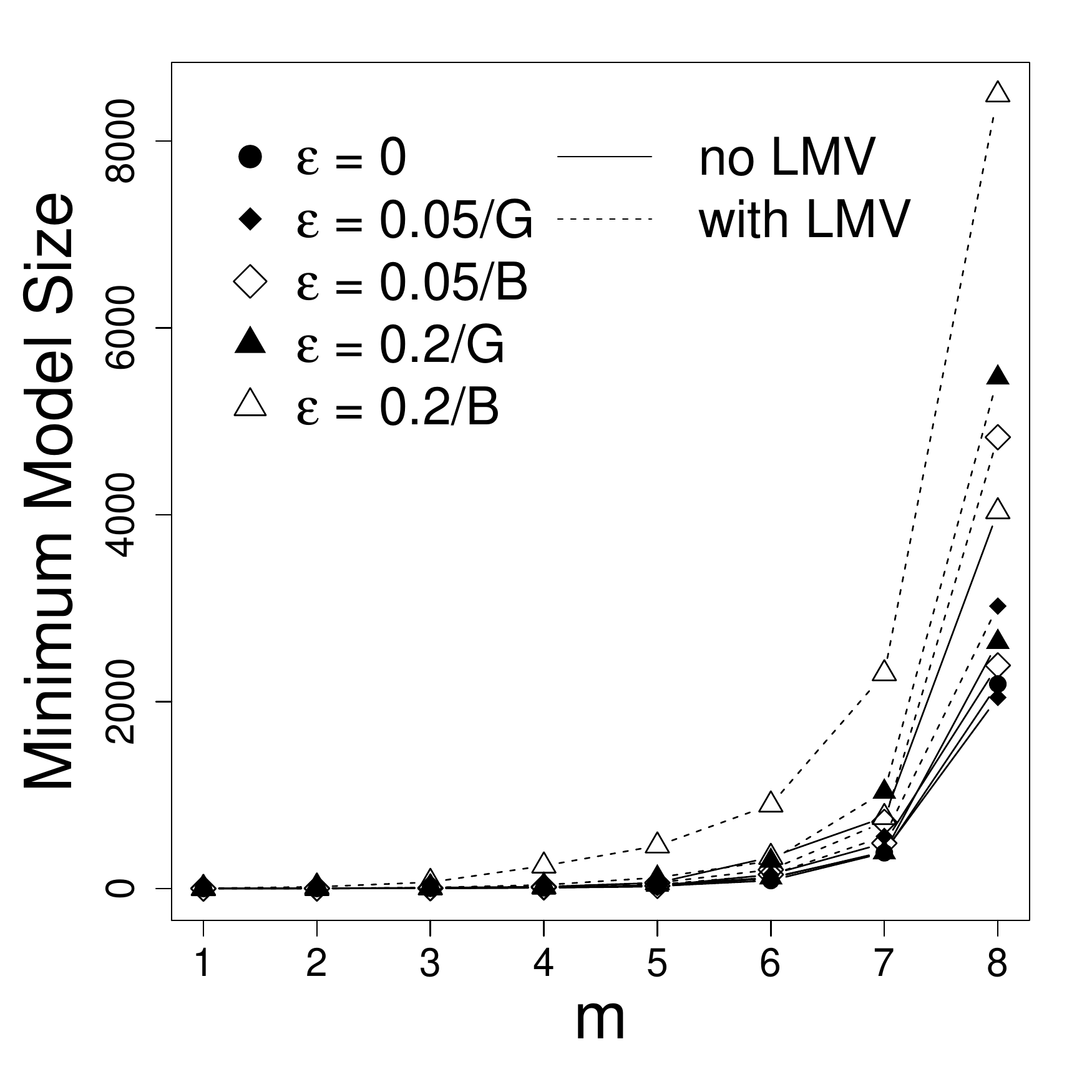}
	\end{minipage}\\

	\begin{minipage}{\textwidth}
		\hfill
	\end{minipage}\\
	
	\begin{minipage}{0.01\textwidth}
		\hfill
	\end{minipage}
	\begin{minipage}{0.32\textwidth}
		\centering
		(e) $n=400$, $c=1$
	\end{minipage}
	\begin{minipage}{0.32\textwidth}
		\centering
		(f) $n=400$, $c=3$
	\end{minipage}
	\begin{minipage}{0.32\textwidth}
		\centering
		(g) $n=400$,  $c=5$
	\end{minipage}\\
	
	\begin{minipage}{0.01\textwidth}
		\rotatebox[]{90}{\footnotesize Median}
	\end{minipage}
	\begin{minipage}{0.32\textwidth}
		\centering
		\includegraphics[width=3.8 cm]{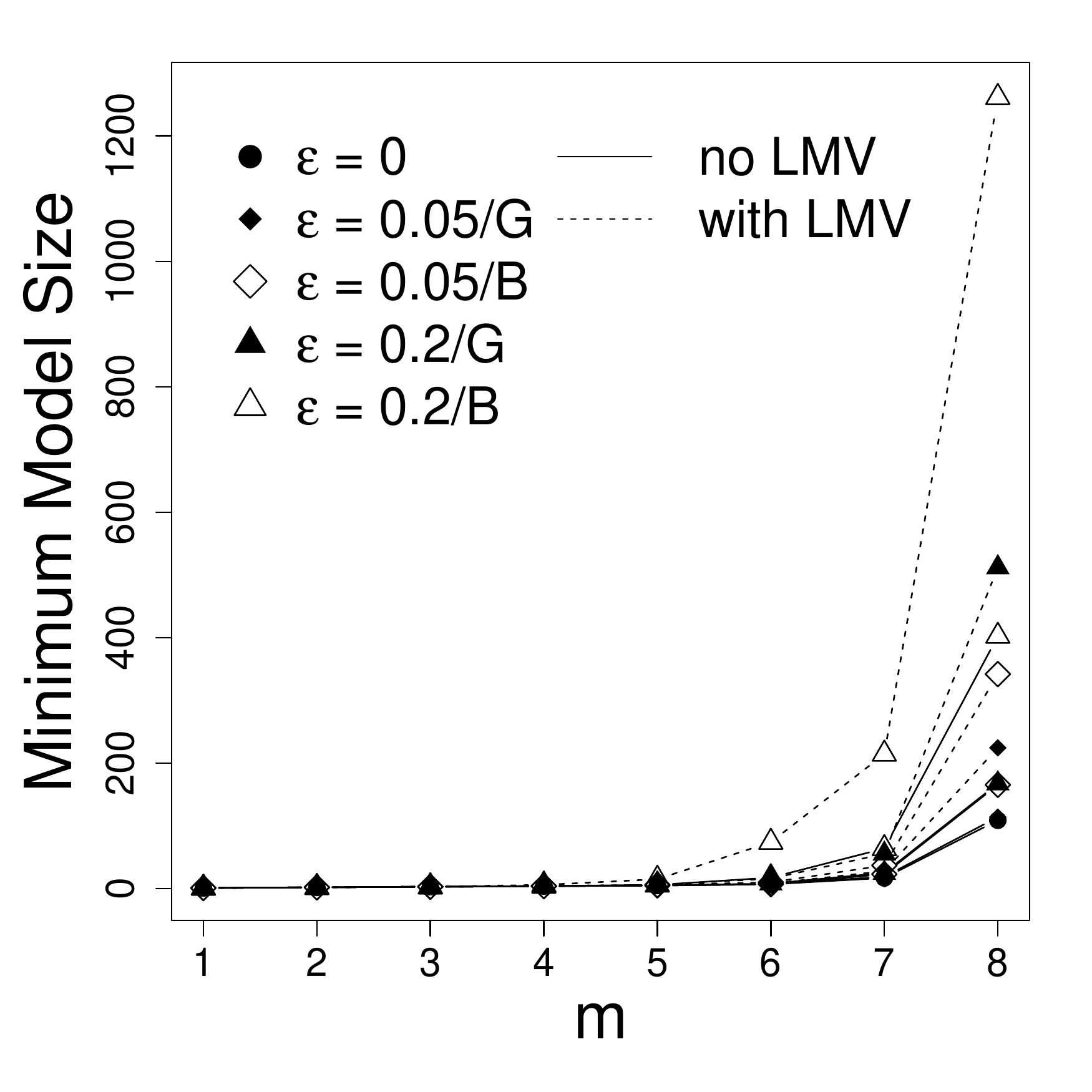}
	\end{minipage}
	\begin{minipage}{0.32\textwidth}
		\centering
		\includegraphics[width=3.8 cm]{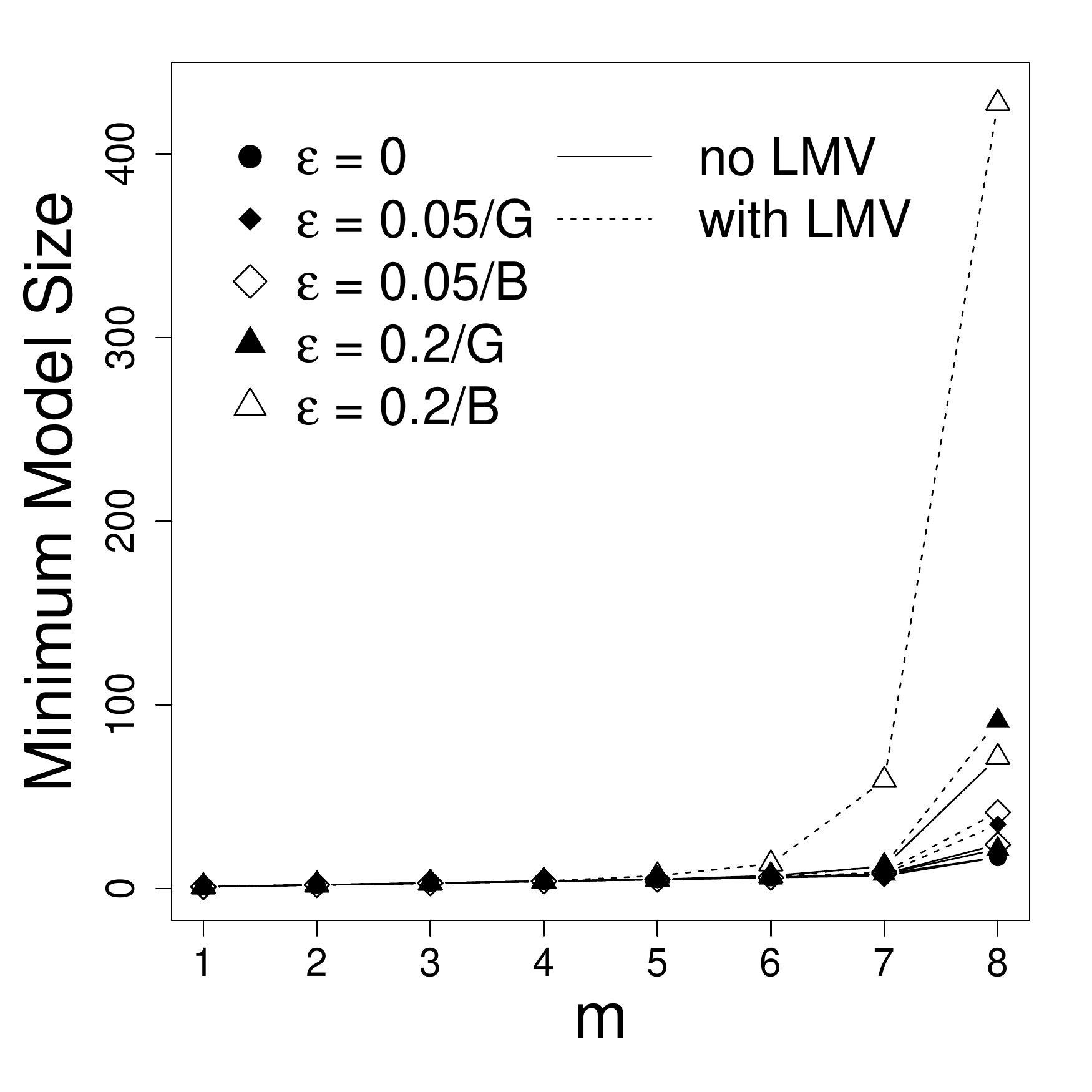}
	\end{minipage}
	\begin{minipage}{0.32\textwidth}
		\centering
		\includegraphics[width=3.8 cm]{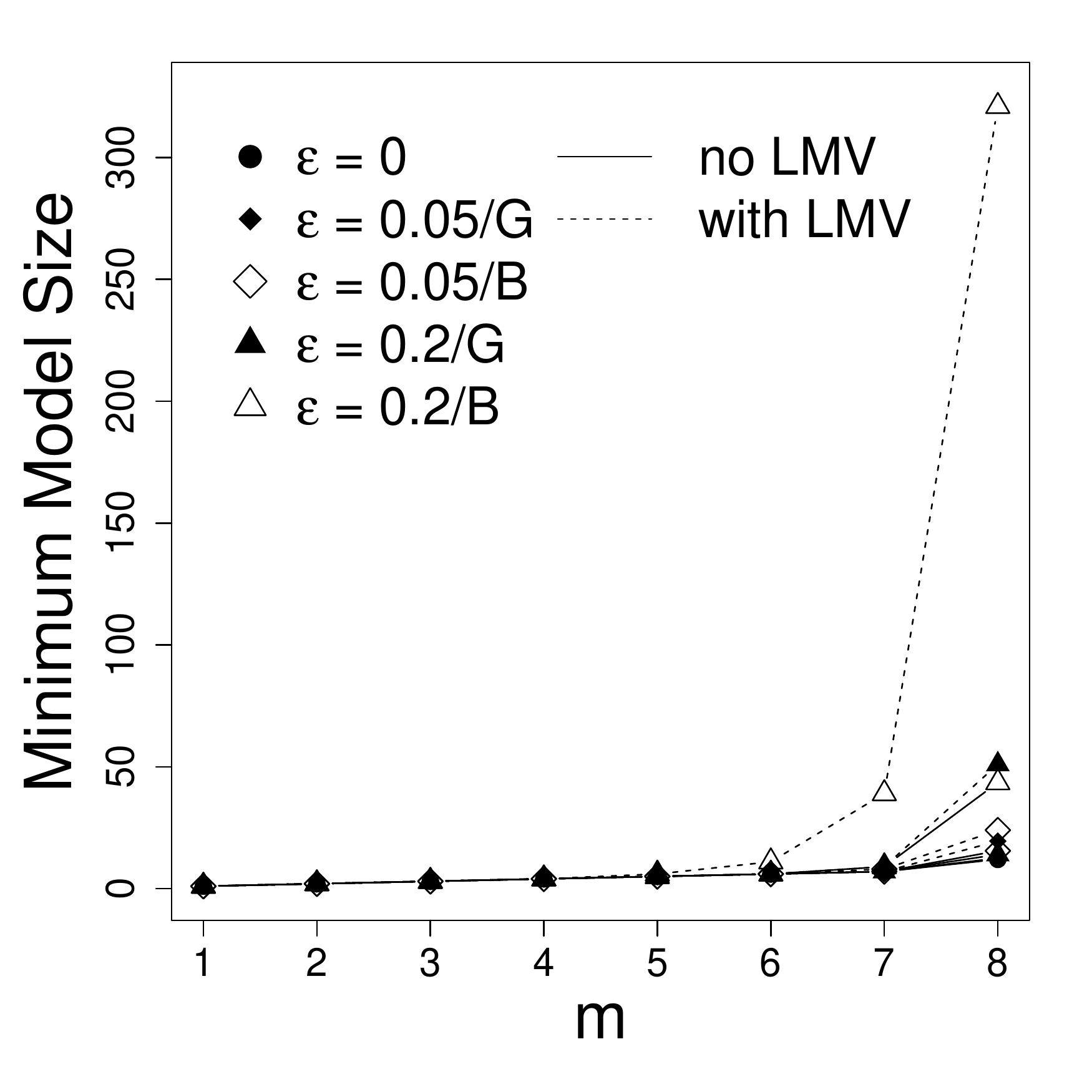}
	\end{minipage}\\
	
	\begin{minipage}{0.01\textwidth}
		\rotatebox[]{90}{\footnotesize $95\%$ Quantile}
	\end{minipage}
	\begin{minipage}{0.32\textwidth}
		\centering
		\includegraphics[width= 3.8 cm]{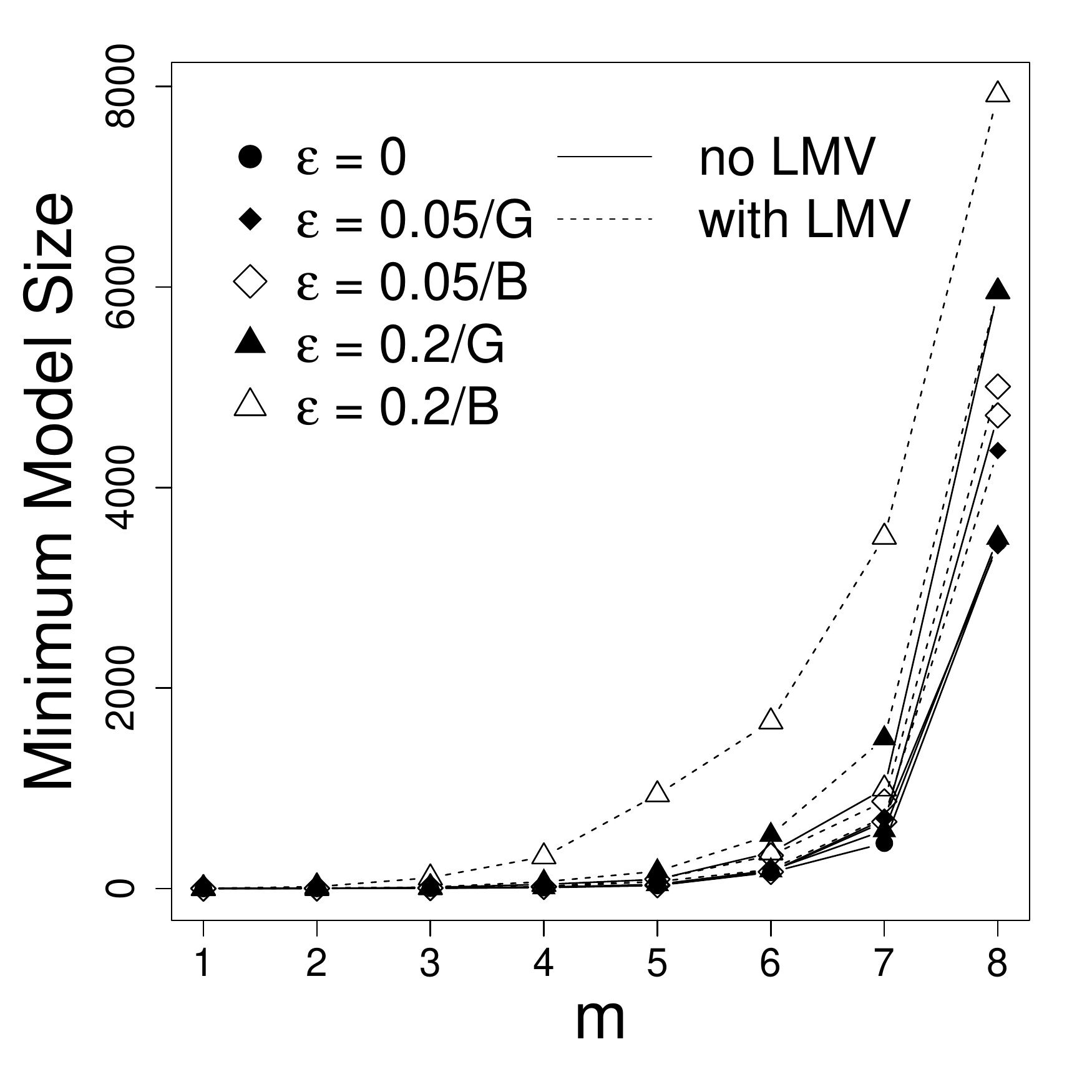}
	\end{minipage}
	\begin{minipage}{0.32\textwidth}
		\centering
		\includegraphics[width= 3.8 cm]{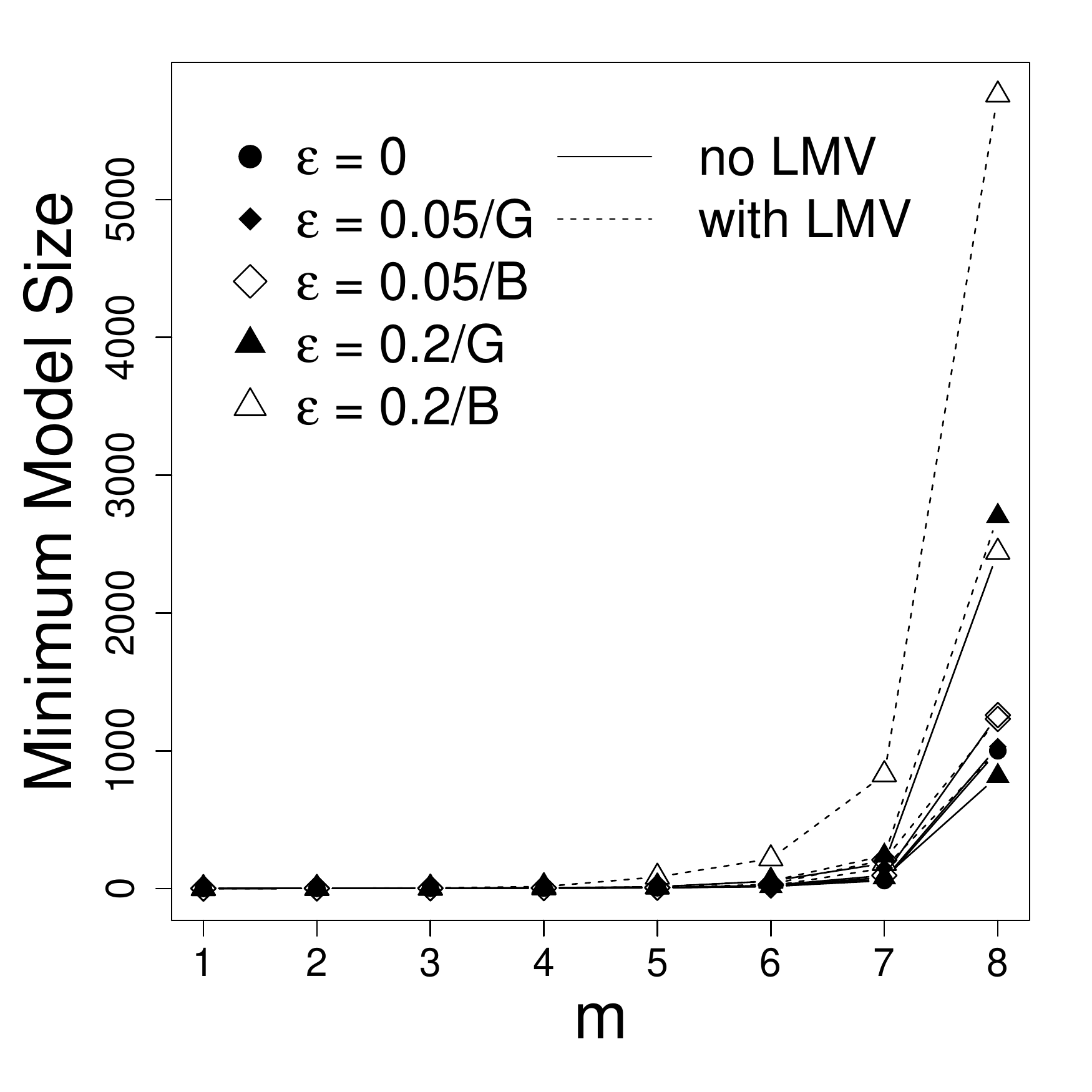}
	\end{minipage}
	\begin{minipage}{0.32\textwidth}
		\centering
		\includegraphics[width= 3.8 cm]{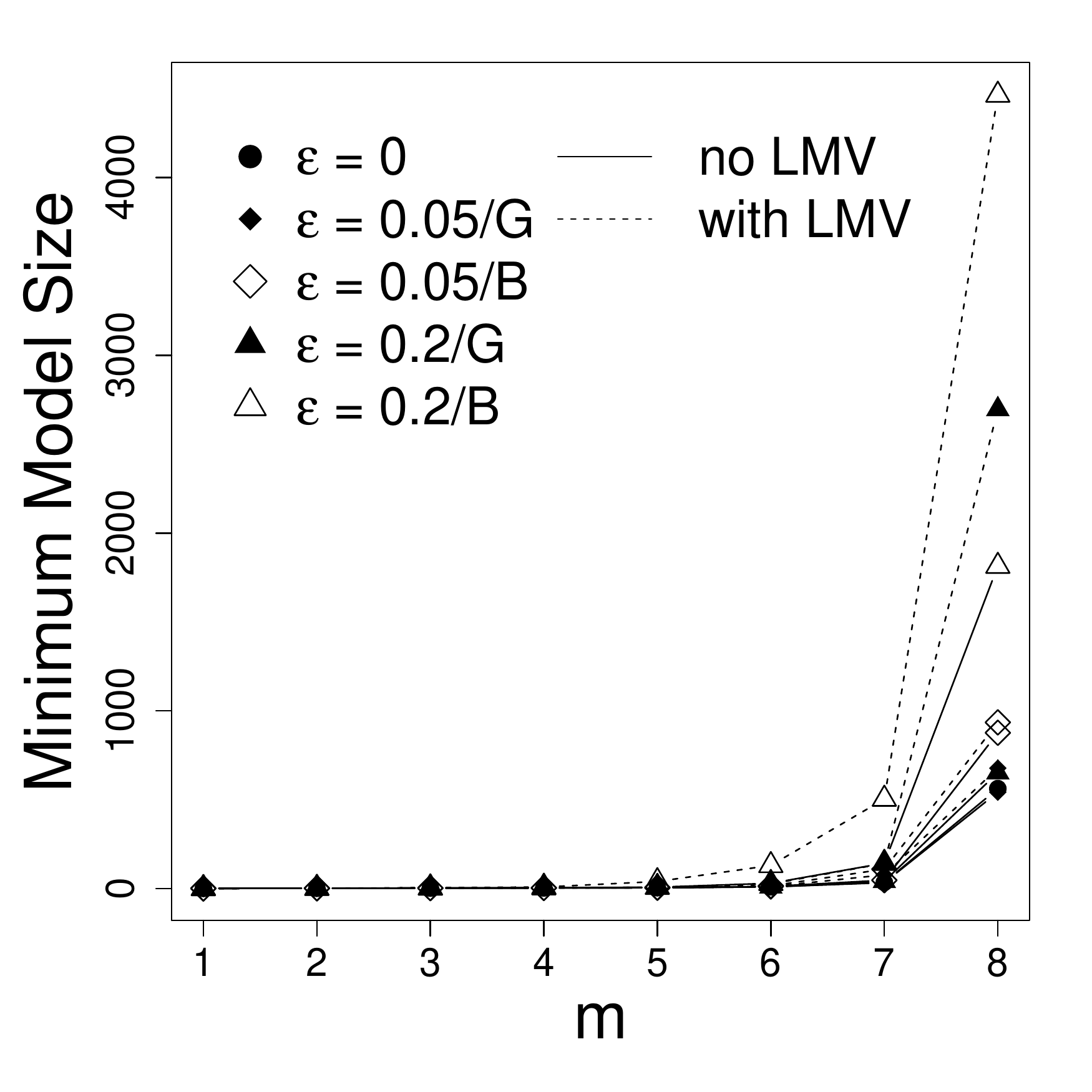}
	\end{minipage}\\
	
	\caption{\small Median and $95\%$ Quantile of the minimal model size needed to capture $m$ important variables by RFPSIS in the case of PC+LMG and PC+LMB for $p=10000$ and $d=2$.}
	\label{location_score}
\end{figure}

\begin{figure}[ht!]
	\centering
	\begin{minipage}{0.01\textwidth}
		\hfill
	\end{minipage}
	\begin{minipage}{0.32\textwidth}
		\centering
		(a) $n=200$, $c=1$ 
	\end{minipage}
	\begin{minipage}{0.32\textwidth}
		\centering
		(b) $n=200$, $c=3$
	\end{minipage}
	\begin{minipage}{0.32\textwidth}
		\centering
		(c) $n=200$, $c=5$
	\end{minipage}\\
	
	\begin{minipage}{0.01\textwidth}
		\rotatebox[]{90}{\footnotesize Median}
	\end{minipage}
	\begin{minipage}{0.32\textwidth}
		\centering
		\includegraphics[width=3.8 cm]{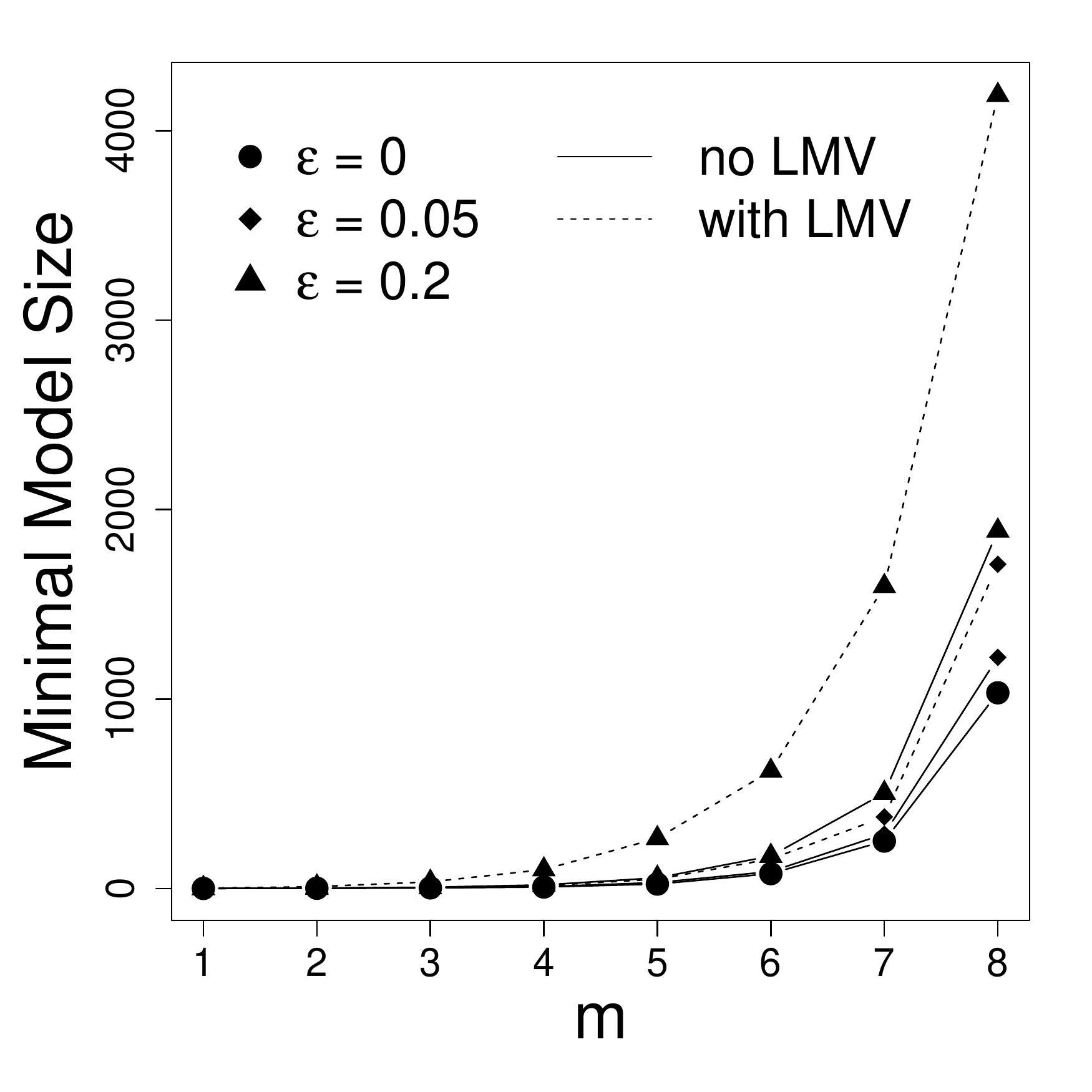}
	\end{minipage}
	\begin{minipage}{0.32\textwidth}
		\centering
		\includegraphics[width=3.8 cm]{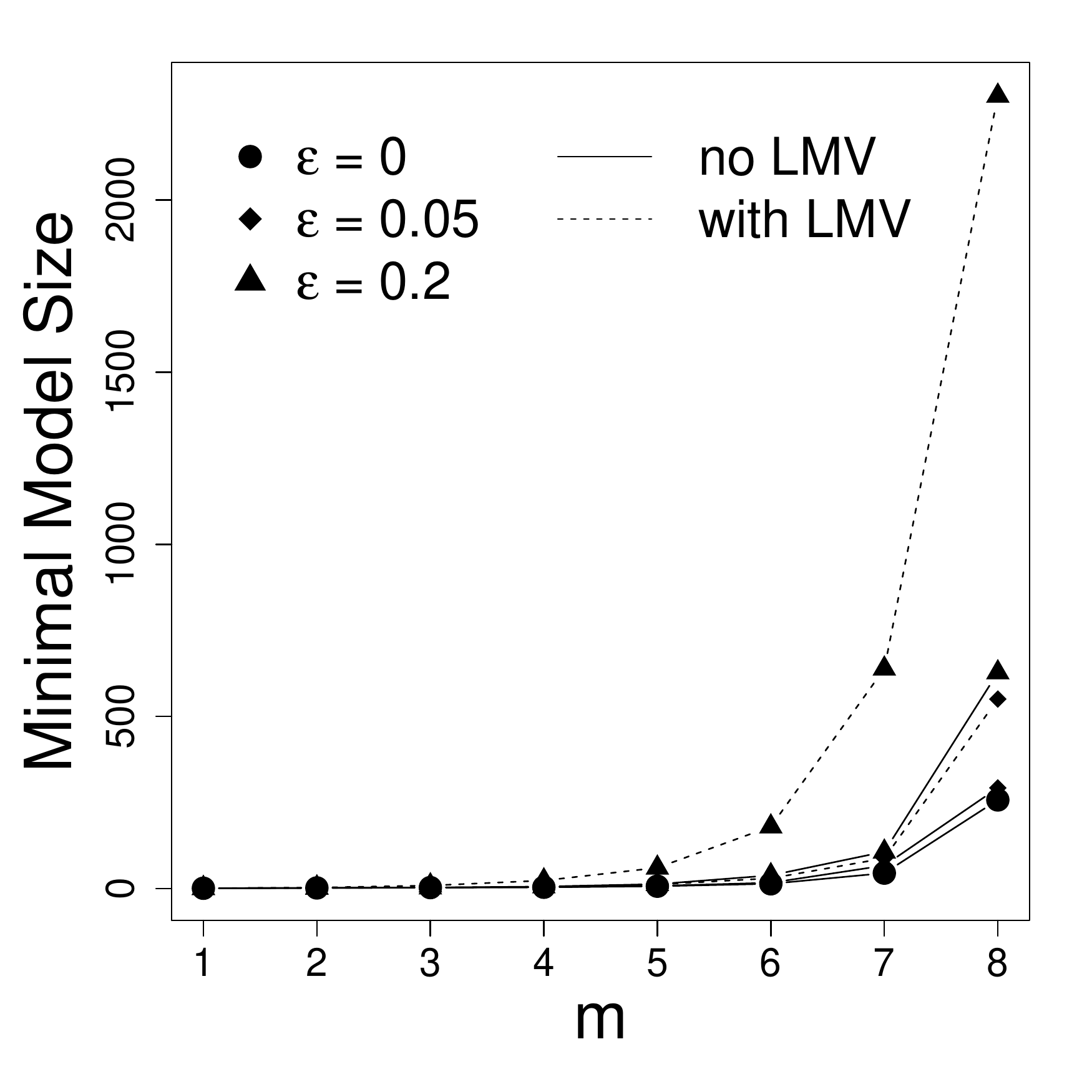}
	\end{minipage}
	\begin{minipage}{0.32\textwidth}
		\centering
		\includegraphics[width=3.8 cm]{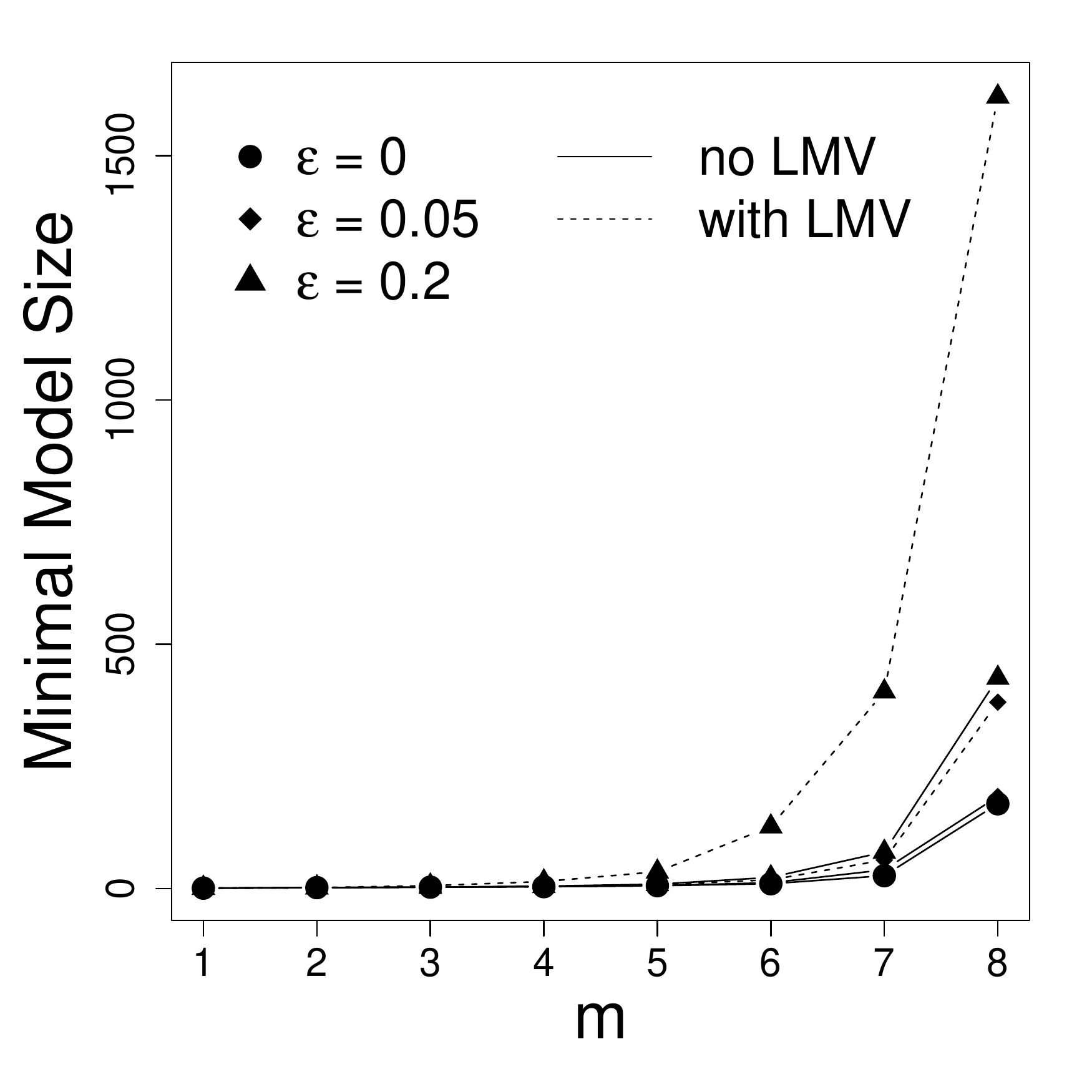}
	\end{minipage}\\
	
	\begin{minipage}{0.01\textwidth}
		\rotatebox[]{90}{\footnotesize $95\%$ Quantile}
	\end{minipage}
	\begin{minipage}{0.32\textwidth}
		\centering
		\includegraphics[width= 3.8 cm]{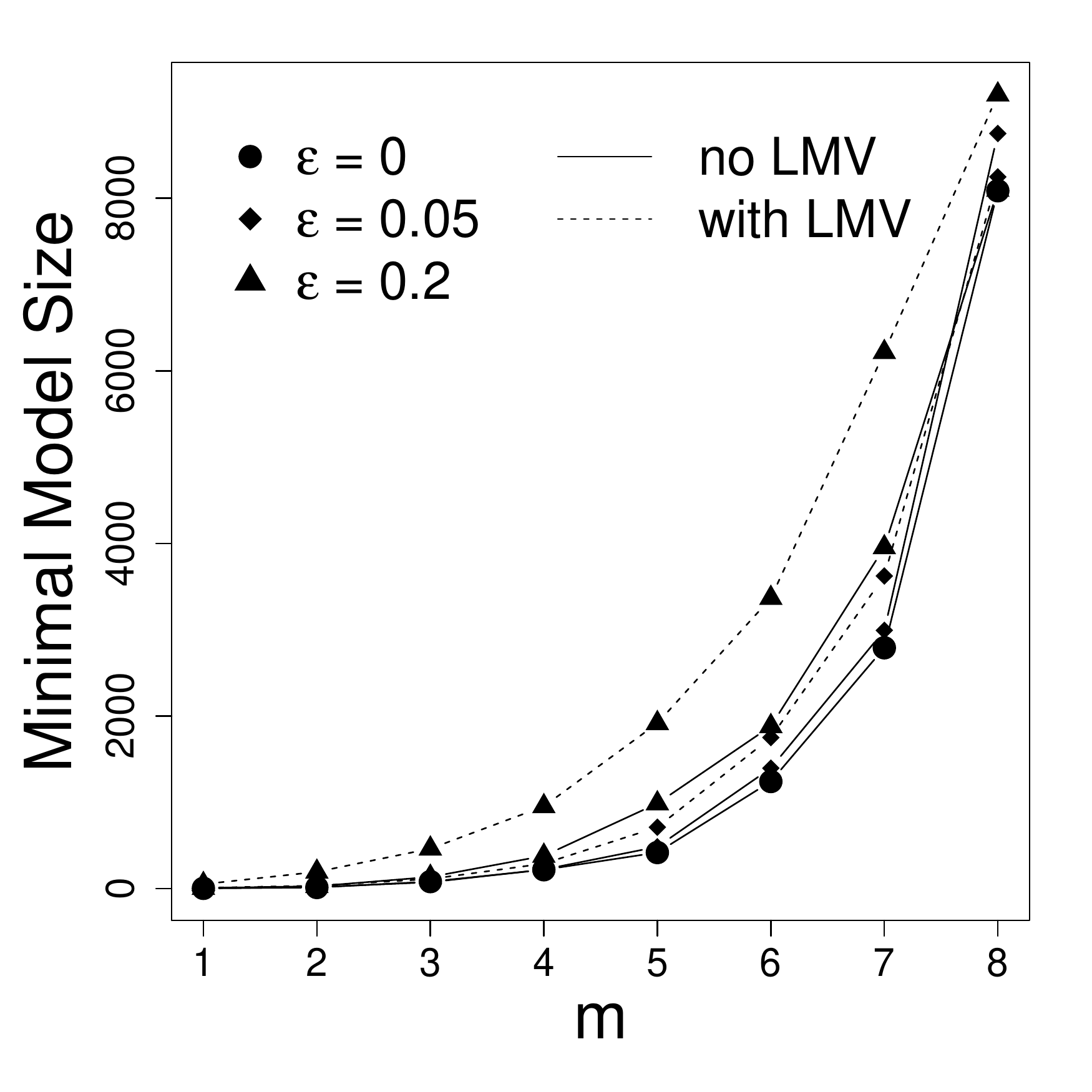}
	\end{minipage}
	\begin{minipage}{0.32\textwidth}
		\centering
		\includegraphics[width= 3.8 cm]{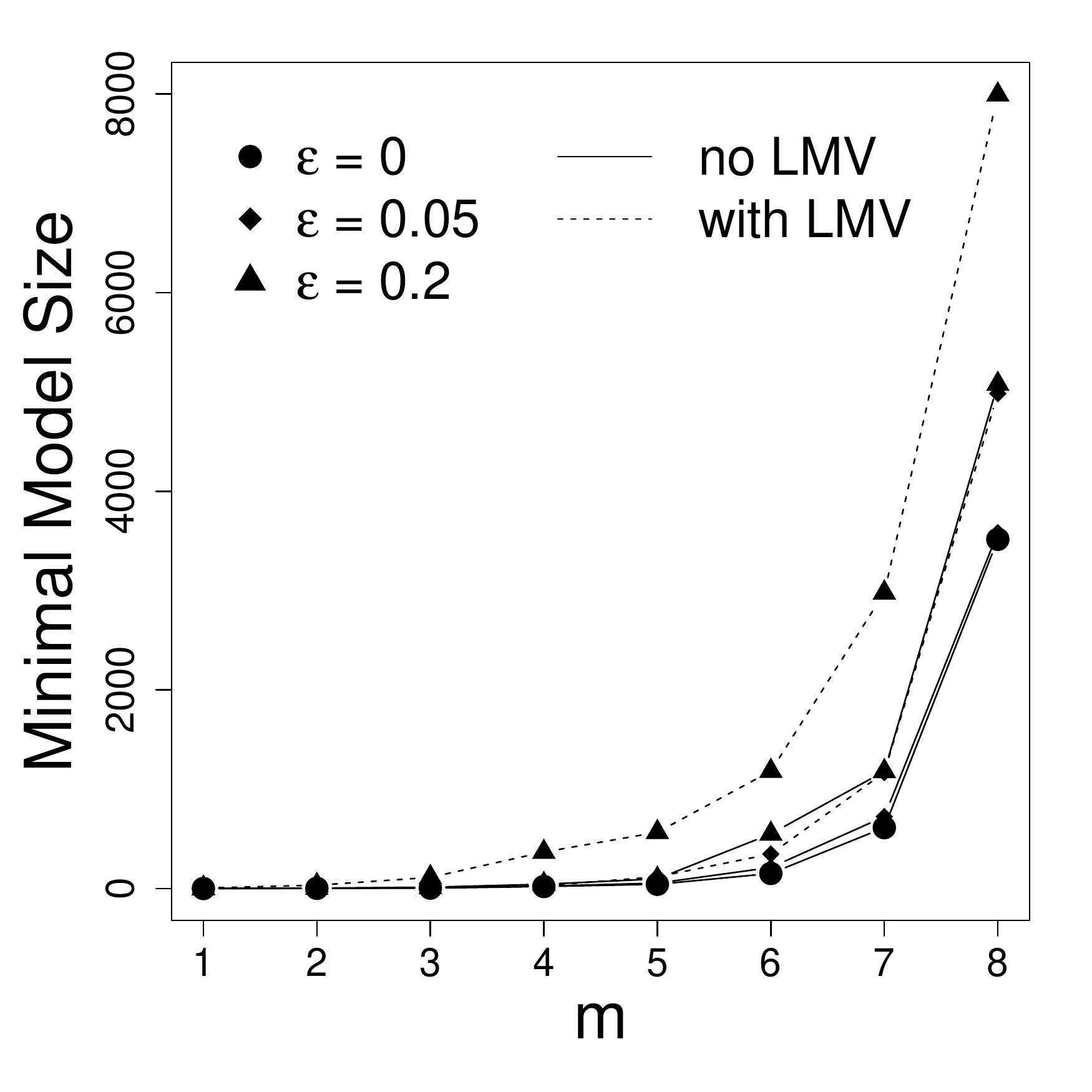}
	\end{minipage}
	\begin{minipage}{0.32\textwidth}
		\centering
		\includegraphics[width= 3.8 cm]{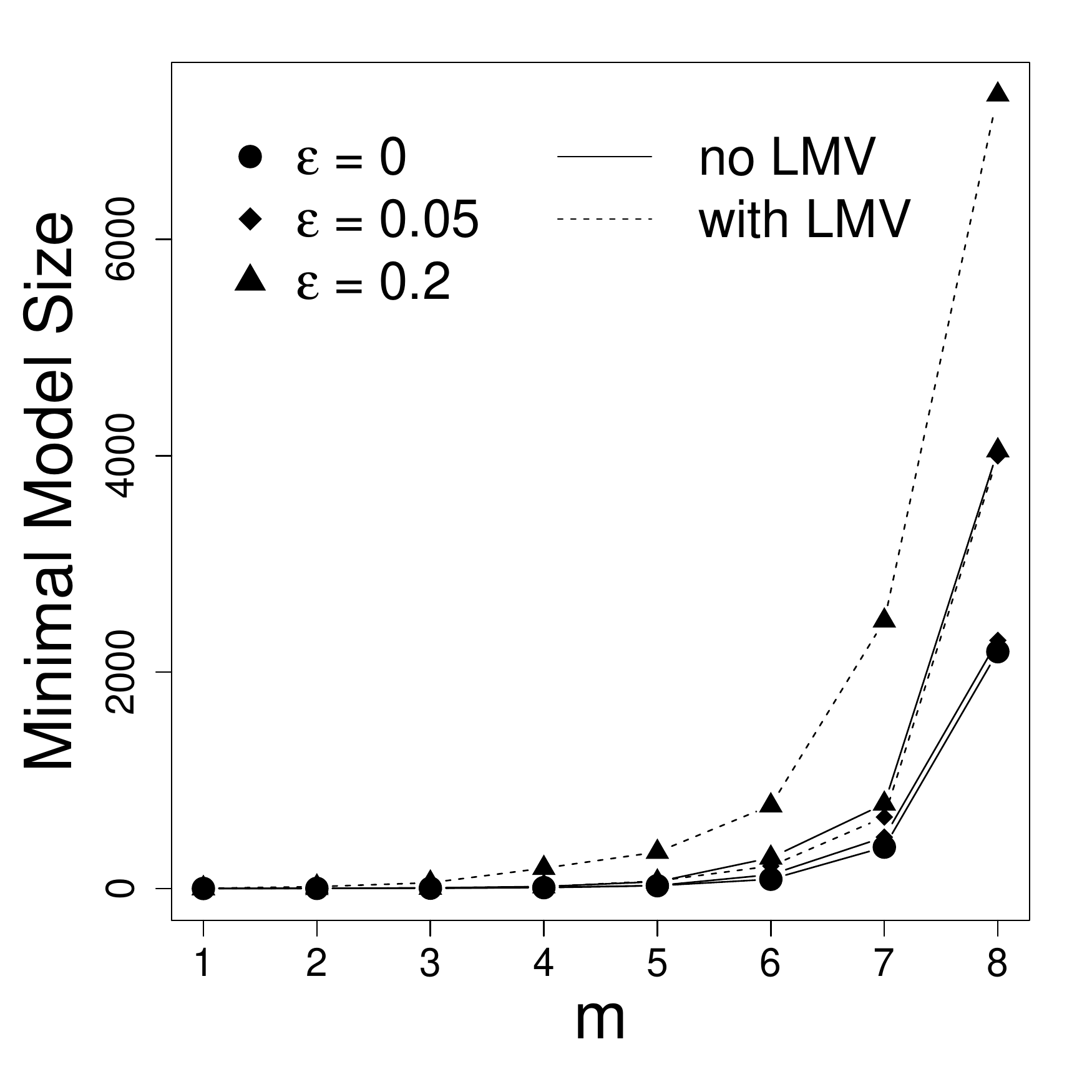}
	\end{minipage}\\

	\begin{minipage}{\textwidth}
		\hfill
	\end{minipage}\\
	
	\begin{minipage}{0.01\textwidth}
		\hfill
	\end{minipage}
	\begin{minipage}{0.32\textwidth}
		\centering
		(e) $n=400$, $c=1$
	\end{minipage}
	\begin{minipage}{0.32\textwidth}
		\centering
		(f) $n=400$, $c=3$
	\end{minipage}
	\begin{minipage}{0.32\textwidth}
		\centering
		(g) $n=400$,  $c=5$
	\end{minipage}\\
	
	\begin{minipage}{0.01\textwidth}
		\rotatebox[]{90}{\footnotesize Median}
	\end{minipage}
	\begin{minipage}{0.32\textwidth}
		\centering
		\includegraphics[width=3.8 cm]{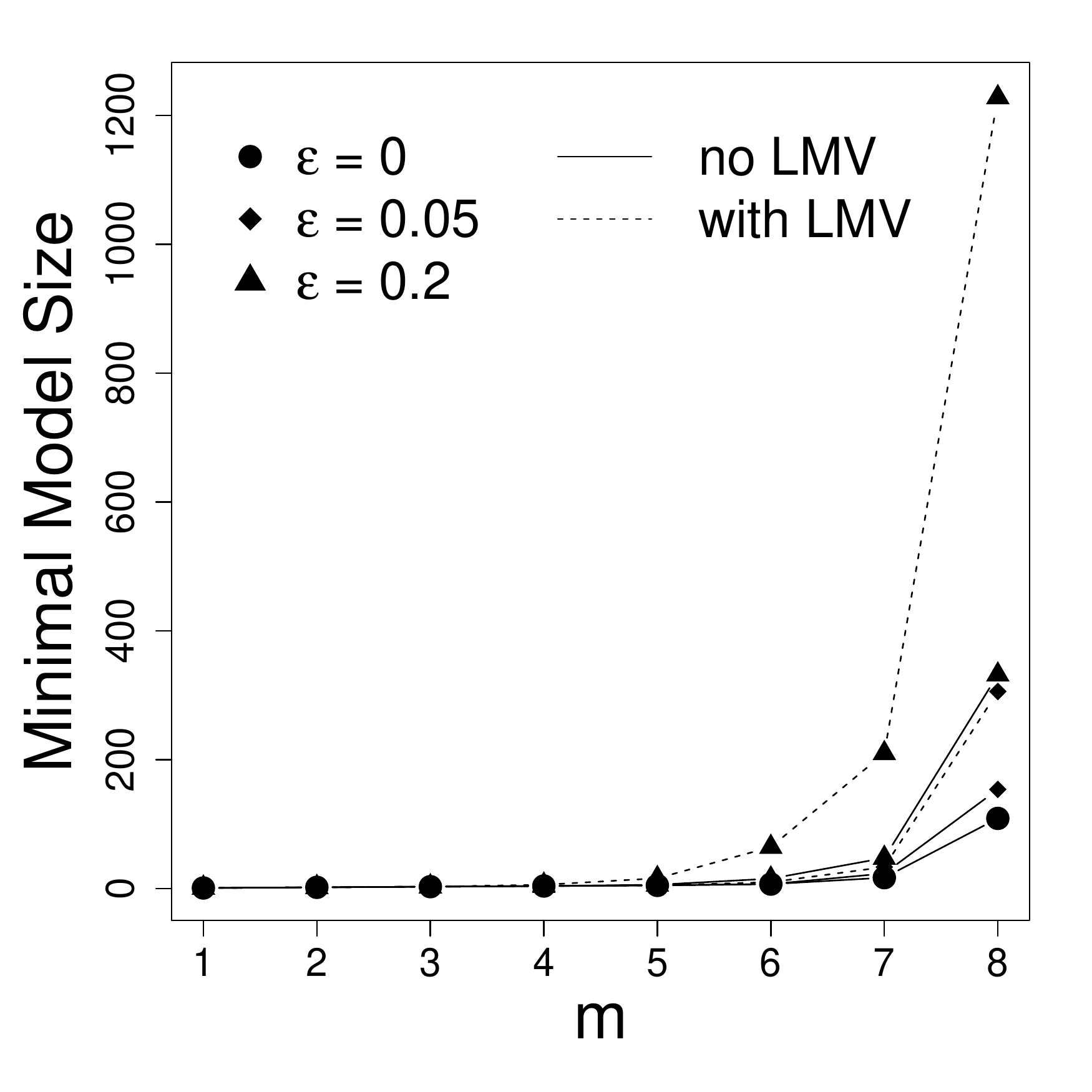}
	\end{minipage}
	\begin{minipage}{0.32\textwidth}
		\centering
		\includegraphics[width=3.8 cm]{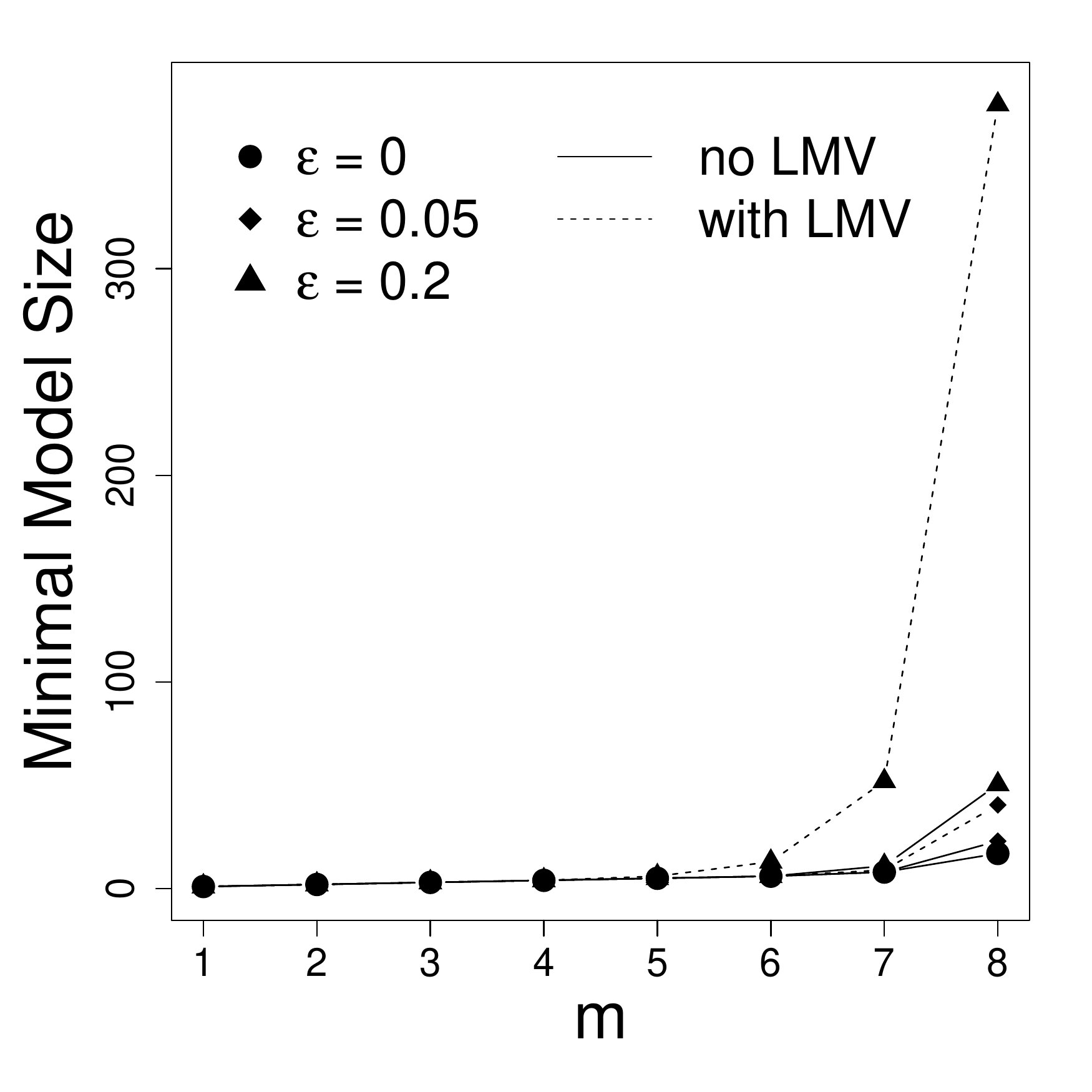}
	\end{minipage}
	\begin{minipage}{0.32\textwidth}
		\centering
		\includegraphics[width=3.8 cm]{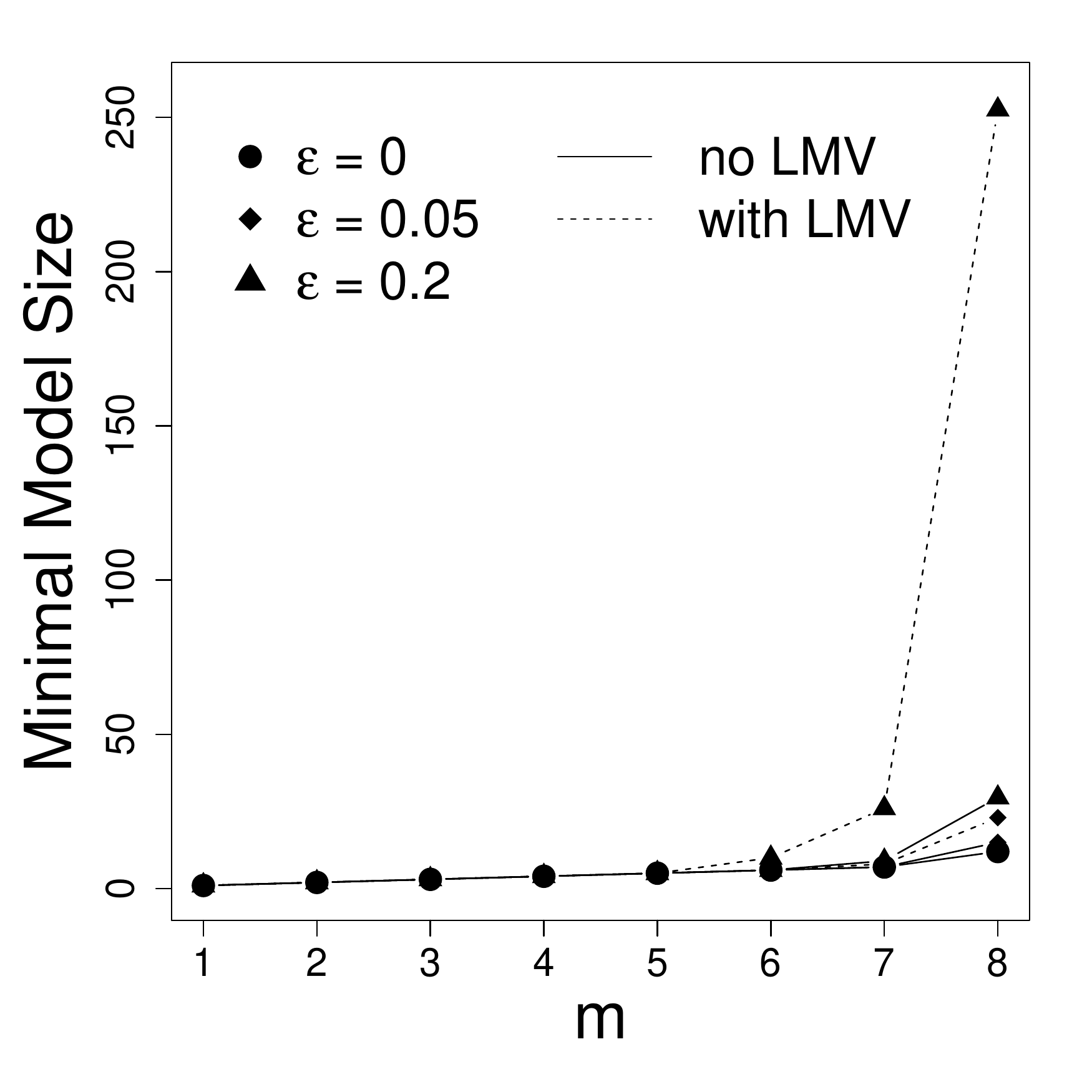}
	\end{minipage}\\
	
	\begin{minipage}{0.01\textwidth}
		\rotatebox[]{90}{\footnotesize $95\%$ Quantile}
	\end{minipage}
	\begin{minipage}{0.32\textwidth}
		\centering
		\includegraphics[width= 3.8 cm]{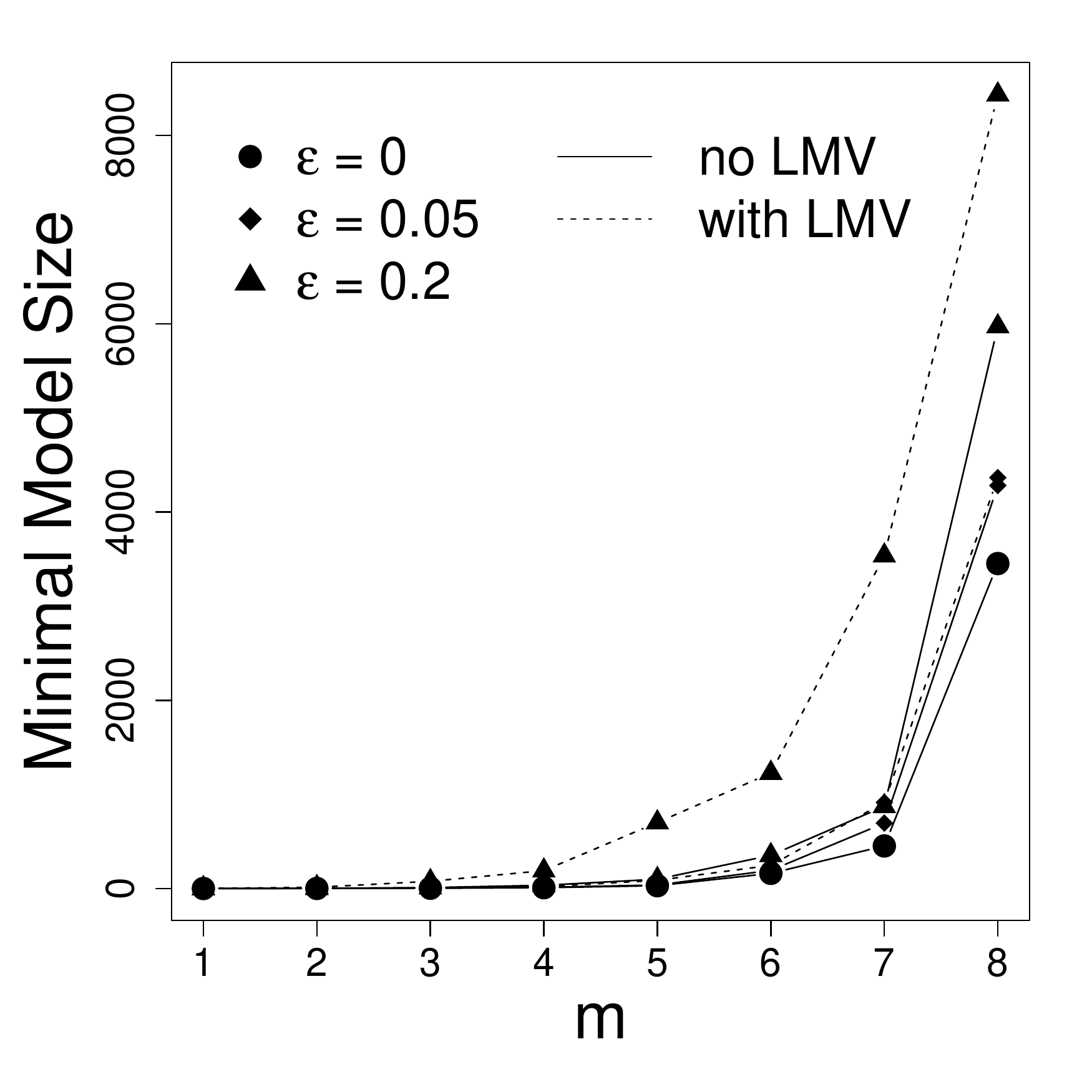}
	\end{minipage}
	\begin{minipage}{0.32\textwidth}
		\centering
		\includegraphics[width= 3.8 cm]{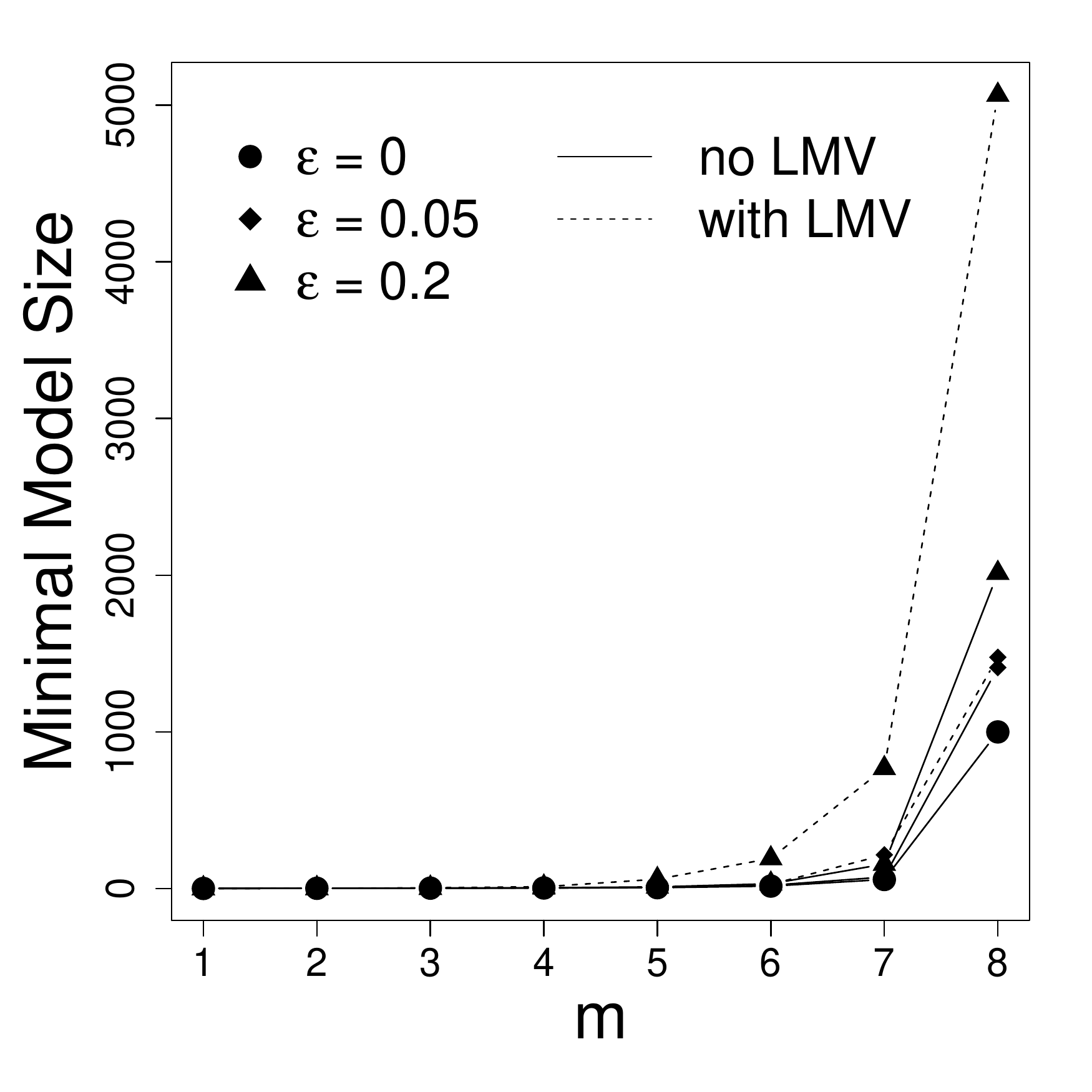}
	\end{minipage}
	\begin{minipage}{0.32\textwidth}
		\centering
		\includegraphics[width= 3.8 cm]{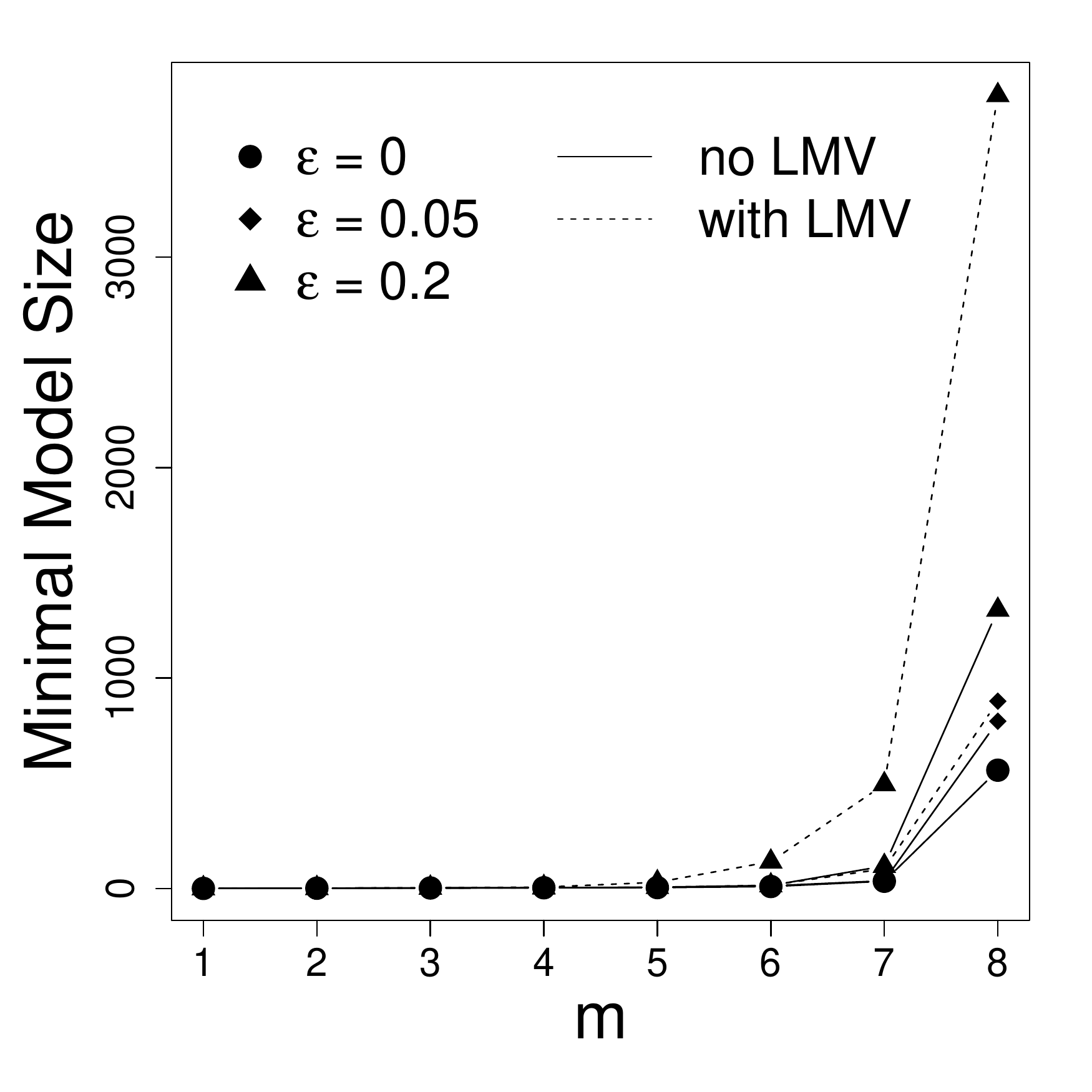}
	\end{minipage}\\
	
	\caption{\small Median and $95\%$ Quantile of the minimal model size needed to capture $m$ important variables by RFPSIS in the case of OC+LMG and OC+LMB for $p=10000$ and $d=2$.}
	\label{location_orth}
\end{figure}

RFPSIS is performed with $h=[(n-d+2)/2]$ for maximal robustness. For RFPSIS we first remark that in our simulation settings the estimate $\hat{d}$ of the factor subspace dimension according to criterion~\eqref{PC_Select} consistently corresponded to the true dimension $d$ that was used to generate the data. The results of RFPSIS in presence of leverage points are shown in Figure~\ref{location_score} for PC outliers, and in Figure~\ref{location_orth} for OC outliers. By comparing the plots in these two figures with those in Figure~\ref{location_classical}, we can see that RFPSIS performs almost as good as FPSIS on regular data.  Moreover, unlike FPSIS, RFPSIS succeeds to reduce the model size to a large extent while keeping all the important predictors for all considered contamination levels and outlier types. Since any OC outliers become bad leverage points in marginal regression models, both good and bad OC leverage points are downweighted in RFPSIS, and hence these two types of outliers lead to the same results. However, for PC outliers, there is a significant difference between good and bad leverage points because they are treated differently by RFPSIS. With good PC leverage points, the screening results of RFPSIS are always close to those obtained on regular data.

By comparing the results for the median of the minimal model size to those for the $95\%$ quantile, we can see that in all cases RFPSIS does pick up 6 to 7 of the important predictors (with the strongest signals) in the beginning of its solution path. The contamination mainly affects the required model size to cover the last one or two important predictors (with the smallest signals), leading to a large variation in the models size needed to pick up these variables. Not surprisingly, the performance decreases for datasets with smaller sample size, lower signal-to-noise ratio and/or higher contamination level. Although RFPSIS overall performs less well for the small sample size case ($n=200$), it is still able to establish a huge dimension reduction when the signal-to-noise ratio is sufficiently high. Including extra vertical outliers in the data also only affects the important variables at the end of the solution path.

	\section{Final Model Selection}
	\label{sec: BIC}
	The RFPSIS procedure above sequences the predictors in order of importance. After sequencing the predictors the goal is now to find a model $\mathcal{M}_{(q)}$ with size $q$ of order $O(n^{\eta})$ ($0<\eta<1$) that ideally covers all the important predictors. A popular criterion to determine the final model size, is the general Bayesian Information Criterion (BIC)
	\begin{equation}
		\text{BIC}(\mathcal{M}) = \log \text{RSS}(\mathcal{M}) + \mathcal{P}(k,n,p)
	\end{equation}
	where $\text{RSS}(\mathcal{M}) = \|\by-\hby\|_E^2$ is the sum of squared residuals corresponding to the fitted model. $\mathcal{P}(k,n,p)$ is a penalty term which depends on the number of predictors $k$ in the model, the sample size $n$ and the dimension $p$. Compared to AIC, BIC includes the sample size dependent factor $\log(n)$ in the penalty term and therefore penalizes more heavily on model complexity, which results in more parsimonious models. Since $\text{RSS}(\mathcal{M})$ involves all observations, the general BIC criterion is not robust. Therefore, we consider robust adaptations of this criterion to select the final model.  
		
For each of the solutions $\mathcal{M}_{(l)}$ ($l=1,\dots,\tilde{k}_{\max}$) in the solution path  $\mathbb{M}$, we robustly regress $\tilde{\by}$ on $\tilde{\bX}_{\mathcal{M}_{(k)}}$, using solely the observations in $\mathcal{I}_2$. Since we already have obtained the marginal slope estimates, we apply a multiple regression M-estimator with these marginal coefficient estimates and the S-scale of the resulting residuals as the initial values rather than fully calculating the MM-estimator from scratch. In this way, we obtain a huge reduction in computation time because we avoid having to calculate the time-consuming initial S-estimator. To avoid the over-identification problem in the multiple regression M-estimator, we set $\tilde{k}_{\max} \leqslant n/2$.	Let us denote the resulting coefficient estimates by $\hth^{(k)}_j$ ($j=1,\ldots,k$, $k=1,\ldots,\tilde{k}_{max}$). For each of these models, we then calculate a weighted sum of squared residuals, given by
	\begin{equation}
	\text{WRSS}_{(k)}=\sum_{i \in \mathcal{I}_2} w^{(k)}_i (\hat{y}_i - \sum_{j=1}^{k}\hth^{(k)}_j\hat{\bx}_{ij})^2,
	\end{equation}
	where $w^{(k)}_i$ is the weight given by the M-estimator for the observations in $\mathcal{I}_2$. Note that observations not in $\mathcal{I}_2$ are thus given weight zero. The final model can then be selected by minimizing either of the following criteria
	\begin{eqnarray}
	\text{BIC}(\mathcal{M}_{(k)})  &=&  \log \text{WRSS}_{(k)} + |\mathcal{M}_{(k)}|n^{-1}\log(n),
	\label{BIC}\\
	\text{EBIC}(\mathcal{M}_{(k)})  &=&    \log \text{WRSS}_{(k)} + |\mathcal{M}_{(k)}|n^{-1}(\log(n) + \log(p)),
	\label{EBIC}\\
	\text{FPBIC}(\mathcal{M}_{(k)})  &=&    \log \text{WRSS}_{(k)} + |\mathcal{M}_{(k)}|n^{-1}\log(n)\log(p),
	\label{FPBIC}
	\end{eqnarray}
	where \eqref{BIC} is a robust adaptation of the original BIC and \eqref{EBIC} belongs to the extended BIC family~\citep{Chen2008} which favors sparser model than BIC. FPBIC uses a penalty term which selects even more parsimonious models than both BIC and EBIC~\citep{FPSIS}. Asymptotically, BIC, EBIC and FPBIC are equivalent when $p=\mathcal{O}(\exp(n^\xi))$ ($0<\xi<1$).

The multiple regression models fitted by M-estimators generally yield more accurate coefficient estimates than the marginal models. Hence, these coefficient estimates can be used to reorder the predictors in order of importance.
For each model $\mathcal{M}_{(k)}$, we thus reorder the coefficient estimates $\hth^{(k)}_j$ in decreasing absolute values. These reordered coefficients and their corresponding predictors are denoted by $\hth^{(k)}_{(j)}$ and $\bX^{(k)}_{(j)}$ ($j=1,\ldots,k$, $k=1,\ldots,\tilde{k}_{\max}$) respectively. Each of the robust general BIC criteria can also be calculated for these reordered sequences, and will be denoted as R-BIC, R-EBIC, R-FPBIC, respectively. That is, for $l=1,\dots,k$ we calculate the weighted sum of squared residuals as
	\begin{equation}
	\text{WRSS}_{(kl)} = \sum_{i \in \mathcal{I}_2} w^{(k)}_i (\hat{y}_i - \sum_{j=1}^{l}\hth^{(k)}_{(j)}\hat{\bX}_{(j)}^{(k)})^2.
	\end{equation}
	The final model is determined by minimizing
	\begin{equation}
	\text{R-BIC}(\mathcal{M}_{(kl)}) = \log \text{WRSS}_{(kl)} + |\mathcal{M}_{(kl)}|n^{-1}\log(n),
	\end{equation}
	or
	\begin{equation}
	\text{R-EBIC}(\mathcal{M}_{(kl)}) = \log \text{WRSS}_{(kl)} + |\mathcal{M}_{(kl)}|n^{-1}(\log(n)+\log(p)),
	\end{equation}
	or
	\begin{equation}
	\text{R-FPBIC}(\mathcal{M}_{(kl)}) = \log \text{WRSS}_{(kl)} + |\mathcal{M}_{(kl)}|n^{-1}\log(n)\log(p).
	\end{equation}

To evaluate the performance of these six criteria, we investigate their average performance over 200 datasets generated according to the designs discussed in Subsection~\ref{sec:perf}. For the model selected by each of these criteria we report both the average number of truly important predictors in the model (TP) and the average number of falsely selected predictors (FP). Tables~\ref{BIC_10000_n200_d2} 
and~\ref{BIC_10000_n400_d2} contain the results for $n=200$ and $n=400$ with $\tilde{k}_{\max}$=100, respectively. From these tables we can see that FPBIC and R-FPBIC select the models with the smallest false positive rate, but these models also miss more important predictors than the other criteria. The penalty term proposed in~\citep{FPSIS} thus tends to select too sparse models in practice. 
The four other criteria generally are able to produce better screening results with a high number of true positives and a small number of false positives for the regular data. Their performance improves for larger sample size and higher signal-to-noise ratio. Among these criteria, R-BIC selects the most important predictors, but at the cost of selecting more noise predictors. Interestingly, R-EBIC not only can get a number of true positives that is similar or larger than BIC/EBIC, but at the same time also a smaller number of false positives when the signal-to-noise ratio is sufficiently high ($c=3$ or $c=5$ in our simulations). This shows that reordering the predictors according to the multiple regression coefficient estimates before computing the selection criterion indeed improves the selection performance. When we have a coherent data set with a strong signal and a sparse model is highly preferred, we recommend to use R-EBIC. However, if only a noisy data set is available, R-BIC may be preferred to avoid missing too many important predictors.

\begin{table}[ht!]
	\footnotesize
	\centering
	\renewcommand\arraystretch{1.25}
	\scalebox{0.85}{

\begin{tabular}{c|c|c|c|cc|cc|cc|cc|cc|cc}
	\hline 
	\multirow{2}{*}{} & \multirow{2}{*}{eps} & \multirow{2}{*}{c} & \multirow{2}{*}{LMV} & \multicolumn{2}{c|}{BIC} & \multicolumn{2}{c|}{EBIC} & \multicolumn{2}{c|}{FPBIC} & \multicolumn{2}{c|}{R-BIC} & \multicolumn{2}{c|}{R-EBIC} & \multicolumn{2}{c}{R-FPBIC}\tabularnewline
	\cline{5-16} 
	&  &  &  & TP & FP & TP & FP & TP & FP & TP & FP & TP & FP & TP & FP\tabularnewline
	\hline 
	\multirow{3}{*}{clean} & \multirow{3}{*}{0} & 1 & \multirow{3}{*}{no} & 3.63 & 4.54 & 2.70 & 1.01 & 0.97 & 0.11 & 4.55 & 19.04 & 2.83 & 0.83 & 0.95 & 0.11\tabularnewline
	&  & 3 &  & 5.45 & 4.36 & 4.51 & 0.70 & 1.52 & 0.01 & 6.36 & 12.60 & 5.29 & 0.33 & 1.54 & 0.01\tabularnewline
	&  & 5 &  & 6.01 & 4.83 & 5.23 & 1.01 & 2.02 & 0.00 & 6.88 & 9.62 & 6.13 & 0.23 & 2.49 & 0.00\tabularnewline
	\hline 
	\multirow{12}{*}{\shortstack{PC \\ +LMG}} & \multirow{6}{*}{5} & 1 & \multirow{3}{*}{no} & 3.55 & 4.71 & 2.67 & 1.04 & 0.93 & 0.12 & 4.51 & 19.84 & 2.88 & 0.80 & 0.92 & 0.11\tabularnewline
	&  & 3 &  & 5.48 & 4.30 & 4.56 & 0.87 & 1.43 & 0.01 & 6.36 & 13.27 & 5.22 & 0.31 & 1.47 & 0.01\tabularnewline
	&  & 5 &  & 6.06 & 5.44 & 5.28 & 1.03 & 1.98 & 0.01 & 6.88 & 10.44 & 5.97 & 0.21 & 2.34 & 0.00\tabularnewline
	\cline{4-16} 
	&  & 1 & \multirow{3}{*}{yes} & 3.05 & 4.21 & 2.33 & 1.13 & 1.00 & 0.14 & 3.09 & 4.12 & 2.43 & 0.85 & 0.93 & 0.12\tabularnewline
	&  & 3 &  & 4.63 & 3.19 & 4.00 & 0.79 & 1.50 & 0.01 & 4.79 & 2.86 & 4.30 & 0.34 & 1.49 & 0.00\tabularnewline
	&  & 5 &  & 5.22 & 3.19 & 4.55 & 0.90 & 1.95 & 0.01 & 5.31 & 2.17 & 4.92 & 0.22 & 1.97 & 0.00\tabularnewline
	\cline{2-16} 
	& \multirow{6}{*}{20} & 1 & \multirow{3}{*}{no} & 3.47 & 5.27 & 2.38 & 1.03 & 0.92 & 0.14 & 4.35 & 21.71 & 2.61 & 0.78 & 0.90 & 0.14\tabularnewline
	&  & 3 &  & 5.17 & 4.22 & 4.23 & 0.82 & 1.37 & 0.01 & 6.15 & 14.27 & 4.79 & 0.34 & 1.36 & 0.01\tabularnewline
	&  & 5 &  & 5.74 & 4.60 & 4.83 & 0.89 & 1.71 & 0.00 & 6.65 & 9.92 & 5.66 & 0.20 & 1.99 & 0.00\tabularnewline
	\cline{3-16} 
	&  & 1 & \multirow{3}{*}{yes} & 2.75 & 4.30 & 2.05 & 1.15 & 0.95 & 0.15 & 2.83 & 4.52 & 2.17 & 1.09 & 0.89 & 0.16\tabularnewline
	&  & 3 &  & 4.34 & 3.17 & 3.59 & 0.75 & 1.41 & 0.02 & 4.43 & 3.04 & 3.92 & 0.35 & 1.34 & 0.01\tabularnewline
	&  & 5 &  & 4.90 & 2.91 & 4.31 & 0.87 & 1.65 & 0.01 & 5.02 & 2.39 & 4.62 & 0.26 & 1.67 & 0.00\tabularnewline
	\hline 
	\multirow{12}{*}{\shortstack{PC \\ +LMB}}  & \multirow{6}{*}{5} & 1 & \multirow{3}{*}{no} & 3.49 & 5.20 & 2.65 & 1.13 & 0.93 & 0.12 & 4.33 & 17.94 & 2.72 & 0.92 & 0.90 & 0.12\tabularnewline
	&  & 3 &  & 5.34 & 4.81 & 4.34 & 0.83 & 1.37 & 0.01 & 6.26 & 14.26 & 5.00 & 0.33 & 1.36 & 0.01\tabularnewline
	&  & 5 &  & 5.90 & 5.38 & 4.96 & 0.97 & 1.82 & 0.00 & 6.77 & 11.44 & 5.91 & 0.23 & 2.06 & 0.00\tabularnewline
	\cline{3-16} 
	&  & 1 & \multirow{3}{*}{yes} & 2.28 & 4.34 & 1.77 & 1.35 & 0.83 & 0.29 & 2.33 & 4.61 & 1.83 & 1.18 & 0.78 & 0.26\tabularnewline
	&  & 3 &  & 3.77 & 2.96 & 3.14 & 0.80 & 1.41 & 0.09 & 3.81 & 2.65 & 3.41 & 0.52 & 1.23 & 0.04\tabularnewline
	&  & 5 &  & 4.22 & 2.62 & 3.72 & 0.85 & 1.66 & 0.07 & 4.28 & 2.23 & 4.00 & 0.33 & 1.60 & 0.02\tabularnewline
	\cline{2-16} 
	& \multirow{6}{*}{20} & 1 & \multirow{3}{*}{no} & 2.64 & 4.94 & 1.82 & 1.14 & 0.79 & 0.22 & 3.41 & 23.61 & 1.91 & 1.03 & 0.76 & 0.25\tabularnewline
	&  & 3 &  & 4.20 & 3.87 & 3.11 & 0.74 & 1.07 & 0.04 & 5.44 & 18.80 & 3.56 & 0.46 & 1.07 & 0.04\tabularnewline
	&  & 5 &  & 4.71 & 3.87 & 3.74 & 0.69 & 1.16 & 0.02 & 5.98 & 16.58 & 4.34 & 0.30 & 1.17 & 0.02\tabularnewline
	\cline{4-16} 
	&  & 1 & \multirow{3}{*}{yes} & 1.32 & 4.74 & 0.93 & 1.62 & 0.49 & 0.54 & 1.31 & 5.47 & 0.95 & 1.47 & 0.51 & 0.49\tabularnewline
	&  & 3 &  & 2.28 & 3.20 & 1.82 & 1.15 & 0.88 & 0.27 & 2.35 & 3.48 & 1.96 & 0.92 & 0.89 & 0.20\tabularnewline
	&  & 5 &  & 2.68 & 2.84 & 2.23 & 1.06 & 1.06 & 0.21 & 2.72 & 3.12 & 2.41 & 0.70 & 1.01 & 0.12\tabularnewline
	\hline 
	\multirow{12}{*}{\shortstack{OC \\ +LMG\\/LMB}} & \multirow{6}{*}{5} & 1 & \multirow{3}{*}{no} & 3.48 & 5.06 & 2.46 & 0.96 & 0.92 & 0.14 & 4.34 & 21.02 & 2.68 & 0.81 & 0.90 & 0.14\tabularnewline
	&  & 3 &  & 5.26 & 4.68 & 4.24 & 0.81 & 1.42 & 0.01 & 6.28 & 14.31 & 4.90 & 0.29 & 1.40 & 0.01\tabularnewline
	&  & 5 &  & 5.94 & 5.77 & 4.99 & 1.15 & 1.78 & 0.00 & 6.85 & 11.53 & 5.94 & 0.22 & 2.23 & 0.00\tabularnewline
	\cline{4-16} 
	&  & 1 & \multirow{3}{*}{yes} & 2.84 & 4.47 & 2.09 & 1.08 & 0.96 & 0.15 & 2.93 & 4.93 & 2.21 & 0.93 & 0.91 & 0.14\tabularnewline
	&  & 3 &  & 4.38 & 3.14 & 3.69 & 0.76 & 1.46 & 0.02 & 4.53 & 3.03 & 3.96 & 0.30 & 1.37 & 0.01\tabularnewline
	&  & 5 &  & 4.99 & 2.95 & 4.32 & 0.78 & 1.85 & 0.01 & 5.12 & 2.20 & 4.72 & 0.24 & 1.93 & 0.01\tabularnewline
	\cline{2-16} 
	& \multirow{6}{*}{20} & 1 & \multirow{3}{*}{no} & 2.79 & 4.76 & 1.84 & 1.04 & 0.82 & 0.21 & 3.72 & 25.64 & 1.96 & 0.95 & 0.81 & 0.21\tabularnewline
	&  & 3 &  & 4.53 & 4.36 & 3.35 & 0.69 & 1.14 & 0.05 & 5.80 & 17.51 & 3.96 & 0.35 & 1.14 & 0.04\tabularnewline
	&  & 5 &  & 5.19 & 5.37 & 4.02 & 0.85 & 1.29 & 0.02 & 6.26 & 14.73 & 5.06 & 0.28 & 1.35 & 0.02\tabularnewline
	\cline{3-16} 
	&  & 1 & \multirow{3}{*}{yes} & 1.36 & 4.31 & 1.01 & 1.48 & 0.53 & 0.52 & 1.34 & 5.71 & 1.05 & 1.32 & 0.55 & 0.45\tabularnewline
	&  & 3 &  & 2.44 & 2.99 & 1.90 & 0.99 & 0.92 & 0.25 & 2.49 & 3.90 & 2.09 & 0.79 & 0.88 & 0.19\tabularnewline
	&  & 5 &  & 2.85 & 2.67 & 2.33 & 0.97 & 1.11 & 0.17 & 2.89 & 3.03 & 2.50 & 0.58 & 1.07 & 0.08\tabularnewline
	\hline 
\end{tabular}
	}
	\caption{\small The model selection performance of the robust modified BIC criteria  for $p=10000$, $n=200$, $d=2$, and all the contamination schemes.}
	\label{BIC_10000_n200_d2}
\end{table}

\begin{table}[ht!]
	\footnotesize
	\centering
	\renewcommand\arraystretch{1.25}
	\scalebox{0.85}{

\begin{tabular}{c|c|c|c|cc|cc|cc|cc|cc|cc}
	\hline 
	\multirow{2}{*}{} & \multirow{2}{*}{eps} & \multirow{2}{*}{c} & \multirow{2}{*}{LMV} & \multicolumn{2}{c|}{BIC} & \multicolumn{2}{c|}{EBIC} & \multicolumn{2}{c|}{FPBIC} & \multicolumn{2}{c|}{R-BIC} & \multicolumn{2}{c|}{R-EBIC} & \multicolumn{2}{c}{R-FPBIC}\tabularnewline
	\cline{5-16} 
	&  &  &  & TP & FP & TP & FP & TP & FP & TP & FP & TP & FP & TP & FP\tabularnewline
	\hline 
	\multirow{3}{*}{clean} & \multirow{3}{*}{0} & 1 & \multirow{3}{*}{no} & 6.13 & 4.51 & 5.47 & 0.71 & 2.29 & 0.02 & 6.24 & 4.16 & 5.70 & 0.42 & 2.35 & 0.02\tabularnewline
	&  & 3 &  & 7.25 & 2.71 & 6.91 & 0.90 & 5.26 & 0.02 & 7.34 & 1.65 & 7.20 & 0.13 & 5.56 & 0\tabularnewline
	&  & 5 &  & 7.50 & 2.52 & 7.26 & 1.02 & 6.38 & 0.12 & 7.62 & 1.31 & 7.52 & 0.05 & 6.78 & 0\tabularnewline
	\hline 
	\multirow{12}{*}{PC+LMG} & \multirow{6}{*}{5} & 1 & \multirow{3}{*}{no} & 6.05 & 4.12 & 5.41 & 0.85 & 2.30 & 0.02 & 6.17 & 4.08 & 5.60 & 0.50 & 2.40 & 0.01\tabularnewline
	&  & 3 &  & 7.23 & 2.56 & 6.88 & 0.74 & 5.24 & 0.02 & 7.35 & 1.64 & 7.19 & 0.07 & 5.57 & 0.00\tabularnewline
	&  & 5 &  & 7.45 & 2.20 & 7.26 & 0.98 & 6.33 & 0.12 & 7.58 & 1.19 & 7.47 & 0.04 & 6.72 & 0.00\tabularnewline
	\cline{4-16} 
	&  & 1 & \multirow{3}{*}{yes} & 5.81 & 4.47 & 5.20 & 0.94 & 2.46 & 0.02 & 5.86 & 4.15 & 5.39 & 0.51 & 2.41 & 0.02\tabularnewline
	&  & 3 &  & 6.99 & 2.53 & 6.69 & 0.92 & 5.00 & 0.07 & 7.09 & 1.49 & 6.93 & 0.15 & 5.40 & 0.00\tabularnewline
	&  & 5 &  & 7.26 & 2.62 & 7.04 & 1.15 & 6.09 & 0.13 & 7.36 & 1.05 & 7.28 & 0.07 & 6.57 & 0.00\tabularnewline
	\cline{2-16} 
	& \multirow{6}{*}{20} & 1 & \multirow{3}{*}{no} & 5.88 & 4.06 & 5.30 & 0.78 & 2.10 & 0.02 & 5.99 & 4.06 & 5.50 & 0.48 & 2.11 & 0.02\tabularnewline
	&  & 3 &  & 7.07 & 2.27 & 6.70 & 0.75 & 4.99 & 0.04 & 7.21 & 1.60 & 6.99 & 0.10 & 5.20 & 0.00\tabularnewline
	&  & 5 &  & 7.35 & 2.36 & 7.10 & 0.78 & 6.11 & 0.08 & 7.50 & 1.22 & 7.35 & 0.06 & 6.50 & 0.00\tabularnewline
	\cline{3-16} 
	&  & 1 & \multirow{3}{*}{yes} & 5.61 & 4.40 & 4.94 & 0.99 & 2.41 & 0.03 & 5.68 & 4.22 & 5.15 & 0.54 & 2.34 & 0.02\tabularnewline
	&  & 3 &  & 6.82 & 2.51 & 6.48 & 0.76 & 4.72 & 0.02 & 6.92 & 1.47 & 6.78 & 0.14 & 5.06 & 0.00\tabularnewline
	&  & 5 &  & 7.10 & 2.32 & 6.89 & 1.11 & 5.79 & 0.12 & 7.17 & 1.02 & 7.10 & 0.12 & 6.29 & 0.00\tabularnewline
	\hline 
	\multirow{12}{*}{PC+LMB} & \multirow{6}{*}{5} & 1 & \multirow{3}{*}{no} & 6.07 & 4.66 & 5.35 & 0.83 & 2.28 & 0.01 & 6.15 & 4.44 & 5.61 & 0.45 & 2.30 & 0.01\tabularnewline
	&  & 3 &  & 7.09 & 2.27 & 6.80 & 0.86 & 5.20 & 0.04 & 7.31 & 1.85 & 7.11 & 0.16 & 5.56 & 0.00\tabularnewline
	&  & 5 &  & 7.43 & 2.57 & 7.18 & 0.82 & 6.20 & 0.12 & 7.58 & 1.27 & 7.47 & 0.06 & 6.67 & 0.00\tabularnewline
	\cline{3-16} 
	&  & 1 & \multirow{3}{*}{yes} & 5.25 & 4.43 & 4.66 & 1.00 & 2.56 & 0.03 & 5.30 & 4.06 & 4.89 & 0.63 & 2.36 & 0.02\tabularnewline
	&  & 3 &  & 6.49 & 2.65 & 6.16 & 1.03 & 4.77 & 0.08 & 6.58 & 2.03 & 6.40 & 0.12 & 4.98 & 0.00\tabularnewline
	&  & 5 &  & 6.83 & 2.39 & 6.59 & 1.02 & 5.66 & 0.18 & 6.89 & 1.25 & 6.83 & 0.08 & 6.05 & 0.00\tabularnewline
	\cline{2-16} 
	& \multirow{6}{*}{20} & 1 & \multirow{3}{*}{no} & 5.16 & 4.62 & 4.45 & 1.02 & 1.41 & 0.02 & 5.33 & 5.47 & 4.67 & 0.77 & 1.41 & 0.02\tabularnewline
	&  & 3 &  & 6.51 & 2.74 & 6.07 & 0.87 & 3.75 & 0.02 & 6.74 & 2.97 & 6.35 & 0.20 & 3.91 & 0.00\tabularnewline
	&  & 5 &  & 6.91 & 2.93 & 6.53 & 0.96 & 4.88 & 0.03 & 7.04 & 1.97 & 6.87 & 0.13 & 5.31 & 0.00\tabularnewline
	\cline{4-16} 
	&  & 1 & \multirow{3}{*}{yes} & 4.00 & 5.08 & 3.44 & 1.25 & 1.60 & 0.11 & 4.09 & 4.96 & 3.53 & 0.99 & 1.30 & 0.08\tabularnewline
	&  & 3 &  & 5.31 & 3.27 & 4.82 & 0.90 & 2.97 & 0.05 & 5.39 & 2.74 & 5.09 & 0.33 & 3.01 & 0.01\tabularnewline
	&  & 5 &  & 5.65 & 2.81 & 5.34 & 1.08 & 3.79 & 0.07 & 5.71 & 2.14 & 5.55 & 0.26 & 3.99 & 0.00\tabularnewline
	\hline 
	\multirow{12}{*}{OC} & \multirow{6}{*}{5} & 1 & \multirow{3}{*}{no} & 5.96 & 4.20 & 5.33 & 0.75 & 2.16 & 0.02 & 6.09 & 4.14 & 5.53 & 0.46 & 2.21 & 0.02\tabularnewline
	&  & 3 &  & 7.13 & 2.72 & 6.74 & 0.75 & 5.06 & 0.03 & 7.26 & 1.77 & 7.08 & 0.10 & 5.36 & 0.00\tabularnewline
	&  & 5 &  & 7.39 & 2.61 & 7.11 & 0.78 & 6.21 & 0.10 & 7.55 & 1.37 & 7.42 & 0.06 & 6.66 & 0.00\tabularnewline
	\cline{4-16} 
	&  & 1 & \multirow{3}{*}{yes} & 5.62 & 4.56 & 4.92 & 0.86 & 2.39 & 0.03 & 5.70 & 4.22 & 5.14 & 0.52 & 2.37 & 0.02\tabularnewline
	&  & 3 &  & 6.86 & 2.52 & 6.53 & 0.89 & 4.82 & 0.05 & 6.97 & 1.47 & 6.82 & 0.14 & 5.07 & 0.00\tabularnewline
	&  & 5 &  & 7.19 & 2.60 & 6.98 & 1.22 & 5.90 & 0.12 & 7.28 & 1.06 & 7.20 & 0.08 & 6.42 & 0.00\tabularnewline
	\cline{2-16} 
	& \multirow{6}{*}{20} & 1 & \multirow{3}{*}{no} & 5.33 & 4.49 & 4.54 & 0.84 & 1.61 & 0.03 & 5.52 & 5.72 & 4.79 & 0.65 & 1.60 & 0.03\tabularnewline
	&  & 3 &  & 6.74 & 3.58 & 6.23 & 0.85 & 4.17 & 0.03 & 7.07 & 3.23 & 6.67 & 0.16 & 4.48 & 0.00\tabularnewline
	&  & 5 &  & 7.16 & 4.11 & 6.69 & 1.21 & 5.40 & 0.09 & 7.41 & 2.43 & 7.20 & 0.13 & 5.92 & 0.00\tabularnewline
	\cline{3-16} 
	&  & 1 & \multirow{3}{*}{yes} & 4.10 & 4.12 & 3.61 & 1.14 & 1.83 & 0.12 & 4.13 & 3.86 & 3.72 & 0.81 & 1.59 & 0.07\tabularnewline
	&  & 3 &  & 5.39 & 2.76 & 5.06 & 0.88 & 3.43 & 0.07 & 5.46 & 2.07 & 5.29 & 0.26 & 3.45 & 0.00\tabularnewline
	&  & 5 &  & 5.80 & 2.47 & 5.51 & 1.02 & 4.16 & 0.13 & 5.90 & 1.65 & 5.74 & 0.11 & 4.56 & 0.00\tabularnewline
	\hline 
\end{tabular}
	}
	\caption{\small The model selection performance of the robust modified BIC criteria  for $p=10000$, $n=400$, $d=2$, and all the contamination schemes.}
	\label{BIC_10000_n400_d2}
	
\end{table}
	
\section{Real Data Analysis}
	
We analyze a dataset which contains gene expression measurements of 31099 genes on eye tissues from 120 12-week-old male F2 rats. The data is available at \url{https://www.ncbi.nlm.nih.gov/geo/query/acc.cgi?acc=GSE5680}. The gene coded as TRIM32 is of particular interest for its causal effect on the Bardet-Biedl syndrome. As in~\citep{GSE2006}, the 18976 genes which exhibit at least a two-fold variation in expression level are included for analysis. It is believed that TRIM32 is associated with a small number of other genes. We consider a multiple regression with TRIM32 as response to identify these genes, which results in an ultra-high dimensional regression problem. 
	
\begin{figure}[ht!]
	\centering
	\begin{minipage}{0.45\textwidth}
		\small
		(a)	
	\end{minipage}
	\begin{minipage}{0.45\textwidth}
		\small
		(b)
	\end{minipage}
	\begin{minipage}{0.45\textwidth}
		\centering
		\includegraphics[width= 6.5 cm]{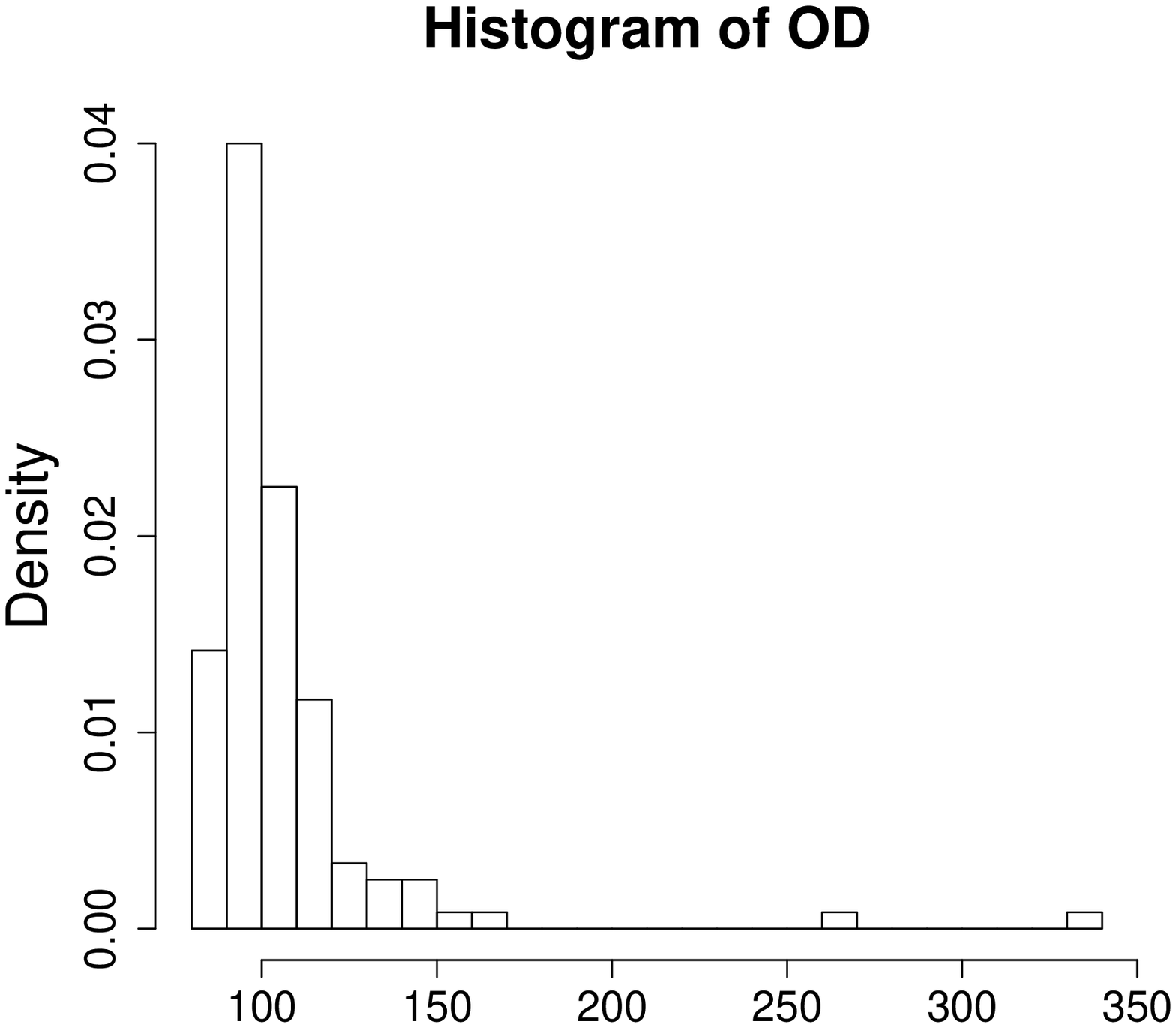}
	\end{minipage}
	\begin{minipage}{0.45\textwidth}
		\centering
		\includegraphics[width= 6.5 cm]{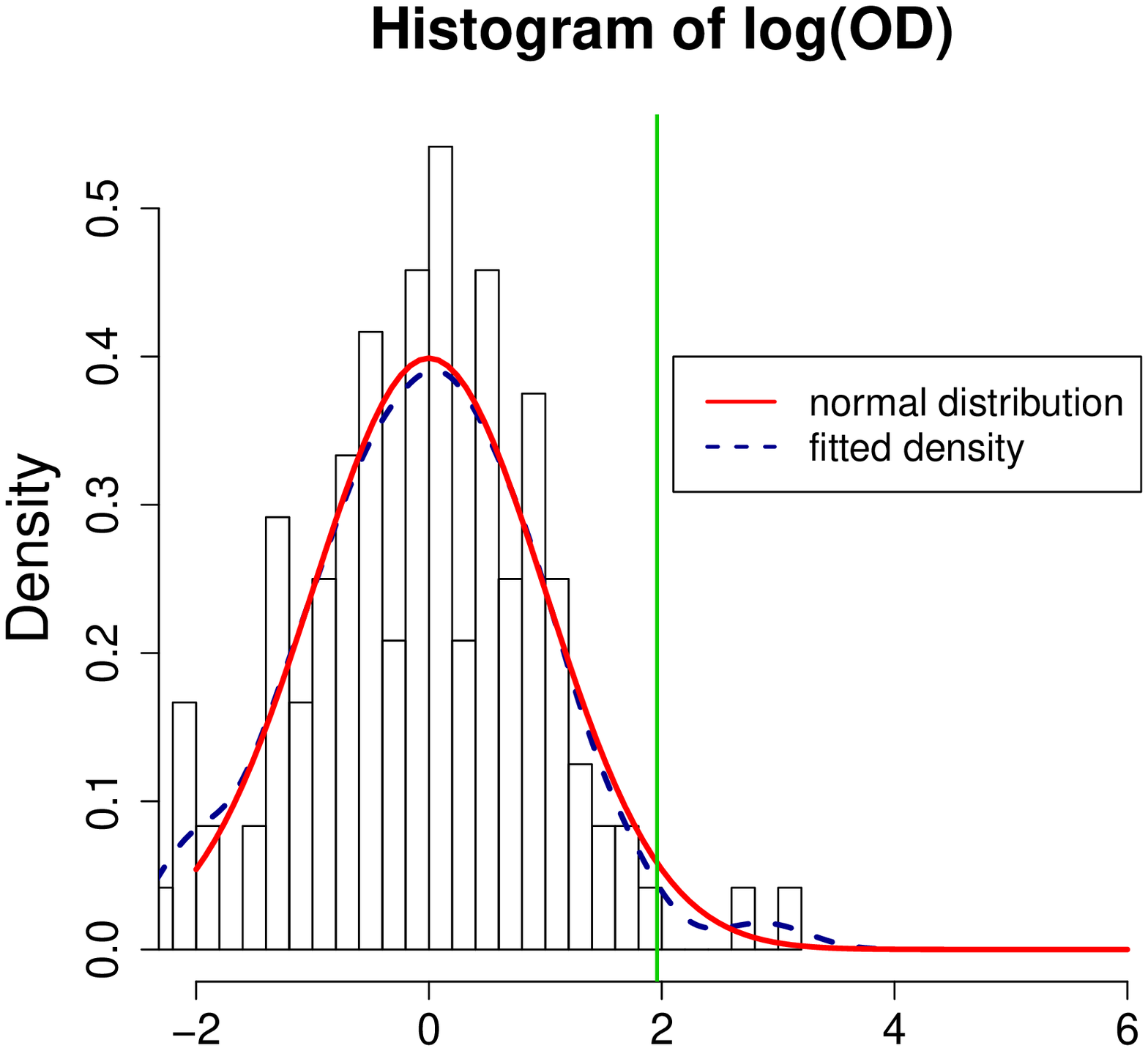}
	\end{minipage}
	\caption{\small The histogram of $\text{\small OD}$, and 
		$\protect \log(\text{\small OD})$ for the rat genome data.}
	\label{hist_t}
\end{figure}

\begin{figure}[ht!]
	\centering
	\includegraphics[width = 6.5 cm]{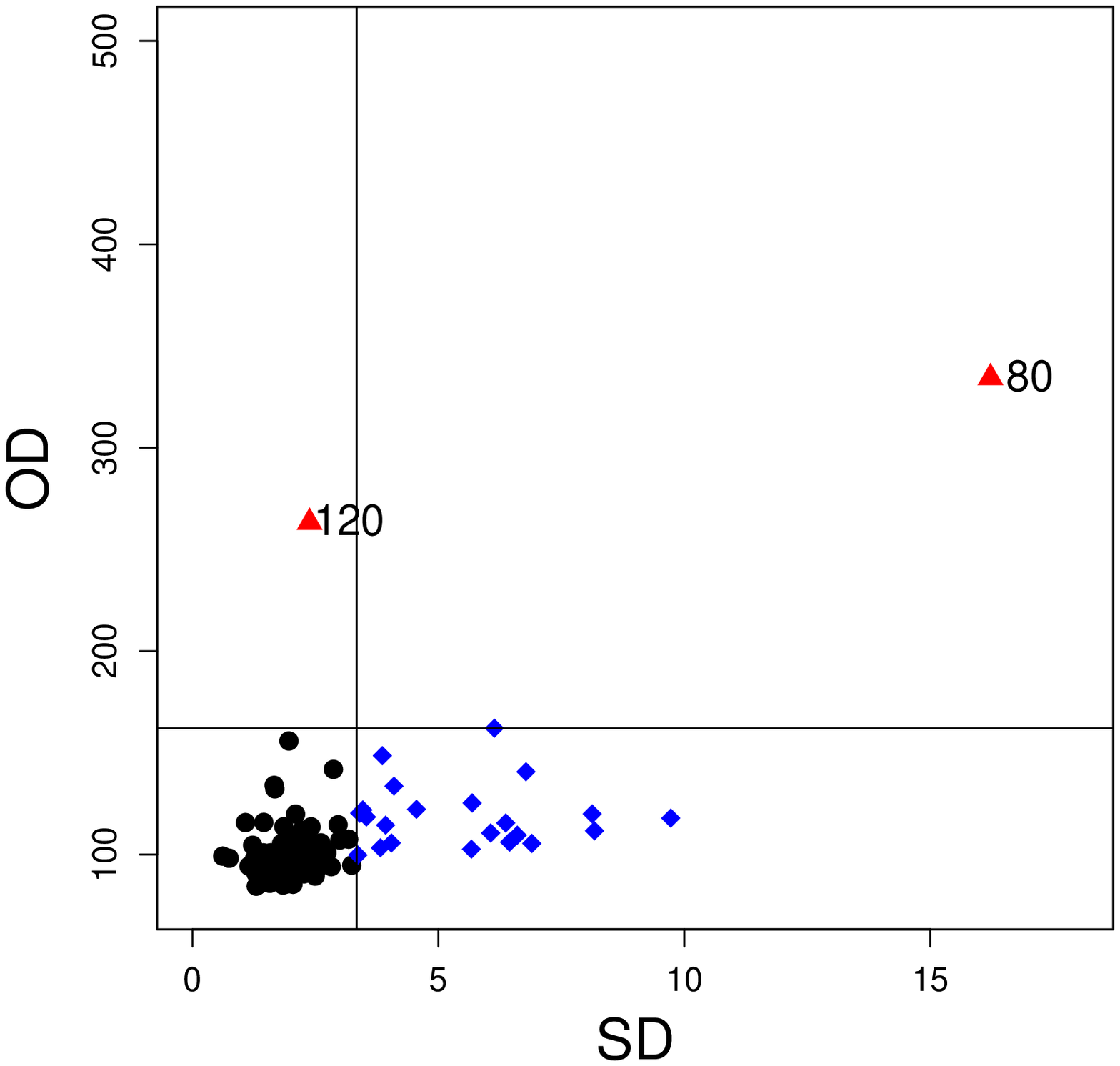}
	\caption{\small The diagnostic plot of the rat genome data showing the clean observations ($\protect \bullet$), the PC outliers ({$\protect \MyDiamond[draw=blue,fill=blue]$}) the OC outliers ({\color{red} $\protect \blacktriangle$}).} 
	\label{diagnostic_t}
\end{figure}

\begin{figure}[ht!]
	\centering
	\begin{minipage}{0.45\textwidth}
		\centering
		\includegraphics[width=4.5 cm]{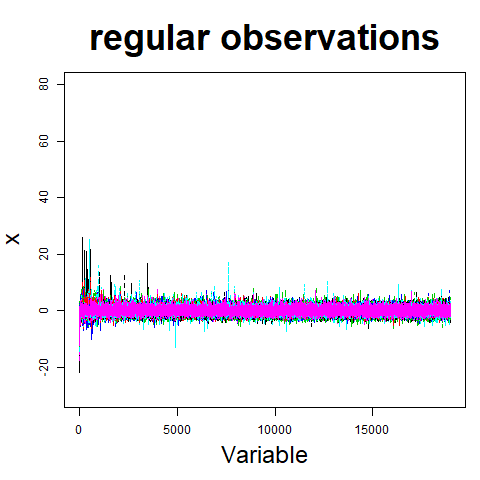}
	\end{minipage}
	\begin{minipage}{0.45\textwidth}
		\centering
		\includegraphics[width=4.5 cm]{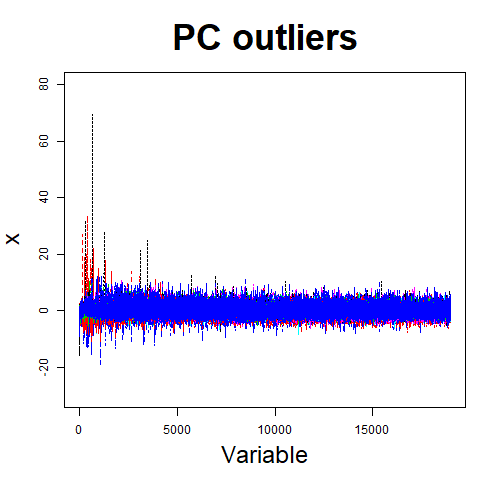}
	\end{minipage} \\
	\begin{minipage}{0.45\textwidth}
		\centering
		\includegraphics[width=4.5 cm]{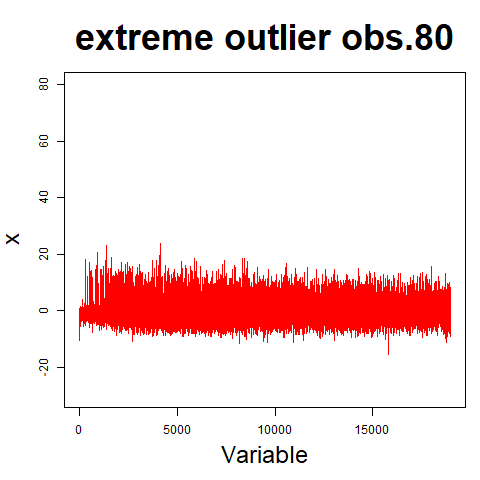}
	\end{minipage} 
	\begin{minipage}{0.45\textwidth}
		\centering
		\includegraphics[width=4.5 cm]{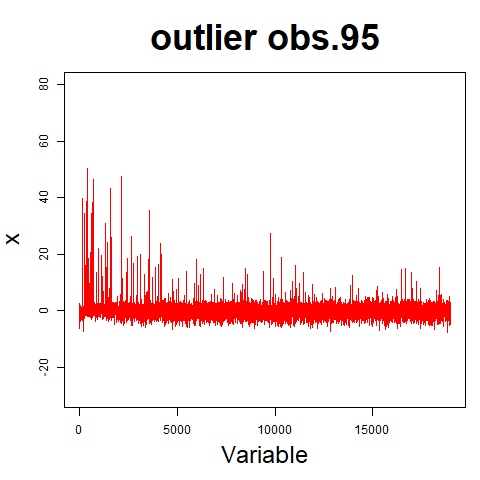}
	\end{minipage}
	\caption{\small The plot of the original variables for the clean observations, the PC outliers, and the OC outliers (obs. 80 and 95) in the rat genome data.}
	\label{matplot_95}
\end{figure}

To identify the most important genes, we apply the RFPSIS method of Section~\ref{sec:RFPSIS} with $h=[(n-d+2)/2]$ for maximal robustness. The variables are first standardized using their median and $Q_n$ scale estimate. Based on criterion \eqref{PC_Select}, the number of factors is estimated to be 4. The robust Yeo-Johnson transformation selects $\lambda=0$, so a logarithmic transformation is applied on the orthogonal distances. The histogram of both the $\text{\small OD}_i$ and $\log(\text{\small OD}_i)$ are shown in Figure~\ref{hist_t}. After applying the logarithmic transformation, the orthogonal distances can clearly be approximated much better by a normal distribution. 
Based on the corresponding diagnostic plot, shown in Figure \ref{diagnostic_t}, we can see that observations
80 and 95 are identified as OC outliers while there are also 21 observations identified as PC outliers. To examine these outliers further, we compare the measurements of all genes in the analysis for the clean observations to the PC and OC leverage points in Figure \ref{matplot_95}. From these plots we can see that the OC outliers show more variation than the remaining data. Hence, these plots indeed confirm that the OC outliers identified in the diagnostic plot show a behavior that is different from the majority.

RFPSIS applied on the full dataset, denoted by {\it rat1}, identified 11 of the PC outliers as bad leverage points, while the other 10 PC outliers are considered to be good leverage points, and thus are included in the variable screening. For comparison, we also consider two reduced datasets. We call {\it rat2} the dataset which contains all the observations except the extreme outlier (obs. 80) identified in Figure~\ref{diagnostic_t}. Finally, {\it rat3} is the reduced dataset obtained by removing the 2 OC outliers as well as the 11 bad leverage PC outliers identified by RFPSIS. We then apply SIS and FPSIS on all three datasets and compare the results with those of RFPSIS on the full dataset ({\it rat1}). We thus obtained 7 solution paths. For convenience, we denote by (FP)SIS({\it rat1}), (FP)SIS({\it rat2}) and (FP)SIS({\it rat3}) the solution path that is obtained when applying (FP)SIS on dataset {\it rat1}, {\it rat2} and {\it rat3}, respectively.
	 
To compare how successfully SIS, FPSIS and RFPSIS screen out the most relevant predictors, we calculate for each solution path the minimally obtainable median of absolute 10-fold cross-validation prediction error. Note that the 10-fold cross-validation prediction errors, denoted by 10-fold-MAPE, are averages over 100 random splits of the data. Hence, for each of the 7 solution paths, we regress the response, TRIM32, on the first $k$ ($k=1,\ldots,50$) variables in the path using MM-estimators. For each solution path, the smallest mean 10-fold-MAPE among the 50 models is reported in Table \ref{gse_cross_mse_05_t}, and Table \ref{gse_cross_k_05_t} contains the corresponding model size $k$, i.e. the number of predictors in the model with smallest mean 10-fold-MAPE.

\begin{table}[ht!]
	\footnotesize
	\centering
	\renewcommand\arraystretch{1.2}
	\scalebox{0.95}{
 \begin{tabular}{c|ccccccc}
	\hline 
	& RFPSIS & SIS ({\scriptsize \it rat1}) & SIS ({\scriptsize \it rat2}) & SIS ({\scriptsize \it rat3}) & FPSIS ({\scriptsize \it rat1}) & FPSIS ({\scriptsize \it rat2}) & FPSIS ({\scriptsize \it rat3})\tabularnewline
	\hline 
	{\it rat1} & 0.3470  & 0.4710  & 0.4478 & 0.4456  & 0.5775 & 0.5640  & 0.4375  \tabularnewline
	{\it rat2} & 0.3416 & 0.4651  & 0.4369 & 0.3996 & 0.5736 & 0.5396 & 0.4348   \tabularnewline
	{\it rat3} & \bf{0.3359} &  0.4064  & 0.4064 & 0.3608  & 0.4597  & 0.4969 & \bf{0.3375}  \tabularnewline
	\hline 
\end{tabular}
	}
	\caption{\small The smallest mean 10-fold-MAPE fitting the first $k$ ($k=1,\protect \ldots,50$) variables in the 7 solution paths and evaluated on the three rat datasets.}
	\label{gse_cross_mse_05_t}
	
\end{table}

\begin{table}[ht!]
	
	\footnotesize
	\centering
	
	\renewcommand\arraystretch{1.2}
	\scalebox{0.95}{
\begin{tabular}{c|ccccccc}
	\hline 
	& RFPSIS & SIS ({\scriptsize \it rat1}) & SIS ({\scriptsize \it rat2}) & SIS ({\scriptsize \it rat3}) & FPSIS ({\scriptsize \it rat1}) & FPSIS ({\scriptsize \it rat2}) & FPSIS ({\scriptsize \it rat3})\tabularnewline
	\hline 
	{\it rat1} & 8  & 4   & 14   & 7  & 9  & 25 &  7 \tabularnewline
	{\it rat2} & 8 & 13    & 14   & 7  & 18  & 25 &  7   \tabularnewline
	{\it rat3} & {\bf 8} &  4  & 4  & 5  & 10  &  8  & {\bf 12} \tabularnewline
	\hline 
\end{tabular}
	}			
	\caption{\small The model sizes with respect to the smallest mean 10-fold-MAPE fitting the first $k$ ($k=1,\protect \ldots,50$) variables in the 7 solution paths and evaluated on the three rat datasets.}
	\label{gse_cross_k_05_t}
	
\end{table}

Comparing the result of RFPSIS with the results of SIS and FPSIS, we can see from Table \ref{gse_cross_mse_05_t} that RFPSIS and FPSIS({\it rat3}) produce the smallest 10-fold-MAPE's for all three datasets, showing that both methods select the most relevant variables. Since we are particularly interested in predicting well the non-outliers, let us consider the 10-fold-MAPE evaluated on the reduced dataset {\it rat3}. Clearly, RFPSIS gives the best 10-fold-MAPE which is 0.3359 for the regular observations. FPSIS({\it rat3}) gives a very close result which is 0.3375 for the regular observations in {\it rat3}, but the optimal model contains 12 predictors rather than only 8 for the model selected by RFPSIS as can be seen from Table \ref{gse_cross_k_05_t}. 

\begin{table}[ht!]
	\footnotesize
	\centering
	\hspace{6 pt}
	\renewcommand\arraystretch{1.2}
	 \begin{tabular}{c|cccc}
		\hline 
		 &  MM-LASSO-50 & MM-LASSO-full & (R-)BIC & (R-)EBIC/(R-)FPBIC  \tabularnewline
		\hline
		k & 21.66 (3.35) & 62.94 (20.41) & 4 & 1    \tabularnewline
		{10-fold-MAPE} & 0.2934 (0.02) & 0.4814 (0.35) & 0.4894 & 0.4568  \tabularnewline
		\hline 
	\end{tabular}
	\caption{\small The model size and 10-fold-MAPE evaluated on the clean observations ({\it rat3}) of the models selected by MM-LASSO-50, MM-LASSO-full, and the six BIC criteria from the RFPSIS solution path.}
	\label{gse_cross_BIC_05_t}
\end{table}

Comparing (FP)SIS({\it rat3}) with (FP)SIS({\it rat1}) and (FP)SIS({\it rat2}), we can conclude that removing the potential outliers significantly improves the predictions for the regular observations in {\it rat3}. Moreover, the smaller 10-fold-MAPE given by FPSIS({\it rat3}) than SIS({\it rat3}) indicates that there exists correlation among the predictors which allows FPSIS to perform better. When there are outliers,  FPSIS({\it rat1}) and FPSIS({\it rat2}) give much worse results than SIS({\it rat1}) and SIS({\it rat2}) since the outliers in these datasets distort the correlation structure estimated by FPSIS. On the other hand, RFPSIS can correctly estimate the correlation structure of the regular data from the full dataset and thus yields similar results as FPSIS applied to the reduced dataset {\it rat3}. 

We also applied MM-LASSO (\citep[]{Smucler2015}) on the full dataset. First we considered all 18976 variables and then we only considered the first 50 variables from the solution path given by RFPSIS. We denote the two models by MM-LASSO-full and MM-LASSO-50, respectively. Due to the randomness of 5-fold-cross-validation for the selection of the optimal value of the regularization parameter, we run MM-LASSO 50 times for each setting. Then, we compute the 10-fold-MAPE when fitting MM-regression with the selected predictors on {\it rat3}. The average number of selected predictors and the resulting 10-fold-MAPE's, with their standard errors, are displayed in Table~\ref{gse_cross_BIC_05_t}. It can be seen that the MM-LASSO-50 model yields a smaller 10-fold-MAPE than the model with the first 8 variables from the solution path of RFPSIS obtained previously. 
To obtain this result, MM-LASSO-50 selects much larger models with around 24 predictors.  MM-LASSO-50 yields very stable results as can be seen from the small standard error for the 10-fold-MAPE. On the other hand, MM-LASSO-full selects even much more variables which results in much larger and unstable 10-fold-MAPE's. 
Moreover, applying MM-LASSO on the dataset with all 18976 variables is much more time consuming. For example, it took on average (over the 50 runs) 10.28 minutes to run MM-LASSO-full in \texttt{R}~\citep{Rcore}  on an Intel Core i7-4790 X64 at 3.6 GHz, while running  MM-LASSO-50 only required 36.58 seconds on average and the initial RFPSIS screening took 42.84 seconds. This illustrates that for ultrahigh-dimensional data, initial screening also yields a big advantage both in terms of performance and computation time when penalized regression methods such as MM-LASSO are used.

In Section~\ref{sec: BIC} we noticed that the BIC type criteria tend to be too parsimonious when the signal-to-noise ratio in the data is low. When using $\tilde{k}_\text{max} = 50$, EBIC and FPBIC, and their re-ordered versions, only select the first predictor in the solution path for this dataset. BIC and R-BIC yield a bit less parsimonious model consisting of the first four predictors in the solution path. 
We again focus on the prediction errors for the regular observations in the reduced dataset ({\it rat3}). The model size and 10-fold-MAPE for the selected models by the different BIC criteria are shown in Table \ref{gse_cross_BIC_05_t}. It can be seen that the model with only the first predictor produces a smaller 10-fold-MAPE than the model with the first four predictors selected by BIC and R-BIC. Furthermore, we found that the first predictor in the solution path was consistently selected by MM-LASSO-50 across the 50 runs. Therefore, we can conclude the model obtained by (R-)EBIC and (R-)FPBIC identified the most important predictor, which can be a good starting point for further analysis.

\section{Conclusions}
Sure Independence Screening has aroused a lot of research interest recently due to its simpleness and fastness. It has been proven that SIS performs well with orthogonal or weakly dependent predictors and a sufficiently large sample size. However, its performance deteriorates greatly when there is substantial correlation among the predictors. To handle this problem, FPSIS removes the correlations by projecting the original variables onto the orthogonal complement of the subspace spanned by the latent factors which capture the correlation structure. However, FPSIS is based on classical estimators which are nonrobust and thus cannot resist the adverse influence of outliers.
		
In this paper we investigated the effect of both vertical outliers and leverage points in the original multiple regression model. Our proposed RFPSIS estimates the latent factors by an LTS procedure. We considered leverage points due to both orthogonal complement outliers and score outliers in the subspace for the factor model, and examined their effect on the marginal regressions with factor profiled variables. 
It turned out that only good leverage points caused by PC outliers do not affect the variable screening results. Hence, RFPSIS only includes this type of good leverage points in the marginal screening to increase efficiency. Moreover, to reduce the influence of potential outliers, the marginal coefficients are estimated using MM-estimators. Our simulation studies showed that RFPSIS is almost as accurate as FPSIS on regular datasets, and at the same time can resist the adverse influence of all types of outliers, while both SIS and FPSIS fail in presence of outliers. 

In Section~\ref{sec: BIC}, we investigated the performance of six BIC criteria to select a final model from the solution path of RFPSIS. Our results indicate that R-EBIC, the EBIC criterion applied to the reordered variable sequence, generally yields the best model. However, for very noisy datasets it may lead to over-sparsified models. Instead of using these information criteria, regularized robust regression methods can be used to select the final model as shown in the real data analysis. Determining the final model after initial screening to determine the most promising predictors is a problem that deserves more attention to further improve selection results.

Similar as FPSIS, RFPSIS is built on the strong assumption that the correlations among the predictions can be fully modeled by a few latent factors. In this case the correlations among the predictors can be removed by factor profiling. Similar technique has been applied to de-correlate covariates in high-dimensional sparse regression~\citep{Fan2016_FAD} and it was stated that Factor Adjusted Decorrelation (FAD) pays no price in case of weakly or uncorrelated covariates. 
When there are weakly correlated predictors, i.e. weak correlations among the predictors that cannot be removed by factor profiling, similar procedure as those to improve SIS, e.g. Iterative SIS~\citep{SIS} or Conditional SIS~\citep{CSIS}, can be applied on the robustly profiled variables in RFPSIS to improve its performance. This could be an interesting topic for future research.

While RFPSIS can effectively handle all types of outlying observations, it does require a majority of regular observations in the dataset. However, for high-dimensional data it is not always realistic to assume that there is a majority of completely clean observations. Therefore, alternative contamination models can be considered, such as the {\it fully independent contamination model} which assumes that each of the variables is independently contaminated by some fraction of outliers~\citep{Propout}. In high-dimensional data, even a small fraction of such cellwise outliers in each variable leads to a majority of observations that is contaminated in at least one of its components. Similarly as in~\citep{CoLTSPCA}, a componentwise  least trimmed squares objective function can be used to estimate the correlation structure. Such a loss function does not require the existence of a majority of regular observations. In future work, we will extend RFPSIS by combining this estimator of the factor structure with the use of marginal regressions for variable screening to handle data with cellwise outliers. 

In high-dimensional data analysis, another difficult situation might be that the outliers are hard to detect due to the presence of abundant noisy variables or due to the complex correlation structure of the features. Hence, searching for a lower dimensional projection subspace, called High Contrast Subspace (HiCS) by~\citep{HiCS}, in which outliers can be distinguished from the regular data, or selecting features which contribute most to the outlyingness of observations, as done by Coupled Unsupervised Feature Selection (CUFS)~\citep{CUFS}, would be crucial to detect outliers. In these cases, combining feature selection for outlier detection and for sparse estimation can be very challenging, and deserves more research attention. 

\section*{Acknowledgments}
This research was supported by grant C16/15/068 of International Funds KU Leuven and 
COST Action IC1408 CRoNoS. Their support is gratefully acknowledged.

\bibliographystyle{plainnat}
\clearpage
\bibliography{WangVanAelst_RFPSIS_SADM_2018}
%
\end{document}